\documentclass[12pt,titlepage]{article}
\usepackage{graphicx}
\usepackage{graphics}
\usepackage{geometry}
\usepackage{amsmath}
\usepackage{amsfonts}
\usepackage{amssymb}
\usepackage{latexsym}
\usepackage{setspace}
\usepackage{url}

\usepackage{natbib}

\usepackage{color}

\newcommand{\ctn}{\cite}
\newcommand{\bi}[1]{\mbox{\boldmath{$ #1 $}}}
\begin{document}

\title{Nonparametric Dynamic State Space Modeling of Observed Circular Time Series with Circular Latent States\/: 
A Bayesian Perspective}
\author{Satyaki Mazumder\footnote{Indian Institute of Science Education and Research Kolkata} and 
Sourabh Bhattacharya\footnote{Indian Statistical Institute, Kolkata.}}
\date{\vspace{-0.5in}}
\maketitle

\begin{abstract}
Circular time series has received relatively little attention in statistics and
modeling complex circular time series using the state space approach is non-existent in the literature.
In this article we introduce a flexible Bayesian nonparametric approach to 
state space modeling of observed circular time series where even the latent states are circular
random variables. Crucially, we assume that the forms of the observational and evolutionary functions,
both of which are circular in nature, are unknown and time-varying. We model these unknown circular
functions by appropriate wrapped Gaussian processes 
having desirable properties.

We develop an effective Markov chain Monte Carlo strategy for implementing our Bayesian model by judiciously 
combining Gibbs sampling and Metropolis-Hastings methods. Validation of our ideas with a simulation study
and two real bivariate circular time series data sets, where we assume one of the variables to be unobserved, 
revealed very encouraging performance of our model and methods. 


We finally analyse a data consisting of directions of whale migration, 
considering the unobserved ocean current direction as the latent circular process of interest.
The results that we obtain are encouraging, and the
posterior predictive distribution of the observed process correctly predicts the observed whale movement.
\\[2mm]
\end{abstract}

\tableofcontents

\pagebreak

\section{Introduction}
\label{intro}

In contrast with the traditional time series where the variables take values on the real line,
circular time series with angular variables have received much less attention in the statistical
literature. Some of the relatively scarce undertakings on circular time series, all in the classical
statistical framework, are \ctn{Breckling89}, \ctn{Fisher94}, \ctn{Holzmann06}, \ctn{Hughes07} and \ctn{Marzio12}.
The classical approach, however, is limited to models with relatively simple dependence structures, such
as the $p$-th order Markov dependence,
to enable inference. Far more flexible and realistic models can be accommodated within the Bayesian paradigm
without any compromise; the inference can be carried out arbitrarily accurately using powerful Markov 
chain Monte Carlo (MCMC) methods. In this article, we propose and develop a novel and highly flexible 
approach to modeling circular time series
using a Bayesian nonparametric state space approach, and an effective MCMC methodology for Bayesian inference. 

Indeed, versatility of state space structures in modeling complex time series is well-appreciated in various
scientific disciplines such as statistics, engineering, economics, biology, ecology, finance, and most
of the existing time series models admit the state space representation.
Notably, state space models consist of an observational equation for the observed part of the time series
and an evolutionary equation that accounts for the latent, unobserved process that is presumed to affect
the observed process. In this work, we shall concern ourselves with Bayesian state space modeling of complex 
circular time series data, where the latent process is also circular in nature, and where
the observational and evolutionary functions are time-varying, but unknown. 
We model both the unknown functions nonparametrically using appropriate wrapped Gaussian 
process priors developed by \ctn{Mazumder16}. 
For our state space model, the circular observations are conditionally independent given
the latent states and the observational function, and the latent states have the Markov property conditionally
on the evolutionary function. 
Unconditionally, however, the latent circular states are realizations of a complex, intractable stochastic process possessing 
a complex, intractable dependence structure in a way that all the latent states depend upon each other. 
The observed circular process is also intractable unconditionally, with a dependence structure that makes 
all the observational variables interdependent. 
However, the attractive properties of the wrapped Gaussian process prior ensure that the dependence structure,
although complex, is very reasonable.
Thus, our Bayesian state space modeling approach frees the circular
time series paradigm of the limitations of the classical approach, while the MCMC procedure we propose, takes care
of the inference part.

It is worth noting that \ctn{Mazumder16} developed a Bayesian nonparametric state space approach to modeling observed 
linear time series, assuming circular latent process. The obvious difference of our current
work with theirs is that in our case, even the observed time series is circular. 
This circular extension of the observed process, however, is not insignificant, and throws much difficult challenges compared
to the approach of \ctn{Mazumder16}. In particular, because of our Gaussian process approach, 
the joint distribution of the observed data conditional on the latent states and the other parameters, 
is a product of wrapped normal distributions, for which closed forms are not available, and as many discrete
auxiliary variables as the data size are needed to be introduced to express the likelihood conditional on
the auxiliary variables. To make things
harder, these wrapped normal distributions, even conditionally on the auxiliary variables, 
involve the unknown observational function in a way that it 
can not be integrated out as in \ctn{Mazumder16}. Indeed, thanks to the linear nature of the observed process, 
\ctn{Mazumder16} could marginalize over the unknown observational function to arrive at a multivariate normal 
distribution of the observed data conditionally on the latent states and other parameters. 

Combination of our involved conditional distribution of the observed data with the heavily involved
form of the joint distribution of the state variables yields a structure that 
is way beyond the scope of \ctn{Mazumder16}. It is thus clear that in comparison with \ctn{Mazumder16}, 
a substantially different and involved MCMC method is also required for the implementation of our approach. 

Once we build the necessary Bayesian nonparametric model and methods, we evaluate them on a simulated data generated from
a state space model with specific circular observational and circular evolutionary equations, but fitted with our proposed
model and methods. We also consider two real data sets for validation of our ideas. In one case, wind directions
at $6.00$ am and $12.00$ pm on each of 21 days, are recorded. Pretending that the wind directions at $6.00$ am are unobserved,
we obtain the posteriors of these circular latent states and compare it with the actual observed values. 
The second data set for our validation purpose consists of spawning time of a particular fish and the low
tide times, both measured on the 24-hour clock, for 86 days. In this case, we assume that the low tide times are 
unobserved, and fit our Bayesian model, viewing the times as circular variables. In all the three cases, we
obtained excellent predictions of the latent circular states and excellent forecasts of both latent and observed
states. 

We then apply our ideas to a real data set consisting of only the directions associated with movements of a whale
in course of its migration. The directions of ocean current, which are presumed to influence the movements
of the whale, are not observed, and hence, we consider them to be the latent circular states of our interest.  
Our Bayesian model and methods yielded quite sensible results on the latent states and correct prediction 
of the whale movement.

The rest of this article is structured as follows. In Section \ref{Univar} we introduce our Bayesian state
space model with circular observed data and circular latent states, discussing our wrapped Gaussian process prior
and explicitly providing the form of the Bayesian hierarchical model, and the forms of the priors on the parameters.
In Section \ref{inference} we briefly discuss the main technical issues distinguishing our MCMC
method with that of \ctn{Mazumder16}, showing that our present MCMC methodology is far more involved.
The details of our MCMC procedure are provided in the supplement \ctn{Mazumdersupp16}.
We then validate our model and methods with a simulation study in Section \ref{simulation_study}
and two real data sets on pairs of wind directions and spawning/low tide times in Section \ref{real_data_analysis}.
We apply our ideas to the whale data in Section \ref{sec:whale_data}, obtaining the posteriors of the ocean
current directions and forecasting whale movement. Finally, we summarize our work and provide concluding remarks
in Section \ref{conclusion}.

\section{Dynamic state space model with circular latent and observed states}
\label{Univar}
Our proposed state space model is as follows\/:
For $t=1,2,\ldots T$, 
\begin{align}
 y_t &= \left\{f(t,x_t) + \epsilon_t\right\}~[2\pi], ~~ \epsilon_t\sim N(0,\sigma^2_{\mbox{\scriptsize{$\epsilon$}}}),
 \label{eq:eq1}
\\[1ex]
x_t &=  \left\{g(t,x_{t-1}) + \eta_t\right\}~[2\pi], ~~ \eta_t\sim N(0,\sigma^2_{\mbox{\scriptsize{$\eta$}}}),
\label{eq:eq2}
\end{align}
where $\{y_t;~t=1,\ldots,T\}$ is the observed circular time series; $\{x_t;~t=0,1,\ldots,T\}$ are the latent
circular states; $f(\cdot,\cdot)$ and $g(\cdot,\cdot)$ are the unknown evolutionary functions with values on the circular manifold. 
In equations (\ref{eq:eq1}) and (\ref{eq:eq2}), $[2\pi]$ stands for the $\mbox{mod}~2\pi$ operation.
Note that 
\begin{align}
\left\{f(t,x_{t}) + \epsilon_t\right\}~[2\pi]
&=\left\{f^*(t,x_{t}) + \epsilon_t\right\}~[2\pi]
\label{eq:modulo_f}\\
&\mbox{ and }\notag\\
\left\{g(t,x_{t-1}) + \eta_t\right\}~[2\pi]
&=\left\{g^*(t,x_{t-1}) + \eta_t\right\}~[2\pi],
\label{eq:modulo_g}
\end{align}
where $f^*$ and $g^*$ are the linear counterparts of $f$ and $g$, that is, 
$f^*(t,x_{t})$ is the linear random variable
such that $f^*(t,x_{t})~[2\pi]=f(t,x_{t})$ and similarly, $g^*(t,x_{t-1})~[2\pi]=g(t,x_{t-1})$.
For convenience, we shall frequently use the representations (\ref{eq:modulo_f}) and (\ref{eq:modulo_g}). 
For more details on such a representation the reader is referred to \ctn{Mazumder16}. Indeed, 
to derive the distributions of $y_t$ and $x_t$, we shall first obtain the distributions of the linear random variables $f^*(t,x_{t}) + \epsilon_t$ and $g^*(t,x_{t-1}) + \eta_t$ and then apply 
the $\mbox{mod}~2\pi$ operation to these variables 
to derive the distribution of the circular variables $y_t$ and $x_t$.

Note that both the observational and evolutionary functions have the linear argument $t$ and the angular argument $x$. 
The linear argument guarantees that the functions are time-varying, that is, the functions are allowed 
to freely evolve with time. 

\subsection{Wrapped Gaussian process representations of the observational and evolutionary functions}
\label{subsec:gp1}
We now provide a brief discussion on modeling the unknown circular functions $f$ and $g$. 
\ctn{Mazumder16} have constructed a wrapped Gaussian process by convolving a suitable kernel with 
standard Wiener process as a function of time and directions. The created wrapped Gaussian process has 
the desirable properties; for example, the covariance becomes 0 whenever the difference between the angles becomes $\pi/2$, 
implying orthogonality of the directions, irrespective of the value of the time difference. 
Similarly, as the time difference goes to $\infty$, the covariance converges to 0. Moreover, the desired 
continuity and smoothness properties do hold for the constructed Gaussian process (the proofs 
are provided in the supplement of \ctn{Mazumder16}). While constructing such a wrapped Gaussian process the basic idea was 
to first construct a Gaussian process as a function of time and angle and then apply the mod $2\pi$ operation. 
Here also, to model $f$ and $g$, we first model $f^*$ and $g^*$, the linear parts of $f$ and $g$, using the 
Gaussian process, created in \ctn{Mazumder16}. The mean functions of $f^*$ and $g^*$ are of the forms 
$\mu_f(\cdot,\cdot)$ = $\bi{h}(\cdot,\cdot)'\bi{\beta}_f$ and $\mu_g (\cdot,\cdot)$ = $\bi{h}(\cdot,\cdot)'\bi{\beta}_g$, 
where $\bi{h}(t,z)$ = $(1,t,\cos(z),\sin(z))'$; here $z$ is an angular quantity and $\bi{\beta}_f$, 
$\bi{\beta}_g$ are parameters in $\mathbb{R}^4$. For any fixed $(t_1,z_1)$ and $(t_2,z_{2})$, 
where $t_1,t_2$ are linear quantities and $z_1,z_2$ are angular quantities, the covariance structures are 
given by $c_f((t_1,z_{1}),(t_2,z_{2}))$ = $\exp \{-\sigma_f^4(t_1-t_2)^2\}\cos(|z_{1}-z_{2}|)$ 
and $c_g((t_1,z_{1}),(t_2,z_{2}))$ = $\exp \{-\sigma_g^4(t_1-t_2)^2\}\cos(|z_{1}-z_{2}|)$, 
where $\sigma_f$ and $\sigma_g$ are positive, real valued parameters. For details, see \ctn{Mazumder16}. 
After modeling $f^*$ and $g^*$ by the aforementioned Gaussian process, 
we derive the wrapped Gaussian distribution of $f(\cdot,\cdot)+\epsilon_t$ and $g(\cdot,\cdot)+\eta_t$ 
using the mod $2\pi$ operation on the distributions of $f^*(\cdot,\cdot)+\epsilon_t$ and $g^*(\cdot,\cdot)+\eta_t$, respectively. 

\subsection{Bayesian hierarchical structure of our proposed model}
\label{subsec:hierarchy}
Our modeling approach can be described succinctly by the following hierarchical Bayesian representation\/:
\allowdisplaybreaks
{
\begin{align}
[y_t|f,\bi{\theta}_f,x_t]&\sim N\left(f^*(t,x_t),\sigma^2_{\epsilon}\right)~[2\pi];~t=1,\ldots,T,\label{eq:y_t_dist}\\
[x_t|g,\bi{\theta}_g,x_{t-1}]&\sim N\left(g^*(t,x_{t-1}),\sigma^2_{\eta}\right)[2\pi];~t=1,\ldots,T,\label{eq:x_t_dist}\\
[f(\cdot,\cdot)|\bi{\theta}_f]&\sim GP\left(\bi{h}(\cdot,\cdot)'\bi{\beta}_f,\sigma^2_fc_f(\cdot,\cdot)\right)~[2\pi],\label{eq:gp_f}\\
[g(\cdot,\cdot)|\bi{\theta}_g]&\sim GP\left(\bi{h}(\cdot,\cdot)'\bi{\beta}_g,\sigma^2_gc_g(\cdot,\cdot)\right)[2\pi],\label{eq:gp_g}\\
[\bi{\beta}_f,\sigma^2_f,\bi{\beta}_g,\sigma^2_g,\sigma^2_{\epsilon},\sigma^2_{\eta}]
&=[\bi{\beta}_f,\sigma^2_f][\bi{\beta}_g,\sigma^2_g][\sigma^2_{\epsilon},\sigma^2_{\eta}],\label{eq:theta_prior}
\end{align}
}where $\bi{\theta}_f = (\bi{\beta}_f,\sigma_f,\sigma_{\epsilon})'$ and 
$\bi{\theta}_g = (\bi{\beta}_g,\sigma_g,\sigma_{\eta})'.$
In the above, GP stands for ``Gaussian Process". 

For obtaining the joint distribution of the latent circular state variables, 
we utilize the ``look-up" table approach (see \ctn{Bhattacharya07}) along the same line as \ctn{Mazumder16}. 
The idea and the complete details of the look-up table approach in the context of circular latent variable is 
discussed in \ctn{Mazumder16}, and thereby we skip the details in this current article. 
In the next section we provide the forms of the prior distributions of the parameters 
associated with the above hierarchical structure.

\subsection{Prior specifications}
\label{Priors}
Our prior distributions have the following forms.
\begin{align}
 [x_0]&\sim \mbox{von-Mises} (\mu_0,\sigma_0^2);
\\[1ex]
 [\sigma^2_{\mbox{\scriptsize $\epsilon$}}] & \propto (\sigma^2_{\mbox{\scriptsize $\epsilon$}})^{\left(-\frac{\alpha_{\mbox{\tiny $\epsilon$}}+2}{2}\right)}\mbox{ exp }\left\{-\frac{\gamma_{\mbox{\tiny $\epsilon$}}}{2\sigma^2_{\mbox{\scriptsize $\epsilon$}}}\right\}; ~~ \alpha_{\mbox{\tiny $\epsilon$}},\,\gamma_{\mbox{\tiny $\epsilon$}} >0
\\[1ex]
 [\sigma^2_{\mbox{\scriptsize $\eta$}}] & \propto (\sigma^2_{\mbox{\scriptsize $\eta$}})^{\left(-\frac{\alpha_{\mbox{\tiny $\eta$}}+2}{2}\right)}\mbox{ exp }\left\{-\frac{\gamma_{\mbox{\tiny $\eta$}}}{2\sigma^2_{\mbox{\scriptsize $\eta$}}}\right\}; ~~ \alpha_{\mbox{\tiny $\eta$}},\,\gamma_{\mbox{\tiny $\eta$}} > 0
 \\[1ex]
  [\sigma^2_{g}] & \propto (\sigma^2_{g})^{\left(-\frac{\alpha_{\mbox{\tiny $g$}}+2}{2}\right)}\mbox{ exp }\left\{-\frac{\gamma_{\mbox{\tiny $g$}}}{2\sigma^2_{g}}\right\}; ~~ \alpha_{\mbox{\tiny $g$}},\,\gamma_{\mbox{\tiny $g$}} >0
 \\[1ex]
 [\sigma^2_{f}] & \propto (\sigma^2_{f})^{\left(-\frac{\alpha_{\mbox{\tiny $f$}}+2}{2}\right)}\mbox{ exp }\left\{-\frac{\gamma_{\mbox{\tiny $f$}}}{2\sigma^2_{f}}\right\}; ~~ \alpha_{\mbox{\tiny $f$}},\,\gamma_{\mbox{\tiny $f$}} >0
\\[1ex]
 [\bi{\beta}_{f}] &\sim N_{4}(\bi{\beta}_{f,0},\bi{\Sigma}_{\beta_{f,0}});
\\[1ex]
[\bi{\beta}_{g}] &\sim N_{4}(\bi{\beta}_{g,0},\bi{\Sigma}_{\beta_{g,0}}).
\end{align}
Here, by $N_{4}(\cdot,\cdot)$, we mean the 4 variate normal distribution. %
We discuss the choice of prior parameters in Sections \ref{simulation_study}, \ref{real_data_analysis}
and \ref{sec:whale_data}, in the contexts of applications to simulated and real data sets. 

\section{MCMC-based inference}
\label{inference}
The aim of our MCMC-based inference is to address the problem of forecasting $y_{T+1}$ and to learn
about the latent circular states, given 
the observed data set $\bi{D}_{T} = (y_1,\ldots,y_T)'$. Unlike \ctn{Mazumder16} here $y_t$ is 
circular and hence dealing with the conditional distribution of $y_{T+1}$ given the data 
$\bi{D}_{T}$ is rendered far more difficult. As already remarked in Section \ref{intro},
the distribution of the observed variables is circular normal, and does not admit any closed
form expression. 
To deal with such a situation one needs to bring in the auxiliary variables 
$N_t$ = $\langle y_{t}^{*}/2\pi\rangle$, where $y_{t}^{*}$ is the linear part of the circular variable 
$y_t$ and $\langle u \rangle$ denotes the largest integer not exceeding $u$. The wrapped random variable 
$N_t$ can take values in the set $\mathbb{Z}$, where $\mathbb{Z}$ denotes the set of the integers. 
For details on wrapped variable in the context of circular data, 
see, for example, \ctn{Ravindran11}; see also \ctn{Mazumder16}. 
However, even conditionally on the auxiliary variables, $f^*$ can not be marginalized out of the model
to simplify the proceedings, so that our MCMC method must include simulation of $f^*$, in addition to 
the auxiliary variables. In contrast, \ctn{Mazumder16} was able to work with a much simpler form
thanks to joint conditional normality of their linear observed data, which also enabled them to 
integrate out the unknown observational function, further simplifying their proceedings. 

In our case, the posterior distribution of 
$y_{T+1}$ given $\bi{D}_{T}$  
admits the following form after introducing the auxiliary variables $N_1,\ldots,N_{T}$: 
\allowdisplaybreaks
\begin{align}
\label{eqn1:prediction equation}
& [y_{T+1}|\bi{D}_T] \notag
\\[1ex]
& = \sum_{N_1,\ldots,N_{T}} \int \left[y_{T+1}|\bi{D}_T,f^*(1,x_1),\ldots,f^*(T+1,x_{T+1}),N_1,\ldots N_{T},x_0,\ldots,x_{T+1},\bi{\beta}_f,\bi{\beta}_g,\sigma_f, \right. \notag
\\[1ex]
& \quad \left. \sigma_g,\sigma_{\epsilon}, \sigma_{\eta} \right] [f^*(1,x_1),\ldots,f^*(T+1,x_{T+1}),N_1,\ldots N_{T},x_0,\ldots,x_{T+1},\bi{\beta}_f,\bi{\beta}_g,\sigma_f,\sigma_g,\sigma_{\epsilon}, \sigma_{\eta}|\bi{D}_{T}] \notag
\\[1ex]
& \quad df^*(1,x_1) \ldots df^*(T+1,x_{T+1}) dx_0 \ldots dx_{T+1} d\bi{\beta}_f d\bi{\beta}_g d\sigma_f d\sigma_g d\sigma_{\epsilon} d\sigma_{\eta}. 
\end{align}
Therefore, once we have a sample from the joint posterior \allowdisplaybreaks $[f^*(1,x_1),\ldots,f^*(T+1,x_{T+1}),N_1,\ldots, N_{T},x_0,\ldots,\\ x_{T+1},\bi{\beta}_f,\bi{\beta}_g,\sigma_f,\sigma_g,\sigma_{\epsilon}, \sigma_{\eta}|\bi{D}_T]$, we can generate an observation from $[y_{T+1}|\bi{D}_{T}]$ by simply generating an observation from \allowdisplaybreaks $[y_{T+1}|\bi{D}_T,f^*(1,x_1),\ldots,f^*(T+1,x_{T+1}),N_1,\ldots N_{T},x_0,\ldots,x_{T+1}, \bi{\beta}_f,\bi{\beta}_g,\sigma_f, \sigma_g,\sigma_{\epsilon}, \sigma_{\eta}]$. Further, we note that given a sample from \allowdisplaybreaks $[f^*(1,x_1),\ldots,f^*(T+1,x_{T+1}),N_1,\ldots N_{T},x_0,\ldots,x_{T+1}, \bi{\beta}_f,\bi{\beta}_g,\sigma_f,\sigma_g, \\ \sigma_{\epsilon}, \sigma_{\eta} | \bi{D}_T]$, $y_{T+1}$ has a wrapped normal distribution with parameter $f^*(T+1,x_{T+1}) \mbox{ and } \sigma^2_{\epsilon}$, because
\begin{align*}
&[y_{T+1}|\bi{D}_T, f^*(1,x_1),\ldots,f^*(T+1,x_{T+1}),N_1,\ldots N_{T},x_0,\ldots,x_{T+1},\bi{\beta}_f,\bi{\beta}_g,\sigma_f,\sigma_g,\sigma_{\epsilon}, \sigma_{\eta}]
\\[1ex]
=& [y_{T+1}|f^*(T+1,x_{T+1}),\sigma_{\epsilon},x_{T+1}].
\end{align*}
Therefore, we need to obtain a sample from \allowdisplaybreaks
$[f^*(1,x_1),\ldots,f^*(T+1,x_{T+1}),N_1,\ldots N_{T},x_0,\ldots,x_{T+1},\\
\bi{\beta}_f,\bi{\beta}_g,\sigma_f,\sigma_g,\sigma_{\epsilon}, \sigma_{\eta}|\bi{D}_T]$.
Define $\bi{f}^*$ = $(f^*(1,x_1),\ldots,f^*(T+1,x_{T+1}))'$ = $(\bi{f}_{\mbox{\scriptsize $D_{T}$}}^*, f^*(T+1,x_{T+1}))'$.  Note that given $(x_1,\ldots, x_{T+1})',\bi{\beta}_f,\mbox{ and }\sigma_f$, $\bi{f}^*$ follows a $(T+1)$-variate normal distribution with mean \allowdisplaybreaks $E[\bi{f}^*|(x_1,\ldots,\\ x_{T+1})',\bi{\beta}_f]$ = $\bi{H}\bi{\beta}_f$, and covariance matrix $\sigma^2_f \bi{A}_{f}$, where $\bi{H} = \left[\bi{h}(1,x_1)',\ldots, \bi{h}(T+1,x_{T+1})'\right]'$ and 
$\bi{A}_f$ = $(\exp\left(-\sigma_f^4(i-j)^2\right)\cos(|x_i-x_j|);~1\leq i,j \leq (T+1))$. Therefore, $\bi{f}_{\mbox{\scriptsize $D_{T}$}}^*$ given $(x_1,\ldots, x_T)',\bi{\beta}_f$ and $\sigma_f$, follows a $T$-variate normal with mean $\bi{H}_{\mbox{\scriptsize $D_{T}$}}\bi{\beta}_f$ and covariance $\bi{A}_{f,D_{T}}$, where  
$\bi{H}_{\mbox{\scriptsize $D_{T}$}} = \left[\bi{h}(1,x_1)',\ldots, \bi{h}(T,x_{T})'\right]'$ and 
$\bi{A}_{f,D_{T}}$ = $(\exp\left(-\sigma_f^4(i-j)^2\right)\cos(|x_i-x_j|);~1\leq i,j \leq T)$.  Moreover, it is immediate that $[f^*(T+1,x_{T+1})|\bi{f}_{\mbox{\scriptsize $D_{T}$}}^*, x_{T+1}, \bi{\beta}_f,\sigma_f]$ $\sim$ $N(\mu_{\mbox{\scriptsize $f^*(T+1,x_{T+1})$}},\sigma_{\mbox{\scriptsize $f^*(T+1,x_{T+1})$}}^2)$, where \allowdisplaybreaks $\mu_{\mbox{\scriptsize $f^*(T+1,x_{T+1})$}} = \bi{h}(T+1,x_{T+1})'\bi{\beta}_f + \bi{s}_{f,D_{T}}(T+1,x_{T+1})'\bi{A}_{f,D_{T}}^{-1}(\bi{f}_{\mbox{\scriptsize $D_{T}$}}^*-\bi{H}_{\mbox{\scriptsize $D_{T}$}}\bi{\beta}_f)$ and 
\allowdisplaybreaks $\sigma_{\mbox{\scriptsize $f^*(T+1,x_{T+1})$}}^2 = \sigma^2_f(1-\bi{s}_{f,D_{T}}(T+1,x_{T+1})'\bi{A}_{f,D_{T}}^{-1}\bi{s}_{f,D_{T}}(T+1,x_{T+1}))$ with 
$ \bi{s}_{f,D_{T}}(T+1,x_{T+1})' = (\exp (-\sigma_f^4(T+1-j)^2)\cos(|x_{T+1}-x_j|),1\leq j\leq T).$

By introducing the auxiliary variables $K_1,\ldots, K_T$, the wrapped variables for latent circular variables $x_t$, 
defined as $K_t$ = $\langle x_{t}^{*}/2\pi\rangle$, and the set of linear-circular grid points $\bi{D}_{z}$ (defined below), 
we can write  
\begin{align}
\label{eqn2:conditional_densities}
&[\bi{f}^*,N_1\ldots,N_{T},x_0,\ldots,x_{T+1},\bi{\beta}_f,\bi{\beta}_g,\sigma_f,\sigma_g,\sigma_{\epsilon}, \sigma_{\eta}|\bi{D}_T]
\notag \\[1ex]
& \propto \sum_{K_1,\ldots,K_{T}} \int [\bi{f}^*,N_1\ldots,N_{T},x_0,\ldots,x_{T+1},\bi{\beta}_f,\bi{\beta}_g,\sigma_f,\sigma_g,\sigma_{\epsilon}, \sigma_{\eta}, g^*(1,x_0),\bi{D}_z,\bi{D}_{T}] dg^* d\bi{D}_z \notag
\\[1ex]
& = \sum_{K_1,\ldots,K_{T}} \int [\bi{\beta}_f][\bi{\beta}_g][\sigma_f][\sigma_g][\sigma_{\epsilon}] [\sigma_{\eta}] [x_0] [\bi{f}_{\mbox{\scriptsize $D_{T}$}}^*|x_1,\ldots,x_{T},\bi{\beta}_f,\sigma_f] [f^*(T+1,x_{T+1})|\bi{f}_{\mbox{\scriptsize $D_{T}$}}^*, x_{T+1}, \bi{\beta}_f,\sigma_f]  \notag
\\[1ex]
& \quad\times\prod_{t=1}^{T} [N_{t}|f^*(t,x_t),\sigma_{\epsilon},x_t] [g^*(1,x_0)|x_0,\bi{\beta}_g,\sigma_g] [\bi{D}_z|g^*(1,x_0),x_0,\bi{\beta}_g,\sigma_g] [x_1|g^*(1,x_0),\sigma_{\eta},K_1] [K_1|g^*(1,x_0),\sigma_{\eta}]
\notag \\[1ex]
& \quad\times\prod_{t=2}^{T+1} [x_t|\bi{\beta}_g,\sigma_{\eta},\sigma_g,\bi{D}_z,x_{t-1},K_t] \prod_{t=2}^{T+1} [K_{t}|\bi{\beta}_g,\sigma_{\eta},\sigma_g,\bi{D}_z,x_{t-1}] \left[\bi{D}_{T}|x_1,\ldots,x_T,f^*(1,x_1), \ldots, f^*(T,x_T), \right.
\notag \\[1ex]
&\left. \qquad N_1,\ldots,N_T,\sigma_{\epsilon}\right] d\bi{D}_z dg^*(1,x_0).
\end{align}
We note that $[\bi{D}_{T}|x_1,\ldots,x_T,f^*(1,x_1),\ldots,f^*(T,x_T), N_1,\ldots,N_T,\sigma_{\epsilon}]$ 
= $\prod_{t=1}^{T} [y_t|N_t,f^*(t,x_t),\sigma_{\epsilon},x_t]$, and define $\bi{D}_z=(g^*(t_1,z_1),\ldots,g^*(t_n,z_n))'$,
where, for $i=1,\ldots,n$, $(t_i,z_i)$ is a set of linear-circular grid-points with $t_i\geq 0$
and $z_i\in[0,2\pi]$, for $i=1,\ldots,n$. This grid, which we denote by $\bi{G}_{z}$, is 
associated with the look-up table; see \ctn{Mazumder16} for details.
%
Hence, from (\ref{eqn2:conditional_densities}), we obtain
\begin{align}
\label{eqn3:final_conditional_equation_form}
&[\bi{f}^*,N_1\ldots,N_{T},x_0,\ldots,x_{T+1},\bi{\beta}_f,\bi{\beta}_g,\sigma_f,\sigma_g,\sigma_{\epsilon}, \sigma_{\eta}|\bi{D}_T] \notag
\\[1ex]
& \propto
\sum_{K_1,\ldots,K_{T}} \int [\bi{\beta}_f][\bi{\beta}_g][\sigma_f][\sigma_g][\sigma_{\epsilon}] [\sigma_{\eta}] [x_0] [\bi{f}_{\mbox{\scriptsize $D_{T}$}}^*|x_1,\ldots,x_{T},\bi{\beta}_f,\sigma_f] [f^*(T+1,x_{T+1})|\bi{f}_{\mbox{\scriptsize $D_{T}$}}^*, x_{T+1}, \bi{\beta}_f,\sigma_f] \notag
\\[1ex]
& \quad\times\prod_{t=1}^{T} [N_{t}|f^*(t,x_t),\sigma_{\epsilon},x_t] [g^*(1,x_0|x_0,\bi{\beta}_g,\sigma_g)] [\bi{D}_z|g^*(1,x_0),x_0,\bi{\beta}_g,\sigma_g] [x_1|g^*(1,x_0),\sigma_{\eta},K_1] [K_1|g^*(1,x_0),\sigma_{\eta}]
\notag \\[1ex]
& \quad\times\prod_{t=2}^{T+1} [x_t|\bi{\beta}_g,\sigma_{\eta},\sigma_g,\bi{D}_z,x_{t-1},K_t] \prod_{t=2}^{T+1} [K_{t}|\bi{\beta}_g,\sigma_{\eta},\sigma_g,\bi{D}_z,x_{t-1}] \prod_{t=1}^{T} [y_t|N_t,f^*(t,x_t),\sigma_{\epsilon},x_t] d\bi{D}_z dg^*(1,x_0).
\end{align}%
In (\ref{eqn2:conditional_densities}) and (\ref{eqn3:final_conditional_equation_form}) the identity \allowdisplaybreaks $[\bi{f}^*|x_1,\ldots,x_{T+1},\bi{\beta}_f,\sigma_f] = \left[\bi{f}_{\mbox{\scriptsize $D_{T}$}}^*|x_1,\ldots,x_{T},\bi{\beta}_f,\sigma_f\right]
[f^*(T+1,x_{T+1}) |\bi{f}_{\mbox{\scriptsize $D_{T}$}}^*, \\ x_{T+1}, \bi{\beta}_f,\sigma_f]$  is used.
  
The complete details of our MCMC algorithm is presented in the supplement.  
\section{Simulation study}
\label{simulation_study}

\subsection{True model}
\label{subsec:true_model}

We now illustrate the performance of our model and methodologies using a simulation study, considering
the following dynamic nonlinear model:
\begin{align*}
y_{t} &= \{-\tan (x_{t}) + \tan^2 (x_t)/20 + v_{t}\}~[2\pi] ;
\\[1ex]
x_{t} & = \left\{ \alpha x_{t-1}+\frac{\beta x_{t-1}}{1+x_{t-1}^2}+\gamma \cos (1.2(t-2)) + u_t \right\}~[2\pi],
\end{align*}
for $t=1,\ldots,T$ with $T=101$. In the above, $u_{t}$ and $v_{t}$ are normally distributed with means zero and variances $\sigma^2_{\eta}$ and $\sigma^2_{\epsilon}$. We choose the values of $\alpha$, $\beta$ and $\gamma$ to be 
$1$, $0.5$ and $0.2$, respectively, and fix the values of both $\sigma_{\eta}$ and $\sigma_{\epsilon}$ at  $0.1$.
The first $100$ observations of $y_t$ are assumed to be known, and the last observation is set aside for the 
purpose of forecasting. 

The true model can be thought of as the circularized version of a slightly generalized non-linear state-space model
on the real line adopted in Example 3.2 of \ctn{Carlin92}. 
A variant of such a nonlinear model is also considered in \ctn{Ghosh14} and in \ctn{Mazumder16}. 

\subsection{Choices of prior parameters and the grid $\bi{G}_{z}$}
\label{subsec:prior_chocies}
For this simulation experiment we have chosen the four-variate normal prior distribution for $\bi{\beta}_f$ 
with mean $(0,0,0,0)'$ and covariance matrix as $4\times 4$ identity matrix. The prior mean for $\bi{\beta}_g$ 
is taken to be $(1,1,1,1)'$. The choice of these prior parameters facilitated adequate mixing of our MCMC algorithm. 
As in \ctn{Mazumder16}, we throughout fixed the third and fourth components of $\bi{\beta}_g$ to be $1$ 
to tackle identifiability problems. Hence, we take the covariance matrix in the prior distribution of $\bi{\beta}_g$ as 
a diagonal matrix with diagonal elements $(1,1,0,0)'$.     
Let us heuristically explain the identifiability problem and why fixing the values of the last two components 
of $\bi{\beta}_g$ is expected
to solve the problem. First note that both the observational and evolutionary equations are non-identifiable
since the last two components of the unknown parameter $\bi{\beta}_f$ are multiplied with the random quantities 
$\cos(x_t)$ and $\sin(x_t)$ in the observational equation and the last two components of the unknown parameter 
$\bi{\beta}_g$ are multiplied with 
the random quantities $\cos(x_{t-1})$ and $\sin(x_{t-1})$ in the evolutionary equation. 
In other words, multiplying the last two components of the two parameter vectors with arbitrary non-zero constants and dividing
the multiplicative random quantities with the same constant, does not change the equations, implying non-identifiability.
In the case of the observational equation, the identifiability problem is less severe considering the fact that least
$\{y_1,\ldots,y_T\}$ are observed (known), and so, for $t=1,\ldots,T$, the quantity $\bi{h}(t,x_t)'\bi{\beta}_f$ is identifiable
and estimable from the observed data,
although $\bi{\beta}_f$ itself is not identifiable. However, putting a prior on $\bi{\beta}_f$ which assigns low probabilities 
to undesirable regions, essentially solves the identifiability problem of $\bi{\beta}_f$. As is evident
from the simulation based posterior inference, our prior on $\bi{\beta}_f$
indeed ensures that the posteriors of all the components of $\bi{\beta}_f$ are unimodal, for all the simulated and
real examples that we reported, essentially indicating identifiability.
However, for the evolutionary equation, the identifiability problem is more severe. This is because, unlike 
$\{y_1,\ldots,y_T\}$, which are observed, $\{x_0,x_1,\ldots,x_T\}$ are unobserved, and so, the quantity 
$\bi{h}(t,x_{t-1})'\bi{\beta}_g$ can not be directly estimated using $\{x_0,x_1,\ldots,x_T\}$, and does not admit 
straightforward identifiability. This makes the task of choosing appropriate ``identifiability-ensuring" prior
on $\bi{\beta}_g$ infeasible. The only solution available to us, is to fix the last two components of $\bi{\beta}_g$,
which of course solves the identifiability problem.

Like \ctn{Mazumder16}, in our current paper impropriety of the posteriors of 
$\sigma_f$, $\sigma_{\epsilon}$, $\sigma_g$, $\sigma_{\eta}$, $K_t,~ t=1,\ldots, T$ and $N_t,~ t=1\ldots, T$ 
result when they all permitted to be random. To be precise, for any value of $N_t$ (or $K_t$), 
exactly the same of circular variable $y_t$ (or $x_t$) is obtained by the mod $2\pi$ transformation 
applied to $y_t^*$ = $y_t+2\pi N_t$ (or $x_t^*$ = $x_t+2\pi K_t$). Therefore, given $y_t$ (or $x_t$), 
it is not possible to constrain $N_t$ (or $K_t$) unless both $\sigma_f$ and $\sigma_{\epsilon}$  
(or $\sigma_g$ and $\sigma_{\eta}$) are bounded. Boundedness of $\sigma_{\epsilon}$ and $\sigma_f$ 
(or $\sigma_g$ and $\sigma_{\eta}$) would ensure finite variance of $y_t$ (or $x_t$), which in turn 
would imply finite variability of $N_t$ (or $K_t$). Since it is not immediate how to select a bounded 
prior for $\sigma_f$, $\sigma_{\epsilon}$, $\sigma_g$ and $\sigma_{\eta}$ we obtain the maximum likelihood 
estimates (MLEs) of these variances and plug those values in our model. For obtaining the MLEs we used 
the simulated annealing technique (see, for instance, \ctn{Robert04}, \ctn{Liu01}) where at each iteration 
new values of these variances are considered, then all the other parameters are integrated out 
using average of Monte Carlo simulations, given the proposed values of $\sigma_f$, $\sigma_{\epsilon}$, $\sigma_g$ 
and $\sigma_{\eta}$, so that the integrated likelihood is obtained given the proposed variances. 
Then we calculate the acceptance ratio, and finally decrease the temperature parameter of our 
simulated annealing algorithm before proceeding to the next iteration. The values of the MLEs turned out to 
be $\hat{\sigma}_f = 2.64410$, $\hat{\sigma}_{\epsilon} = 0.29289$, $\hat{\sigma}_g = 0.20363$ and $\hat{\sigma}_{\eta} = 0.64201$.   

As regards the von-Mises$(\mu_0,\sigma^2_0)$ prior on $x_0$, we throughout set $\mu_0=\pi$, the mean of the interval $(0,2\pi]$
and $\sigma^2_0=3$, which ensures moderately large variability.

Lastly, we divide the interval $[0,2\pi]$ into $100$ sub-intervals 
of the forms $\left[\frac{2\pi i}{100},\frac{2\pi(i+1)}{100}\right]$, $i=0,\ldots,99$, and choose a random number
uniformly for each of the sub-intervals;
these values constitute the circular second component of the two dimensional grid $\bi{G}_{z}$. For the linear first component 
of $\bi{G}_{z}$, we choose a random number uniformly from each of the 100 subintervals 
$\left[i,i+1\right]$, $i=0,\ldots,99$. 

\subsection{MCMC details}
\label{subsec:mcmc_details}
Our MCMC algorithm updates some parameters using Gibbs steps and the remaining ones using Metropolis-Hastings steps; 
see the supplement for details. We update $x_0$ using von-Mises distribution with the mean being the accepted value 
of $x_0$ in the previous iteration and
with the concentration parameter $\kappa$ = $3$; for updating $x_t$ we use a mixture of two von-Mises distributions 
with $\kappa=0.5$ and $\kappa=3$ for $t=1,\ldots, T$. 
The wrapping variables $N_t;~t=1,\ldots T$ and $K_t;~t=1,\ldots,T$, are updated using the discrete normal random walk 
with variance $1$. All these choices are made very painstakingly 
after cautiously adjudging the mixing properties of many pilot MCMC runs. The remaining parameters are 
updated using Gibbs steps. 

We performed $2.5\times 10^5$ MCMC iterations with a burn-in period consisting of the first $2\times 10^5$ iterations 
using the above choices of the prior parameters and $\bi{G}_{z}$, and with the above MCMC updating 
procedure of the parameters. The time taken to run $2.5\times 10^5$ MCMC simulations in a server machine 
(64 bit, 1247.265 MHz CPU speed and 65 GB memory) is 25 hours and 34 minutes. The usual tests for convergence 
of MCMC iterations are performed and all the tests turned out to be satisfactory. 
All our codes, for all the applications reported in this paper, are written in C.

\subsection{Results of simulation study}
\label{subsec:simulation_study}

\begin{figure}[htp]
\centering
\includegraphics[height=1.5in,width=1.5in]{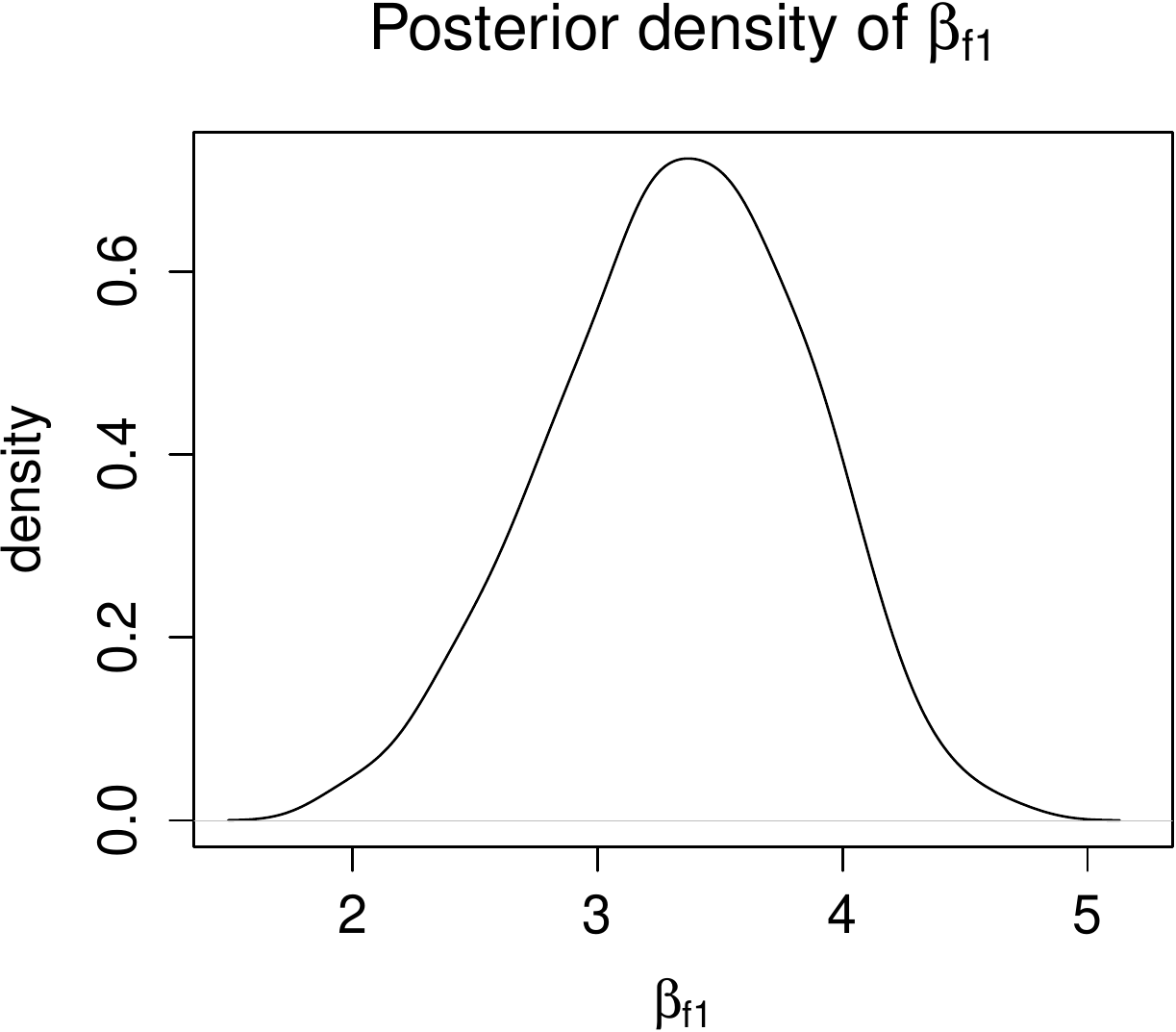}
\includegraphics[height=1.5in,width=1.5in]{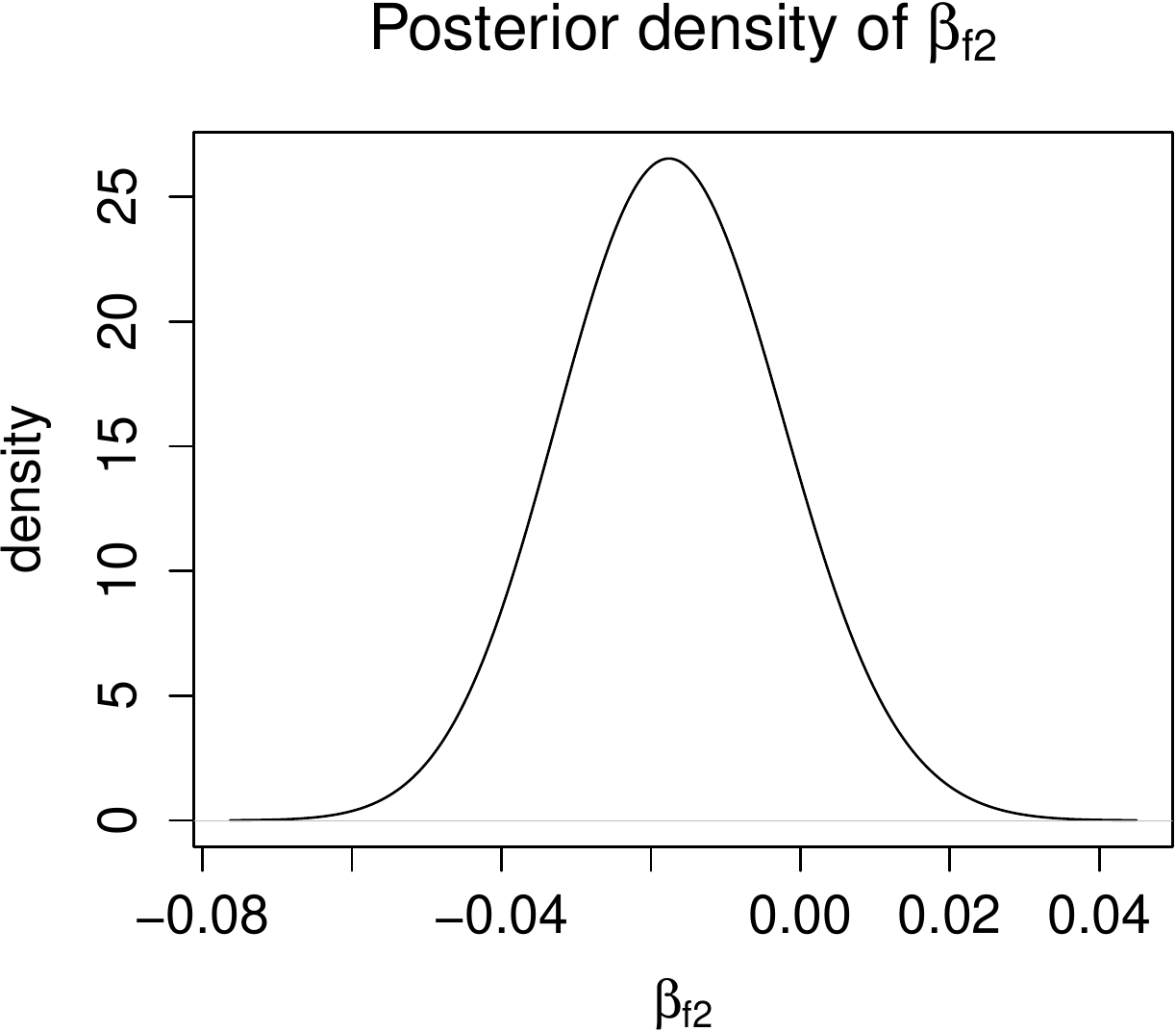}
\includegraphics[height=1.5in,width=1.5in]{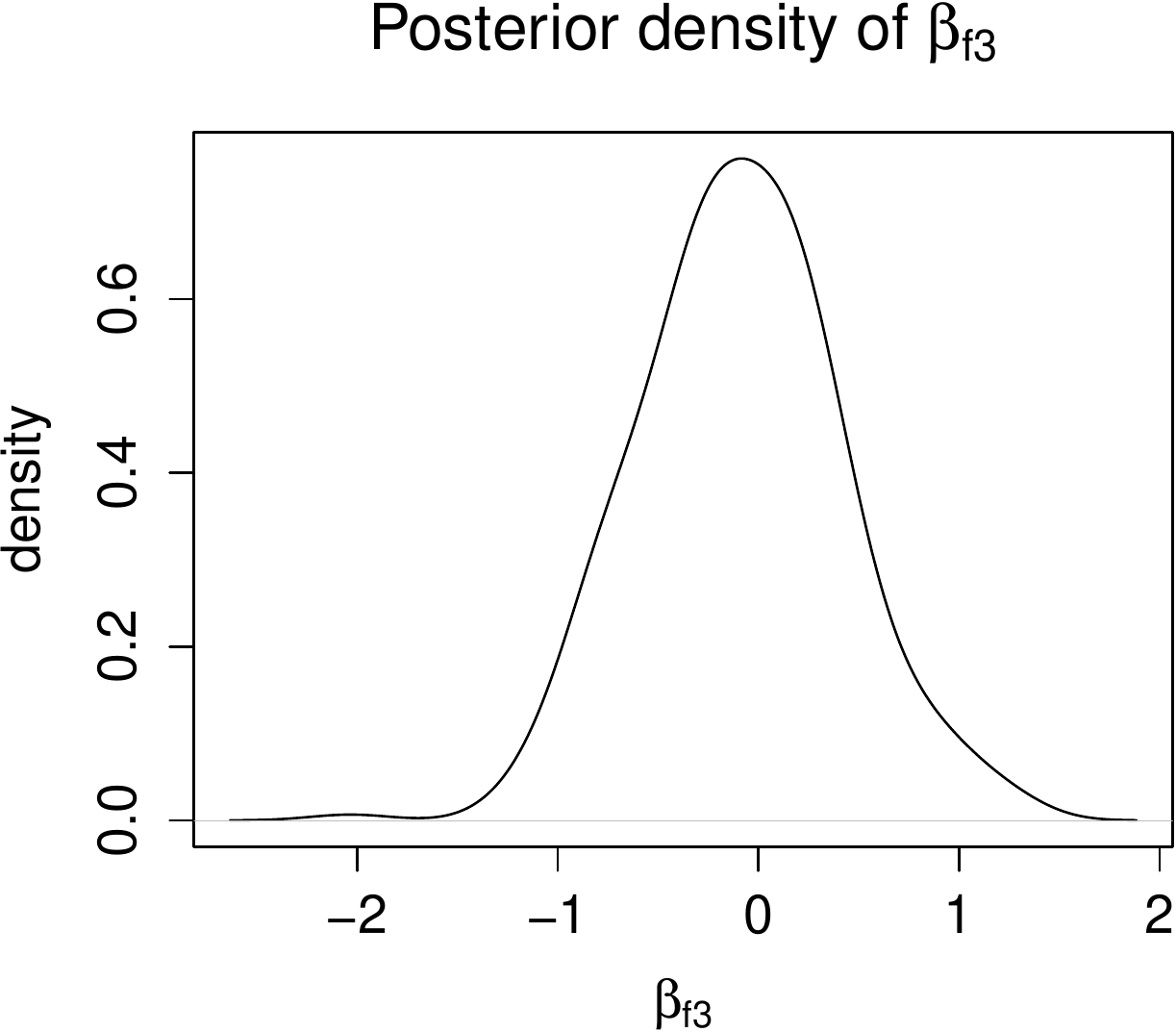}
\includegraphics[height=1.5in,width=1.5in]{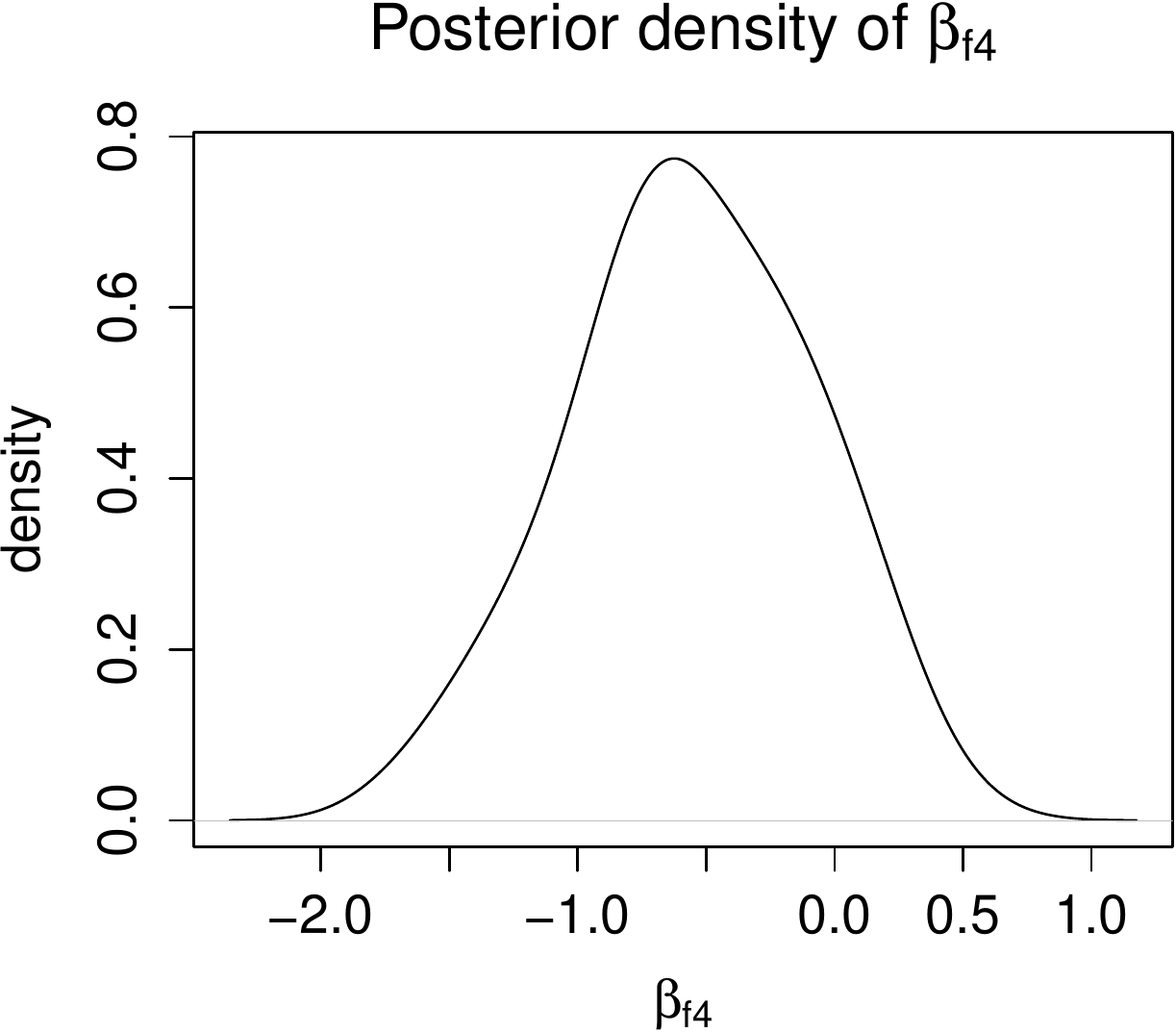}
\caption{Posterior densities of the four components of $\bi{\beta}_f$ for simulated data.}
\label{Fig00:Post_of_beta_f_components_in_sim_data}
\end{figure}

\begin{figure}[htp]
\centering
\includegraphics[height=2in,width=2in]{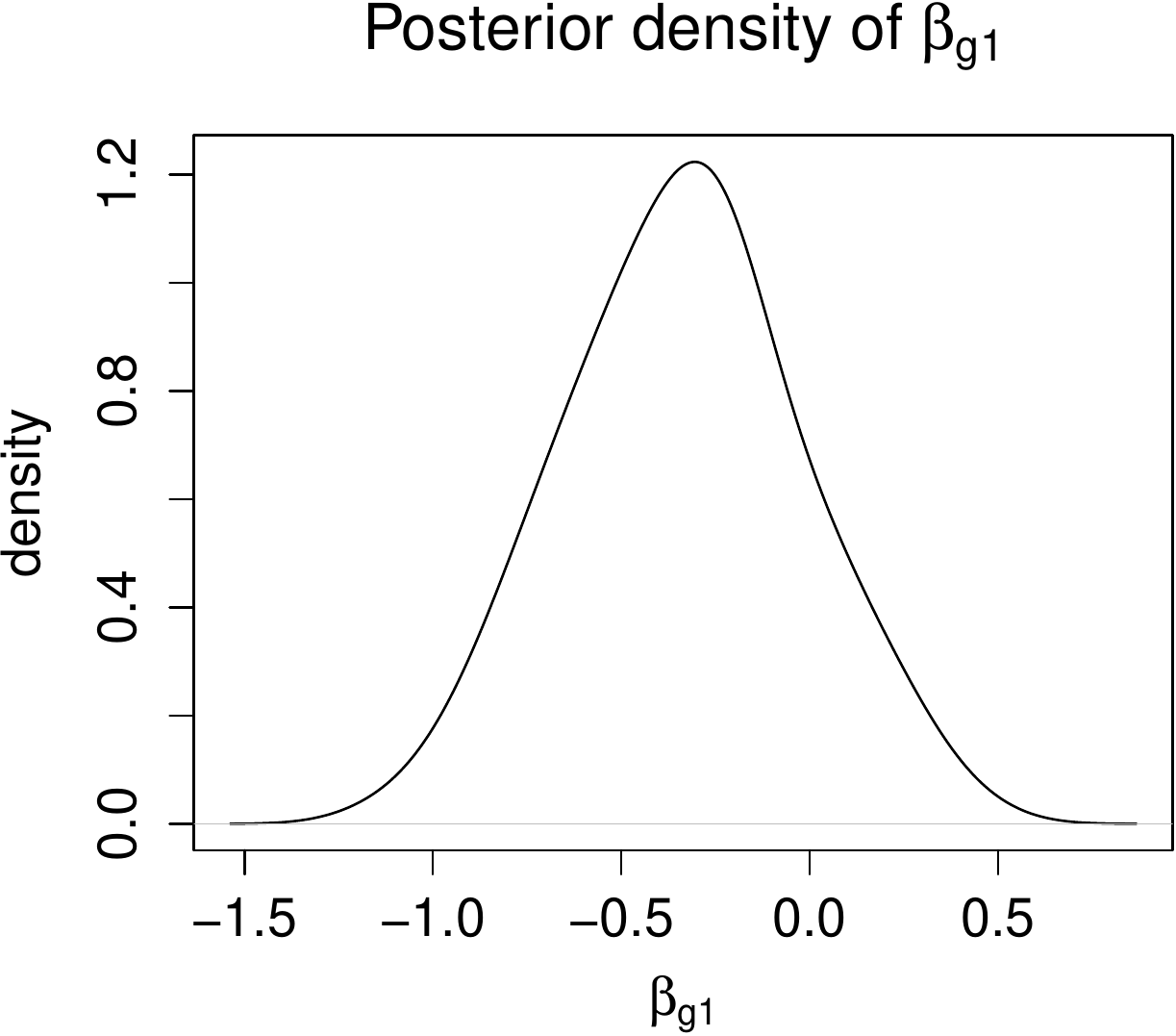}
\includegraphics[height=2in,width=2in]{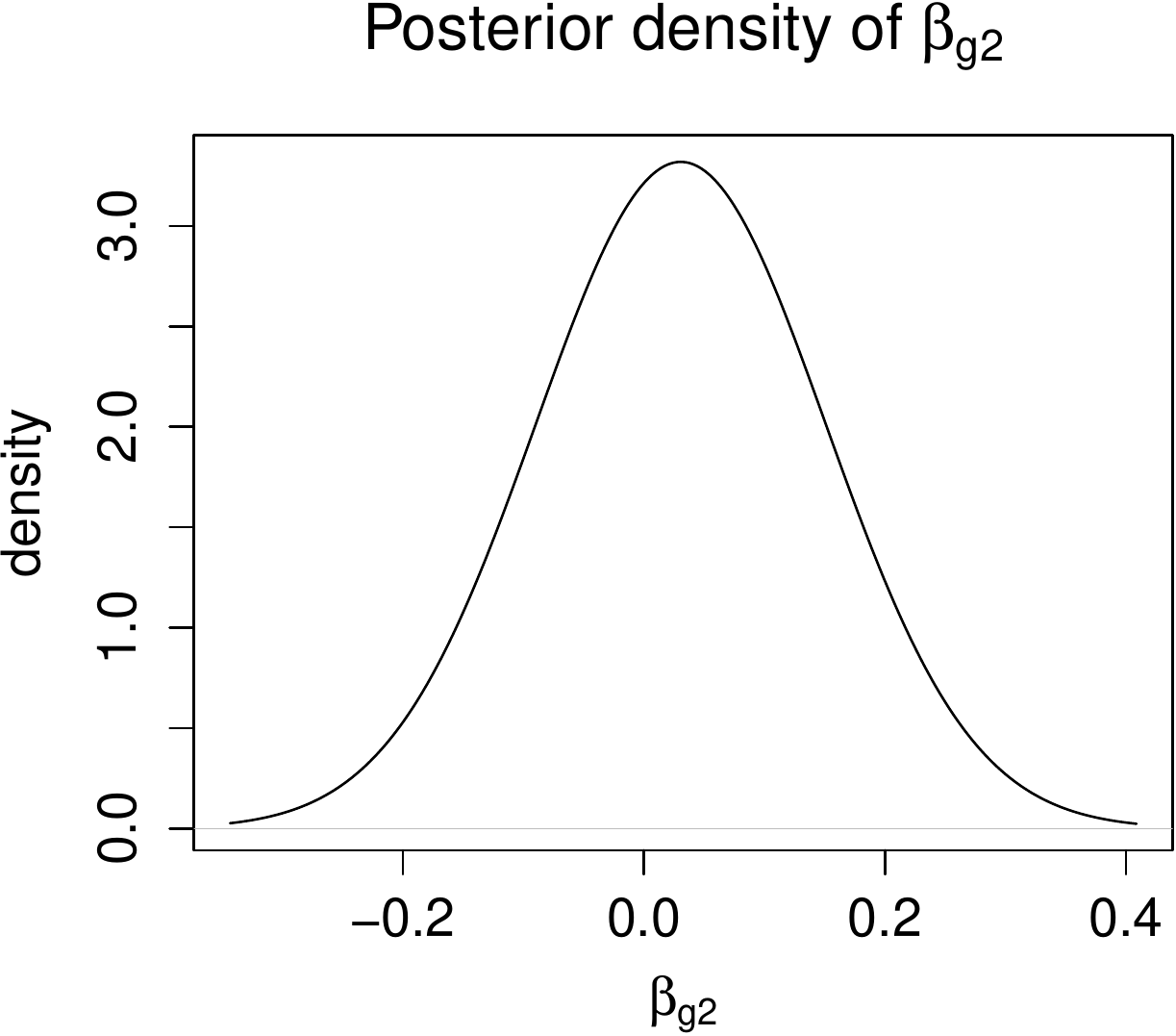}
\caption{Posterior densities of the first and second components of $\bi{\beta}_g$ for simulated data.}
\label{Fig01:Post_of_beta_g_and_sigma_f_in_sim_data}
\end{figure}

\begin{figure}[htp]
\centering
\includegraphics[height=2in,width=2in]{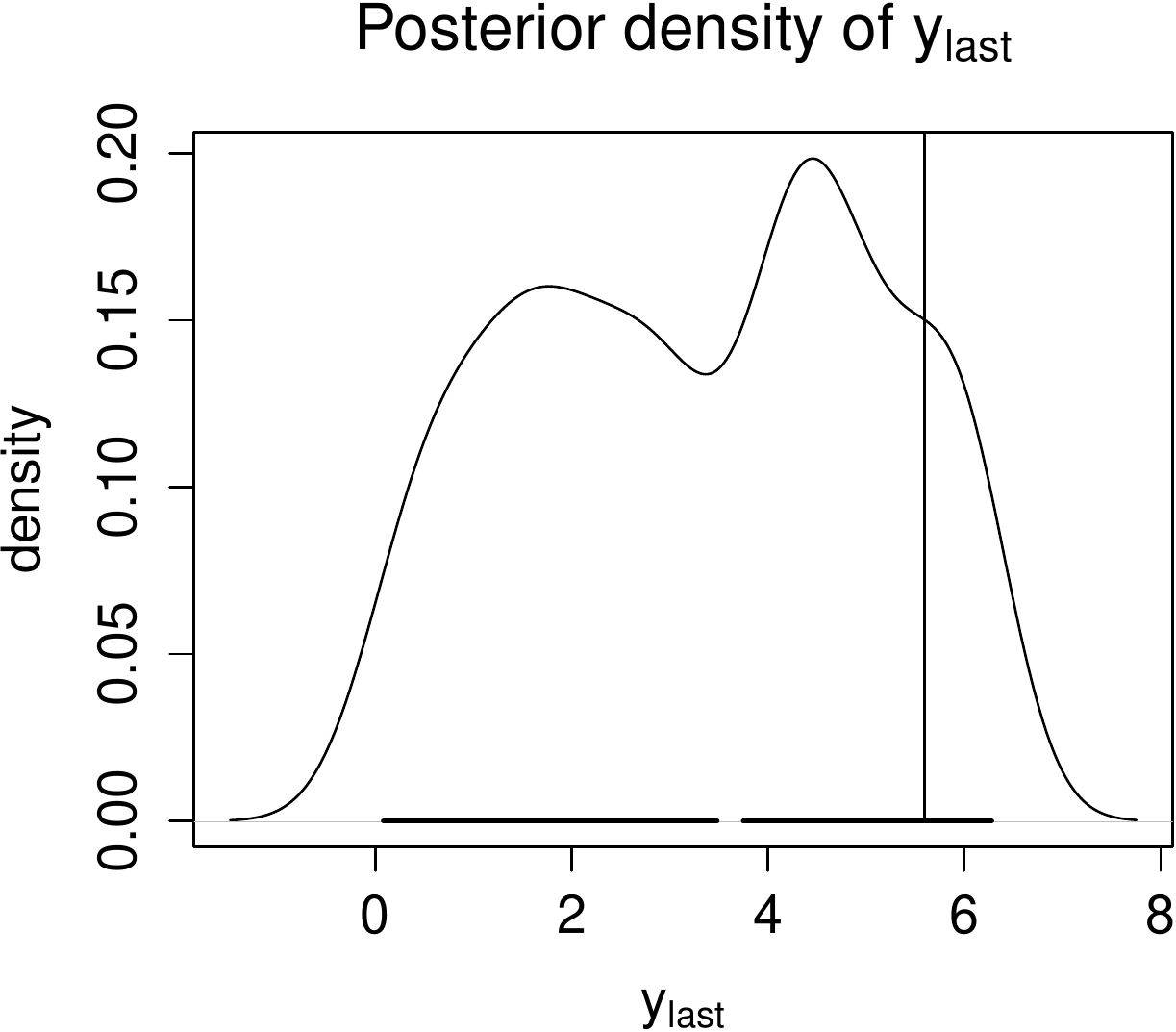} 
\includegraphics[height=2in,width=2in]{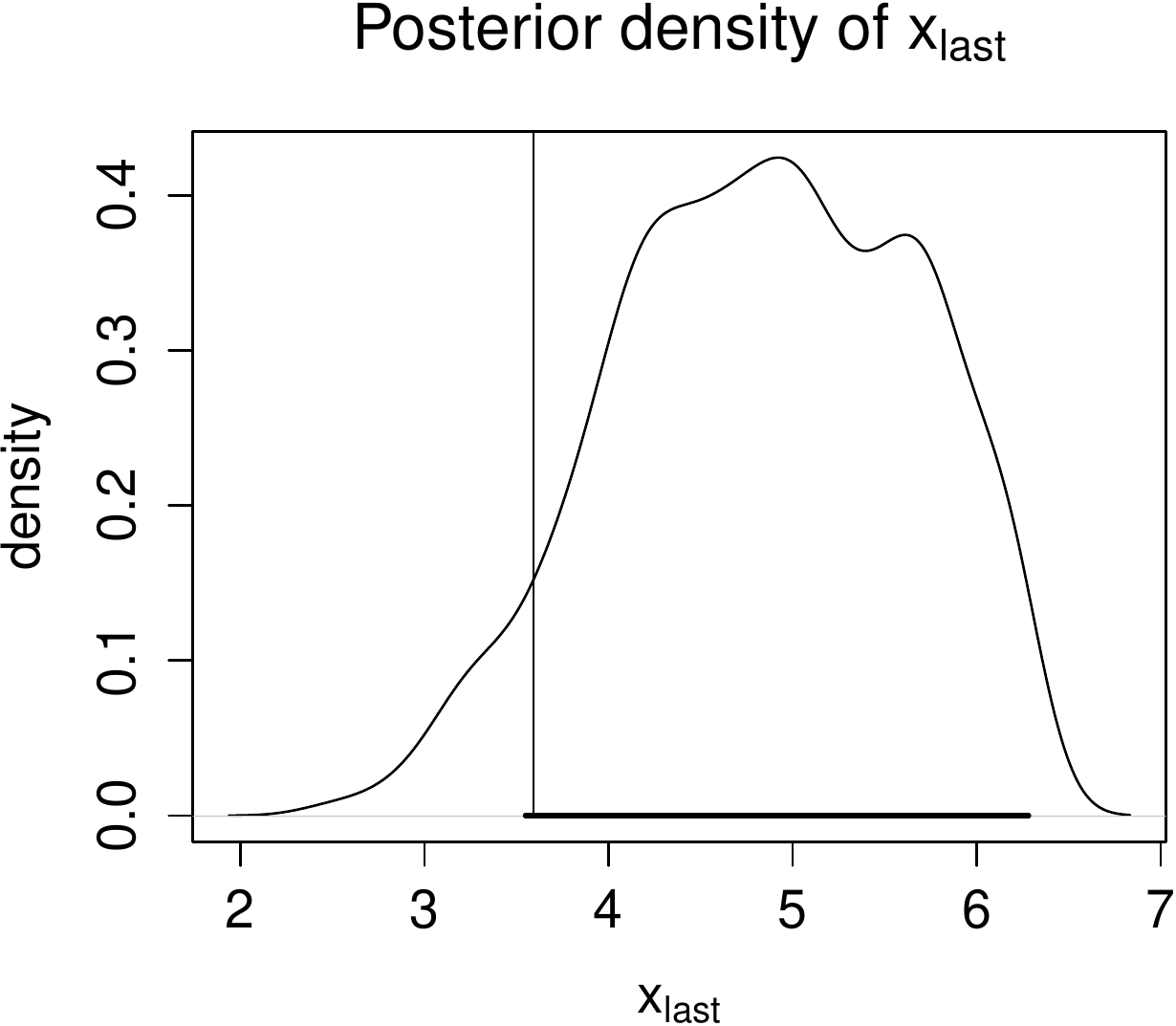}
\caption{From left\/: The first panel displays the posterior predictive density of $y_{101}$ of simulated data. The second plot represents the posterior predictive density of $x_{101}$ (last latent observation). The thick 
horizontal line denotes the 95\% highest posterior density credible interval and the vertical line denotes 
the true value for both the figures.}
\label{Fig1:Post_of_x_last_and_y_last_for_simu_data}
\end{figure}

Figures \ref{Fig00:Post_of_beta_f_components_in_sim_data} and \ref{Fig01:Post_of_beta_g_and_sigma_f_in_sim_data} 
depict the posterior densities of the four components of $\bi{\beta}_f$ and the first two 
components of $\bi{\beta}_g$, respectively. 
The posterior predictive densities of $x_{101}$ and $y_{101}$ are provided in 
Figure \ref{Fig1:Post_of_x_last_and_y_last_for_simu_data}. The horizontal bold black lines denote 
the 95\% highest posterior density credible intervals and the vertical lines denote the true values. 
We observe that the true values of $y_{101}$ and $x_{101}$ fall well within the respective intervals. 

\begin{figure}[htp]
\centering
\includegraphics[height = 3in,width=5in]{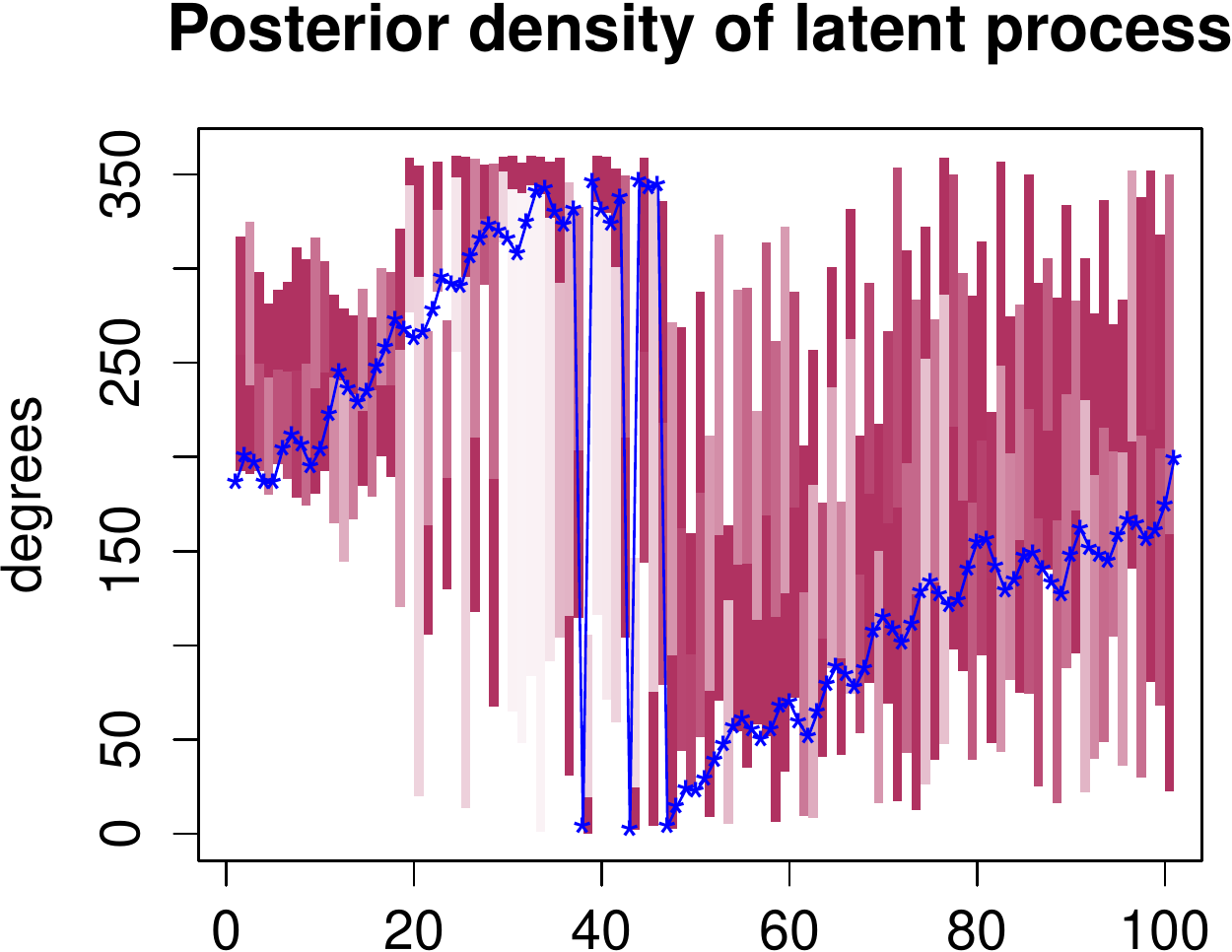}
\caption{Depiction of the posterior densities of the latent circular process $\left\{x_t;t=1,\ldots,T\right\}$; 
higher the intensity of the color, higher is the posterior density. The blue stars denote the true time series.}
\label{Fig2:fitted_latent_process}
\end{figure}

It is seen that the densities of most of $x_t$, $t=1,\ldots,T,$ have multiple modes. So, a plot of the posterior probability 
distribution of the latent process for each time point is more appropriate than ordinary credible regions. 
Such a plot for the latent circular time series $x_1,\dots,x_T$ is displayed in 
Figure \ref{Fig2:fitted_latent_process}. In this plot, regions with progressively higher densities 
are shown by progressively more intense colors. Encouragingly, most of the true values are seen to 
lie in the high probability regions.  


\subsubsection{Cross-validation}
\label{subsubsec:cross validation}
\begin{figure}[htp]
\centering
\includegraphics[height = 3in,width=3in]{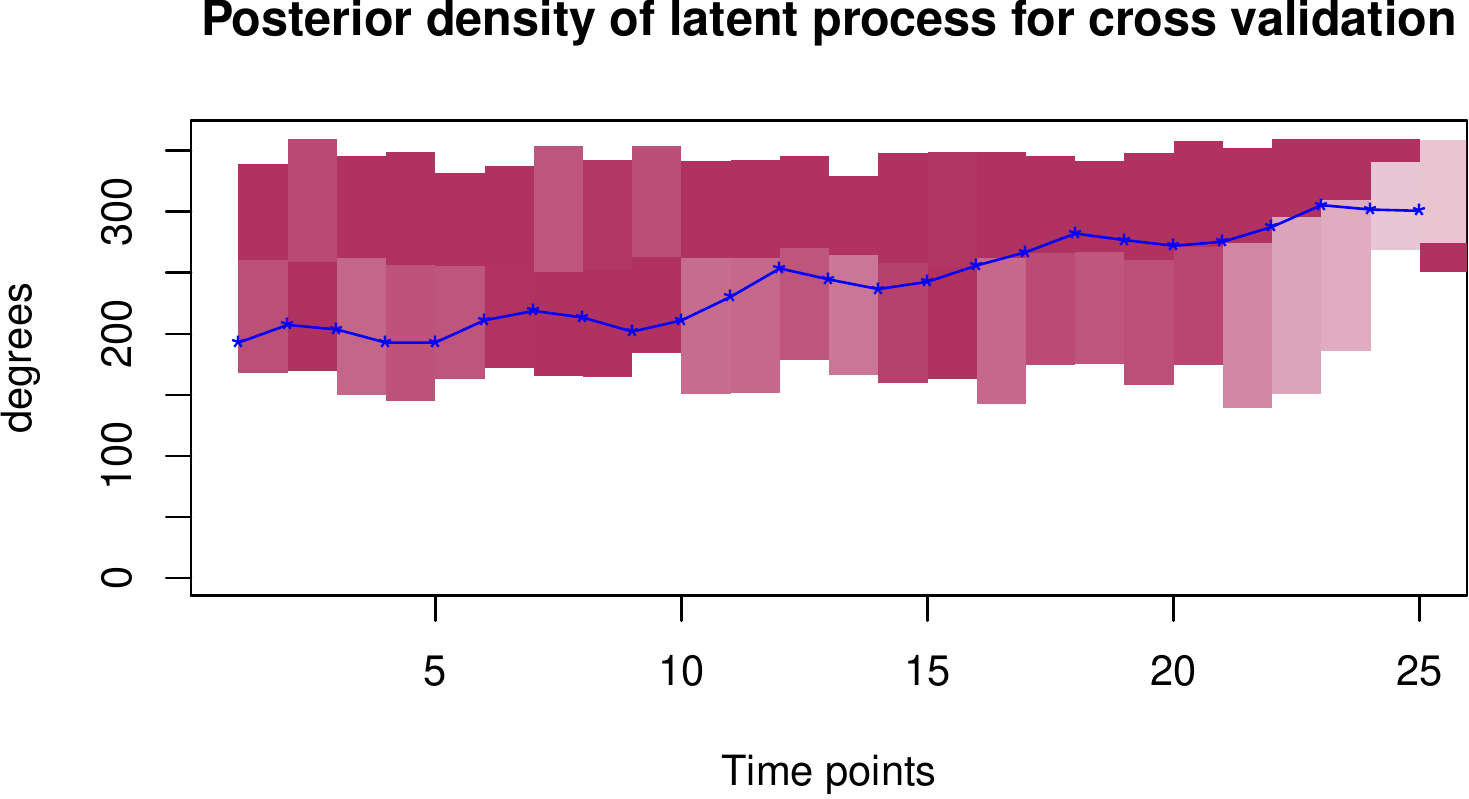}
\includegraphics[height = 3in,width=3in]{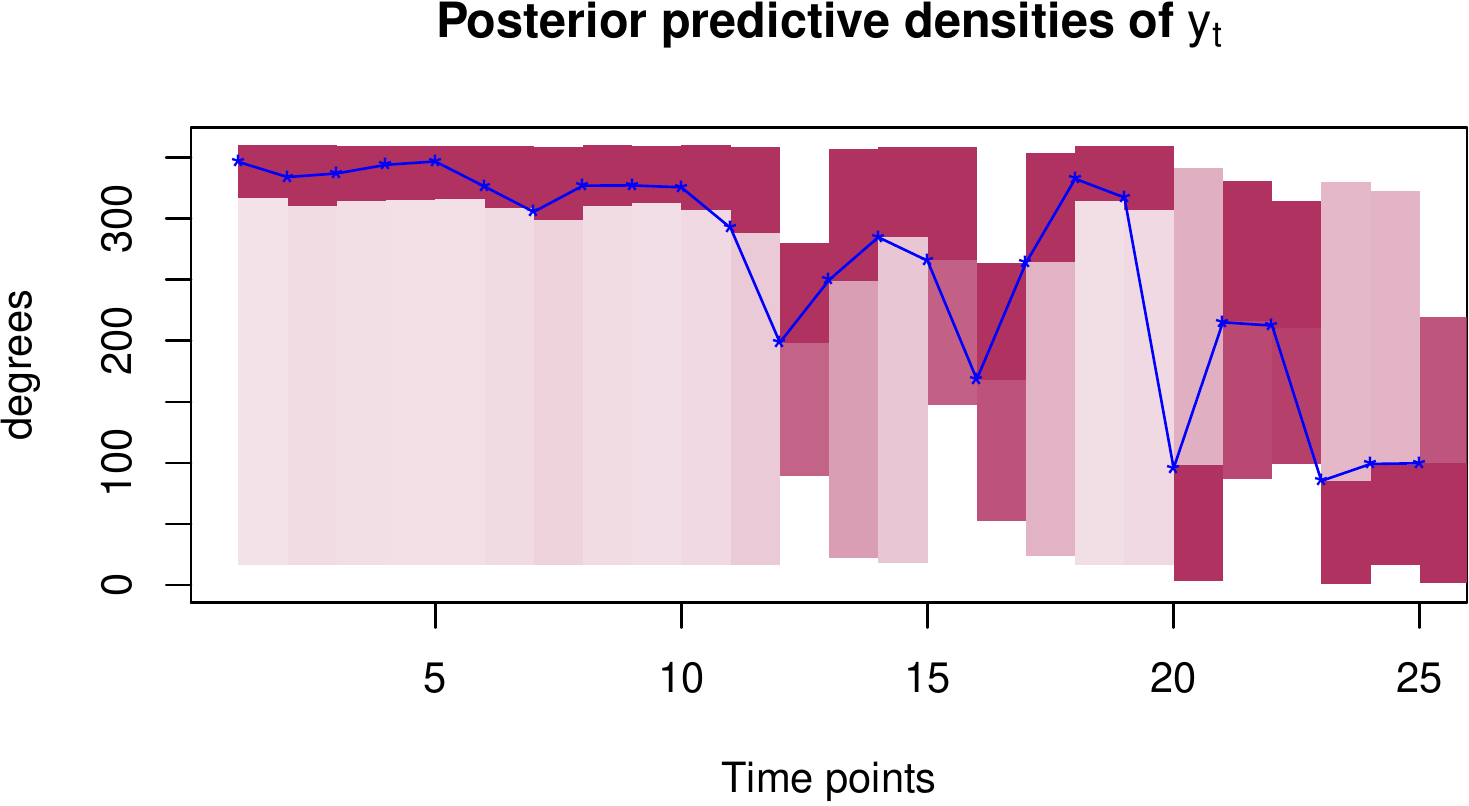}
\caption{Depiction of the posterior densities of the latent circular process $\left\{x_t;t=1,\ldots,25\right\}$ 
and the observed circular process $\left\{y_t;t=1,\ldots,25\right\}$ under the leave-one-out cross-validation scheme; 
higher the intensity of the color, higher is the posterior density. The blue stars denote the true time series 
for $x_t$ and $y_t$.}
\label{Fig:cross valid simu_data size_25}
\end{figure}

\begin{figure}[htp]
\centering
\includegraphics[height = 3in,width=3in]{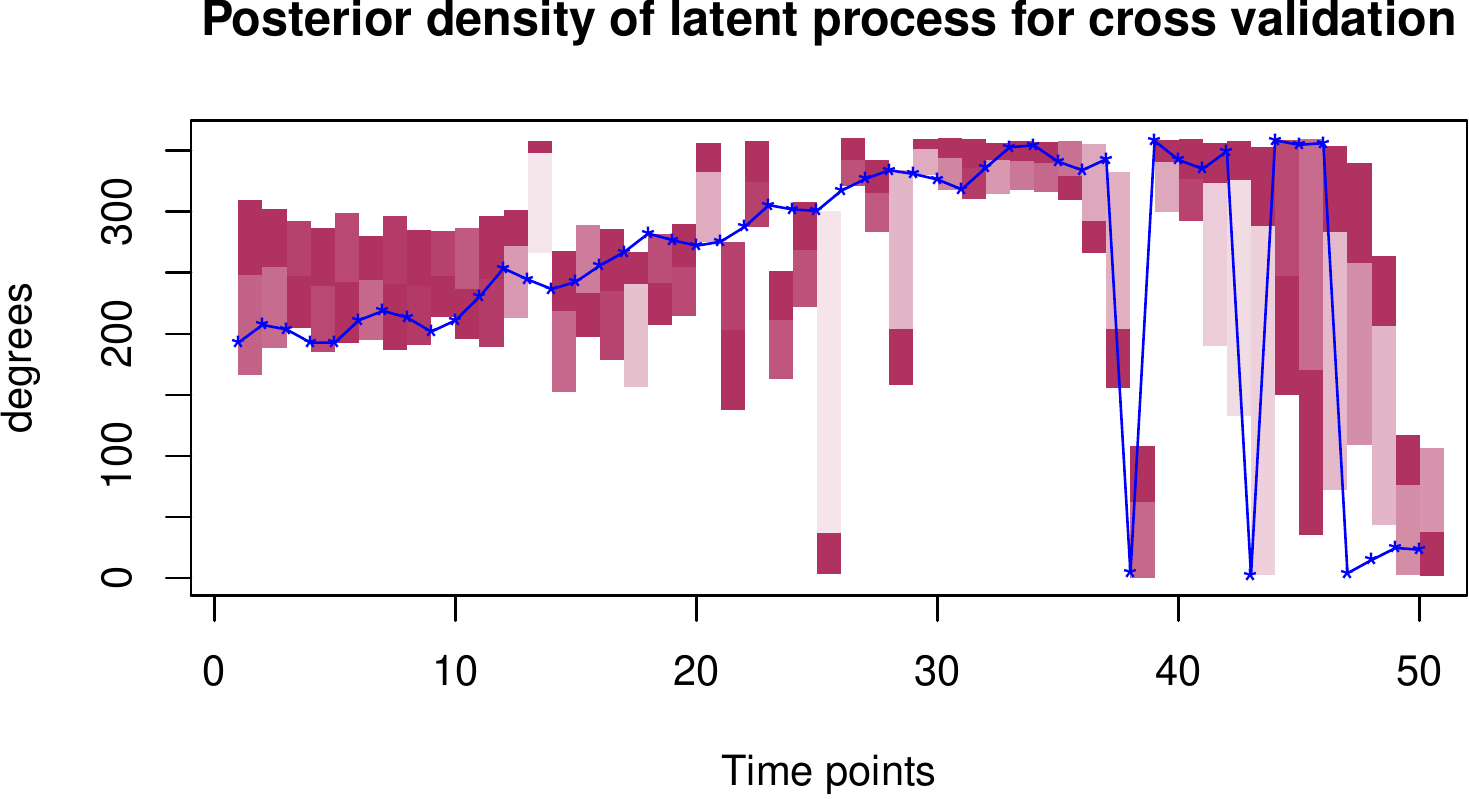}
\includegraphics[height = 3in,width=3in]{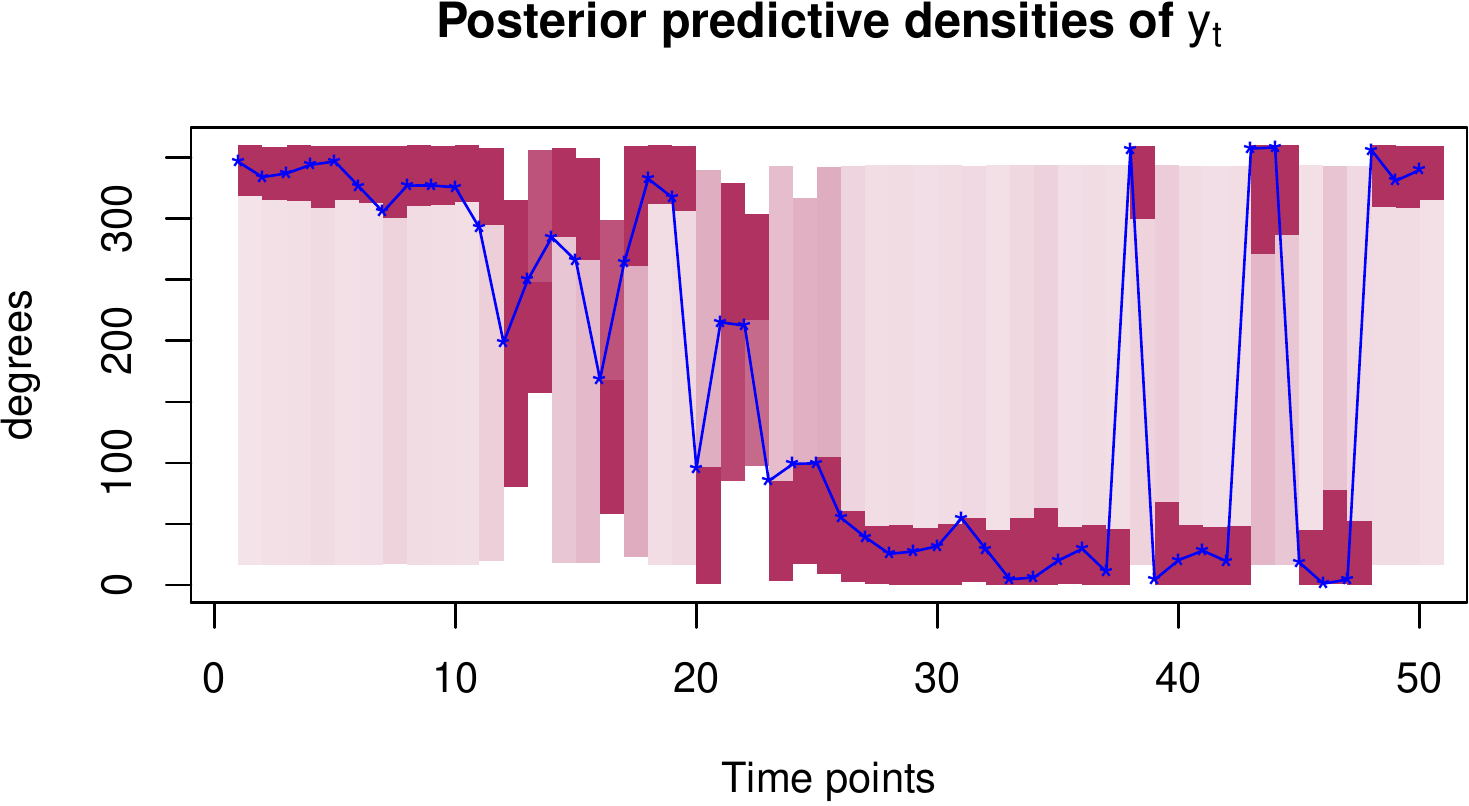}
\caption{Depiction of the posterior densities of the latent circular process $\left\{x_t;t=1,\ldots,50\right\}$ and the observed circular process $\left\{y_t;t=1,\ldots,50\right\}$ under the leave-one-out cross-validation scheme; higher the intensity of the color, higher is the posterior density. The blue stars denote the true time series for $x_t$ and $y_t$.}
\label{Fig:cross valid simu_data size_50}
\end{figure}

\begin{figure}[htp]
\centering
\includegraphics[height = 3in,width=3in]{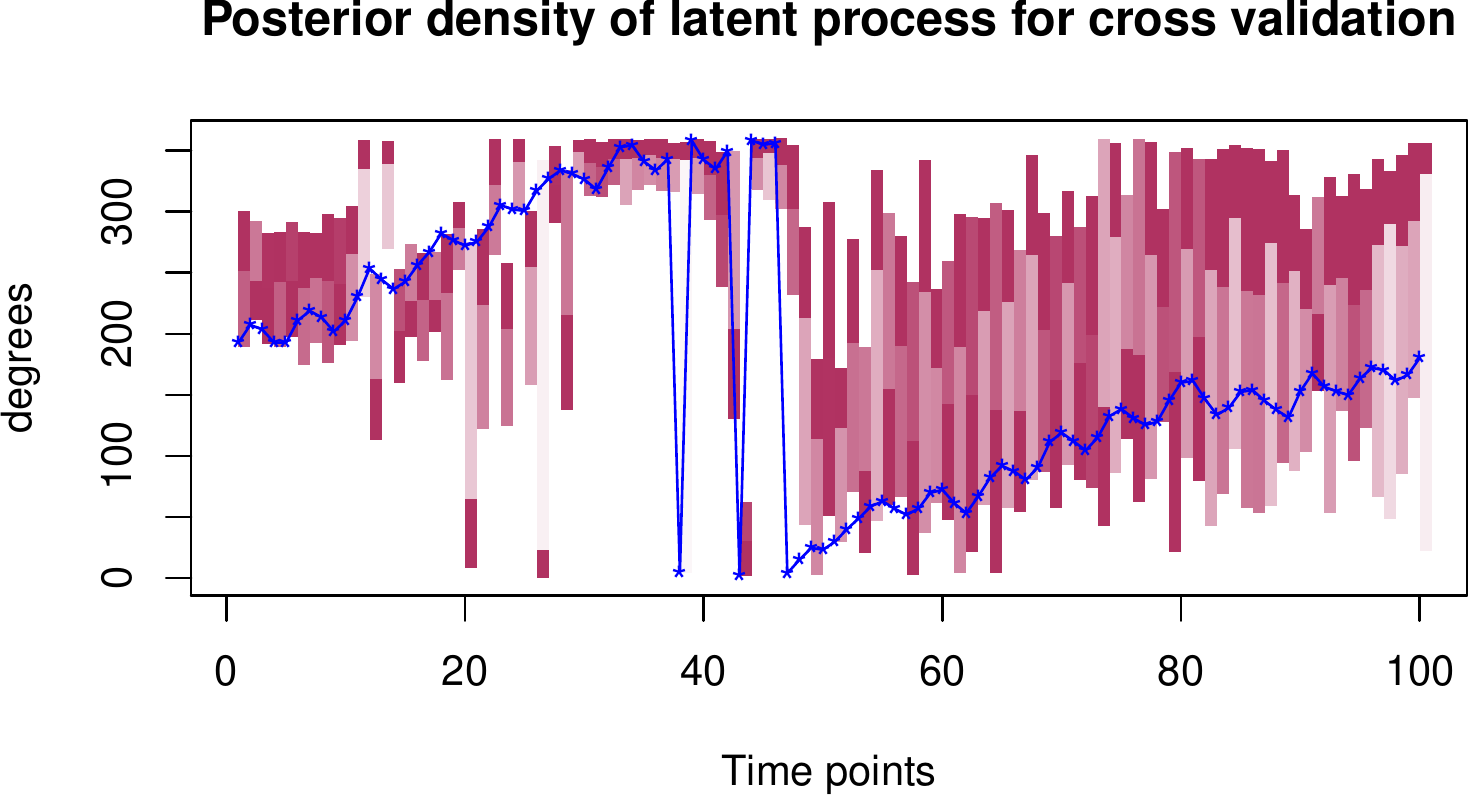}
\includegraphics[height = 3in,width=3in]{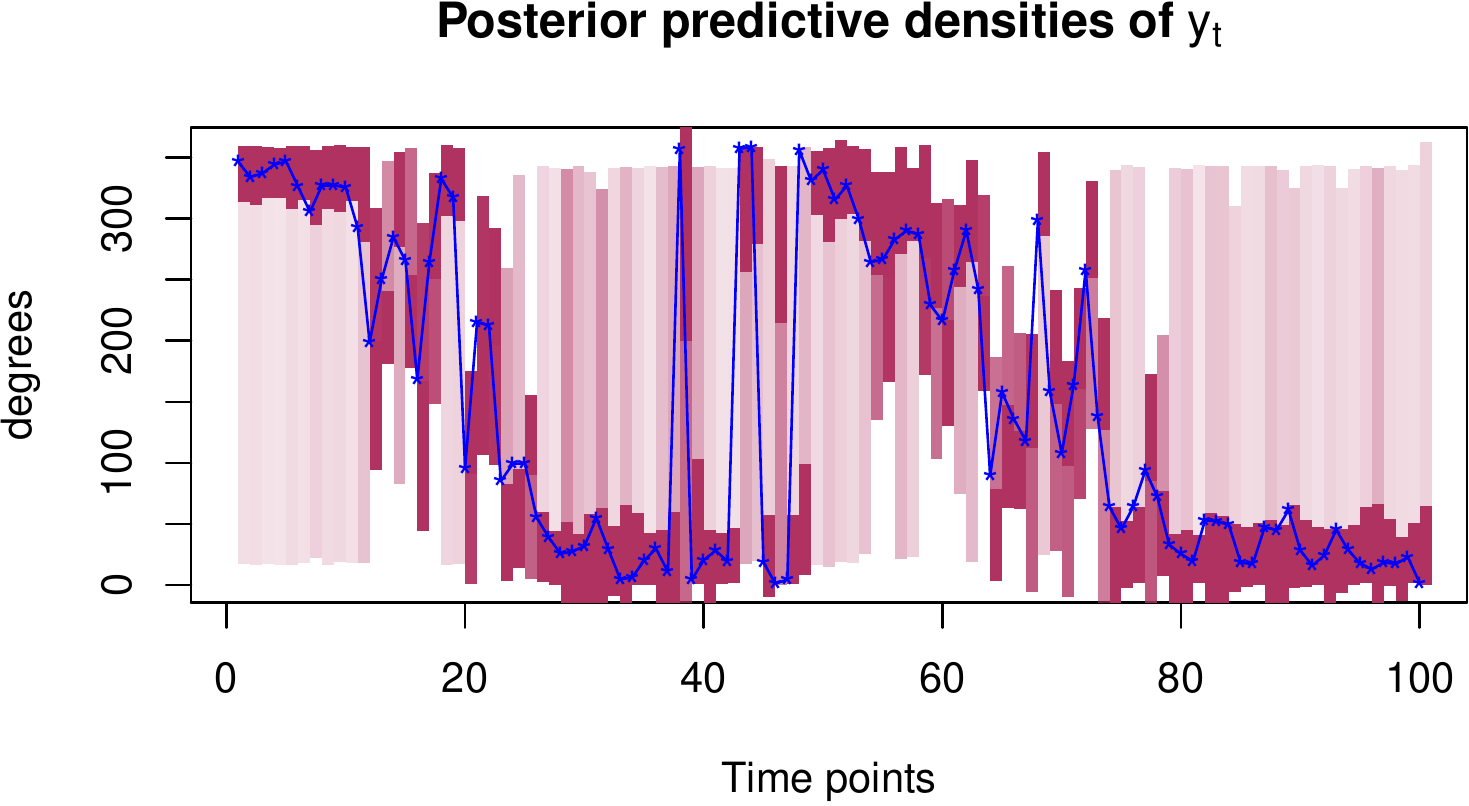}
\caption{Depiction of the posterior densities of the latent circular process $\left\{x_t;t=1,\ldots,100\right\}$ and the observed circular process $\left\{y_t;t=1,\ldots,100\right\}$ under the leave-one-out cross-validation scheme; higher the intensity of the color, higher is the posterior density. The blue stars denote the true time series for $x_t$ and $y_t$.}
\label{Fig:cross valid simu_data size_100}
\end{figure}

For further validation of our model, we consider detailed leave-one-out cross-validation for three simulated data sets 
with sample sizes 25, 50 and 100. In each case, at every time point $t$, we set aside $y_t$ pretending it to be
unobserved and predict the same, for $t=1,\ldots,T$. 
The colored posterior probability plots for the leave-one-out posterior predictive
densities of $\left\{y_t,~ t=1,\ldots,T\right\}$, along with the corresponding latent circular process 
$\left\{x_t,~t=1,\ldots,T\right\}$, 
are displayed in Figures \ref{Fig:cross valid simu_data size_25}, \ref{Fig:cross valid simu_data size_50} 
and \ref{Fig:cross valid simu_data size_100}, respectively, for simulated data with sample sizes 25, 50 and 100. 
It is very heartening to observe that in almost all the cases, $y_t$ and $x_t$ fall in high probability regions 
under the leave-one-out cross-validation scheme. Importantly, one should note that the sample sizes do not have 
any effect on our proposed methodology.    

Thus, in a nutshell, we can claim that our model performs quite optimistically, in spite of the true model 
being highly non-linear and assumed to be unknown. As a consequence, we expect our model and methods 
to perform adequately in general situations.

\section{Model validation with real data analysis}
\label{real_data_analysis}
In this section we validate our methodologies by applying them to two real circular time series data where the 
time-varying and circular covariates are originally known but we assume to be unknown. In other words, we attempt to predict
the circular covariates considering them to be unobserved latent variables. 
\subsection{Wind pair data}
\label{subsec:wind_pair}
\subsubsection{Brief description of the data}
\label{description_of_wind_speed}
\begin{figure}[htp]
\centering
\includegraphics[height=3in,width=3in]{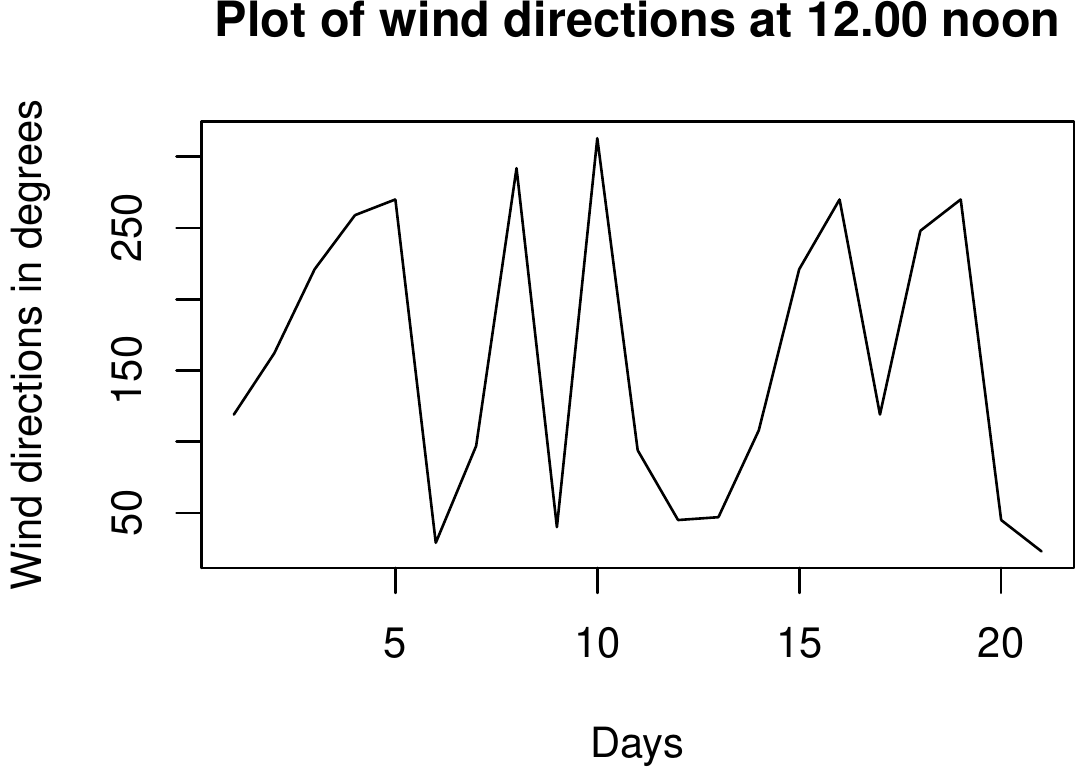}
\includegraphics[height=3in,width=3in]{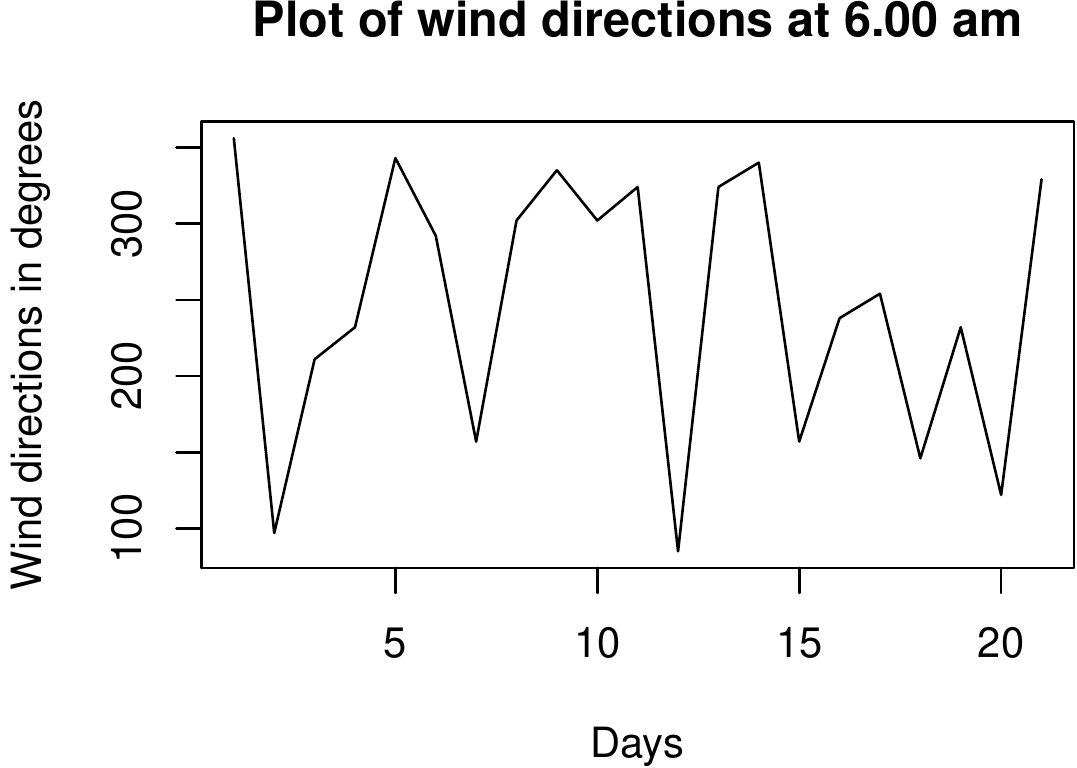}
\caption{Plot of the wind directions at 12.00 noon and 6.00 am  for 21 days.}
\label{Fig:Wind_direction_plot}
\end{figure}

Our first data set consists of a pair of wind directions recorded at $6.00$ am and $12.00$ 
noon for each of twenty one consecutive days at a weather station in Milwaukee. This data set, 
which is taken from \ctn{Fisher93} (see page $253$ of \ctn{Fisher93} for more details of the data),
is originally recorded in degrees; for our convenience, we converted them to radians ranging from 
$0$ to $2\pi$. Here we consider the wind directions at $6.00$ am to be the latent 
circular variables and those at $12.00$ noon, to be the observed variables. Moreover, we use 20 observations 
for our analysis and set aside the last observation for the purpose of forecasting. In this study we are 
mainly interested in demonstrating how well our model is able to capture 
the recorded, real, wind direction data observed at $6.00$ am, considered to be latent relative to our model. 
Plots of the true wind directions at both the time points for 21 days are provided in Figure \ref{Fig:Wind_direction_plot}.   

\subsubsection{Prior choices and MCMC implementations}
\label{prior_for_wind_direction}
We choose the prior means of $\bi{\beta}_f$ and $\bi{\beta}_g$ to be $(0,0,0,0)'$ and $(1,1,1,1)'$, respectively. 
As regards the prior covariance matrices of $\bi{\beta}_f$ and $\bi{\beta}_g$, we consider the identity matrix of 
dimension 4 and a diagonal matrix with diagonal entries $(1,1,0,0)'$, respectively. To avoid
identifiability issues, we fix the last two components of $\bi{\beta}_g$ at $1$.

The MLEs of $\sigma_f$, 
$\sigma_{\epsilon}$, $\sigma_g$ and $\sigma_{\eta}$, which are calculated using simulated annealing and assumed to be known
thereafter, turned out to be $2.74891$, $0.30649$, $0.71260$ and $0.36719$, respectively. 
Using these prior parameters we run $3.5\times 10^5$ MCMC iterations and store the last $50,000$ 
observations for inference, discarding the 
first $3\times 10^5$ as burn in. The time taken by our server machine 
to implement $3.5\times 10^5$ iterations of our MCMC algorithm is 11 hours 47 minutes.  
\subsubsection{Model fitting results}
\label{results_of_wind_direction}
\begin{figure}[htp]
\centering
\includegraphics[height=1.5in,width=1.5in]{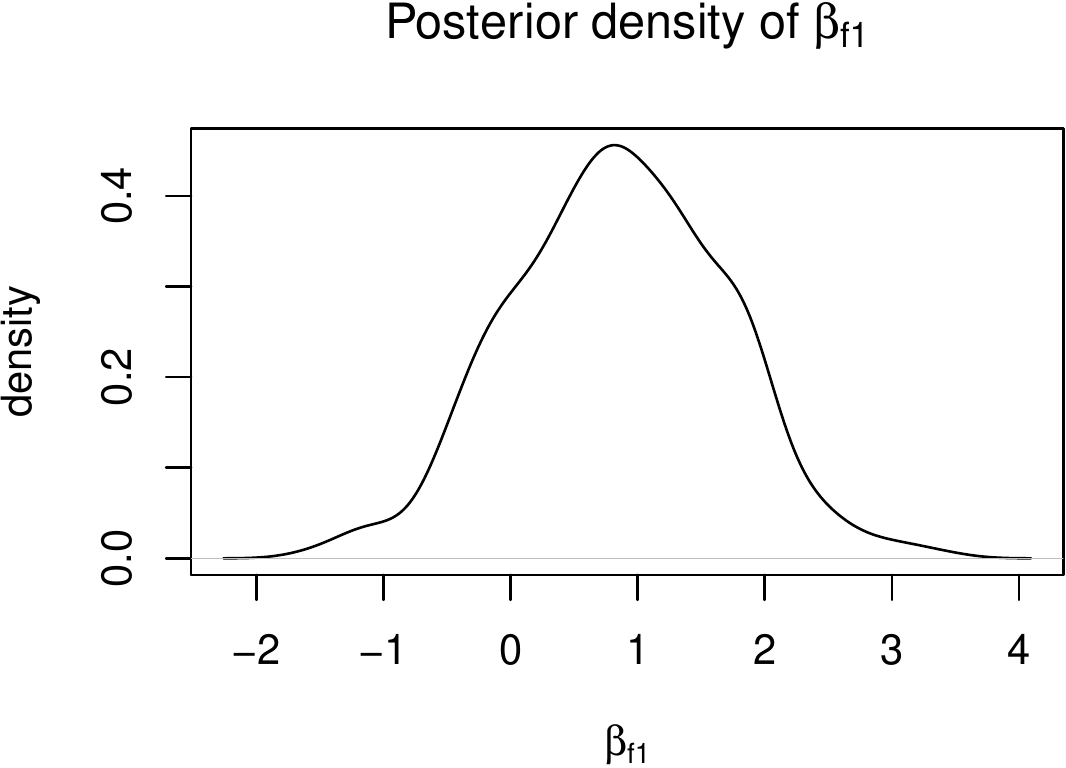}
\includegraphics[height=1.5in,width=1.5in]{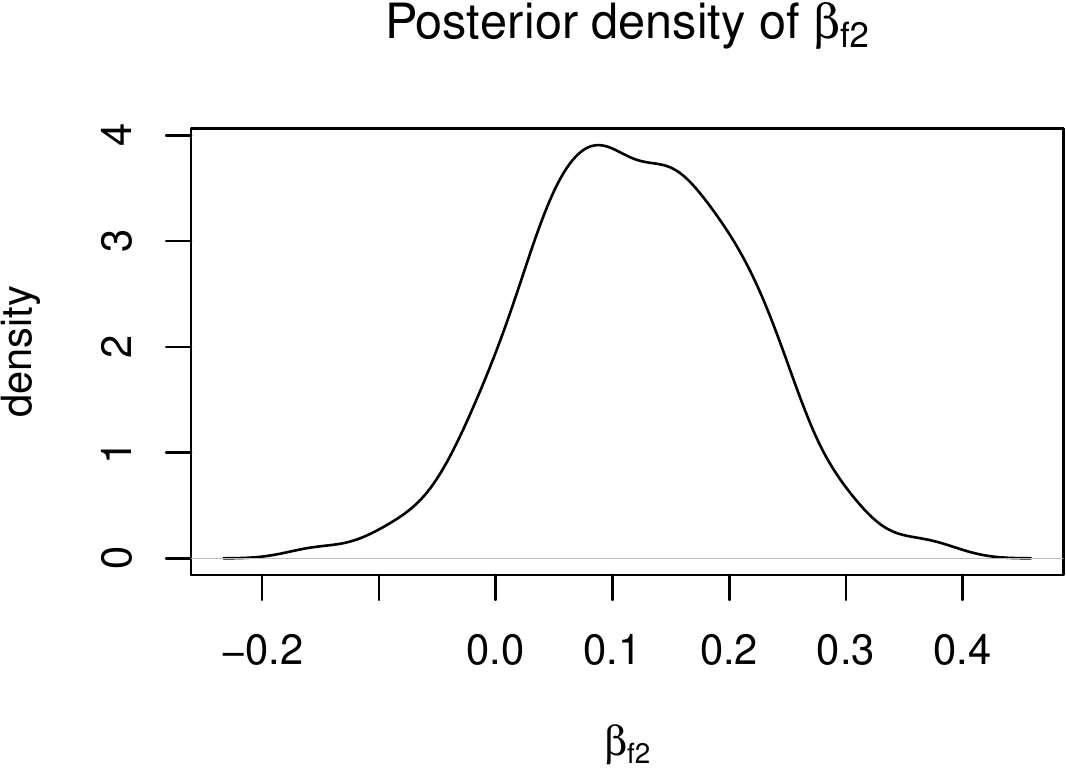}
\includegraphics[height=1.5in,width=1.5in]{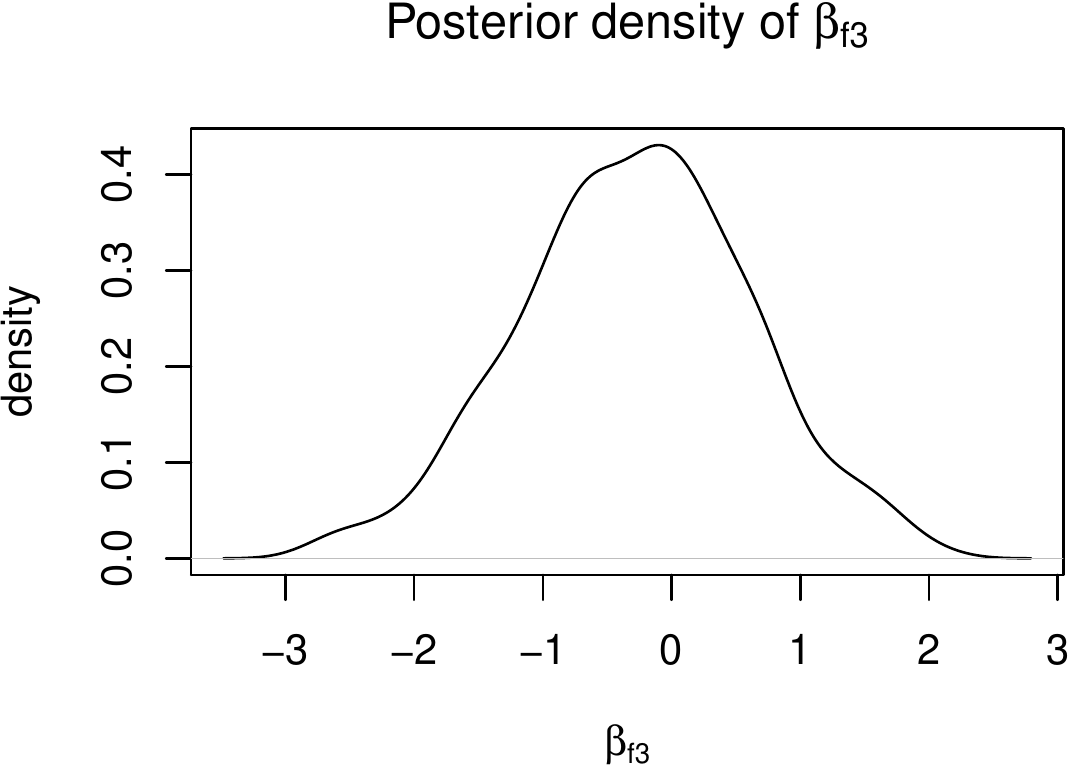}
\includegraphics[height=1.5in,width=1.5in]{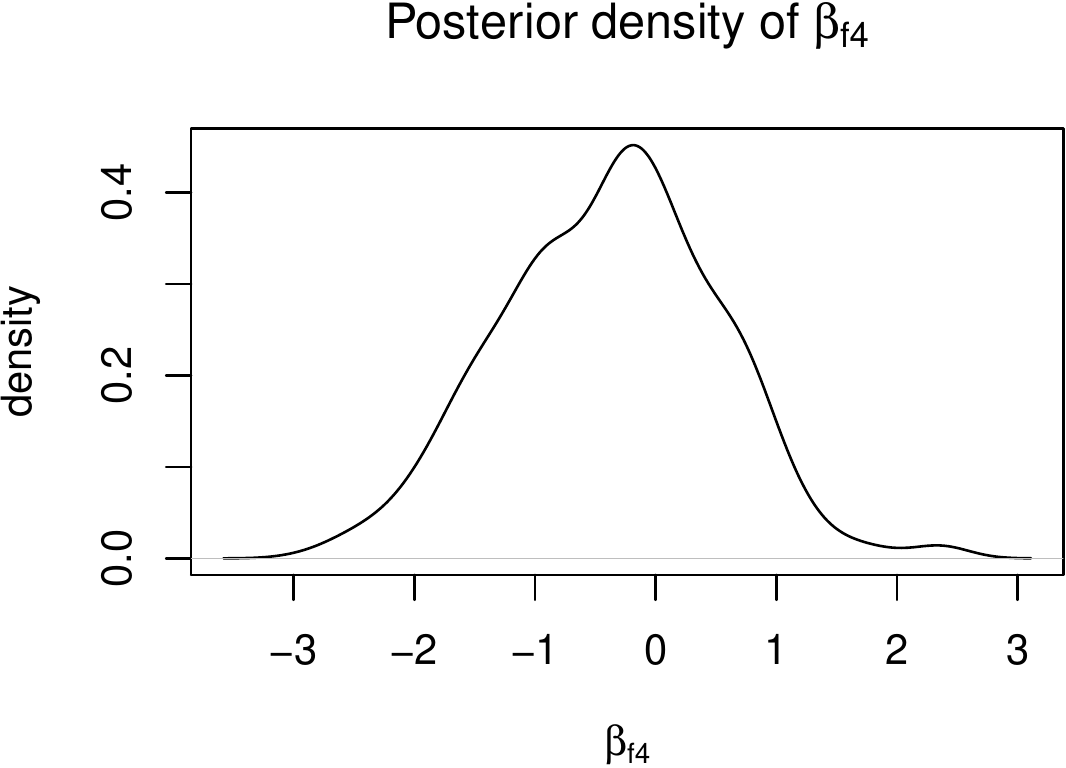}
\caption{Posterior densities of the four components of $\bi{\beta}_{f}$ for the wind pair data.}
\label{Fig:Post_of_beta_f_for_windpair_data}
\end{figure}

\begin{figure}[htp]
\centering
\includegraphics[height=2in,width=2in]{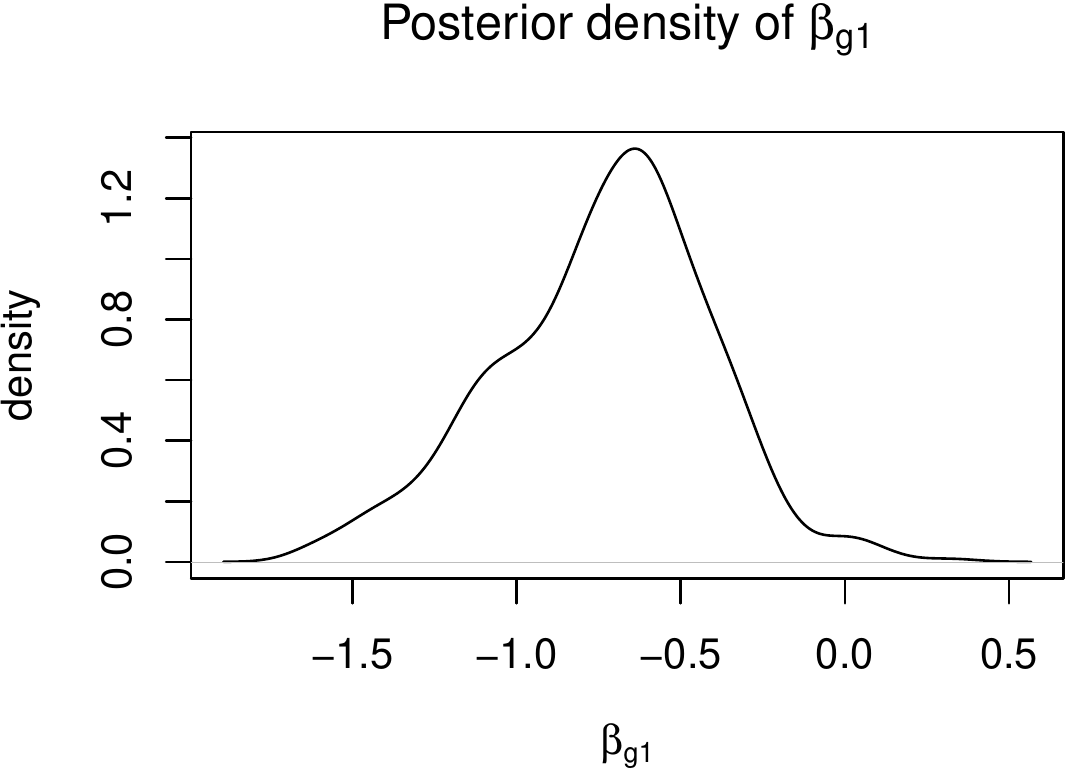}
\includegraphics[height=2in,width=2in]{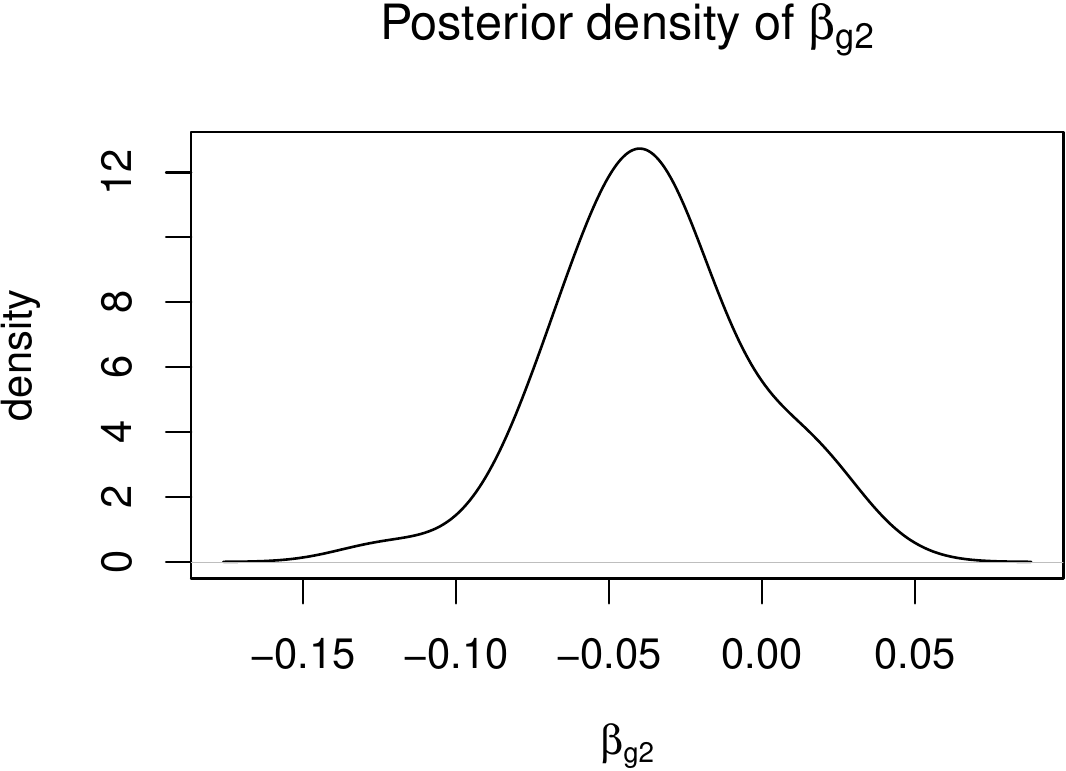}
\caption{Posterior densities of the first two components of $\bi{\beta}_{g}$ for the wind pair data.}
\label{Fig:Post_of_beta_g_for_windpair_data}
\end{figure}

\begin{figure}[htp]
\centering
\includegraphics[height=3in,width=5in]{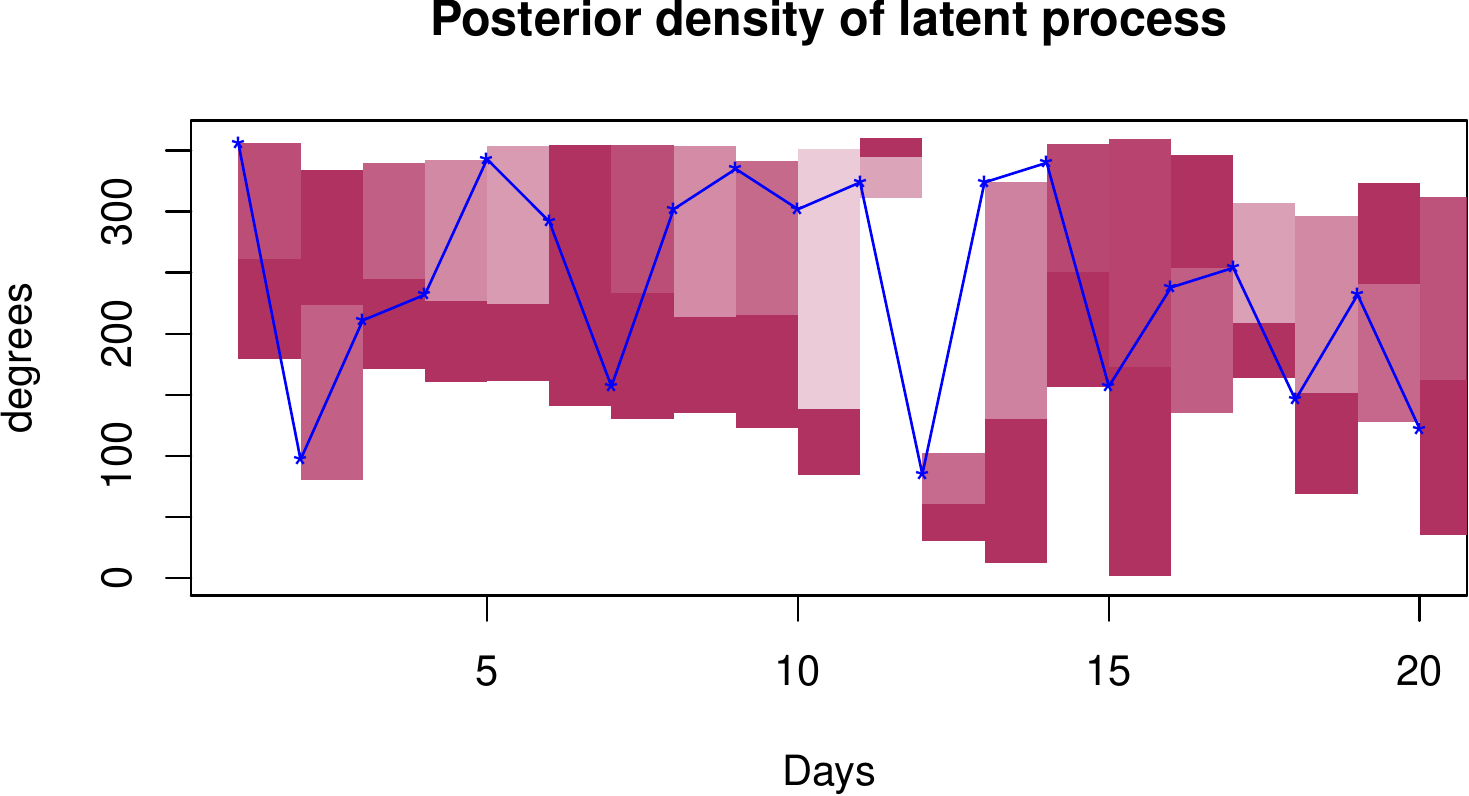}
\caption{Representation of the marginal posterior densities of the latent variables corresponding to
wind directions at 6.00 am as a color plot; progressively higher densities are represented 
by progressively intense colors. The blue stars represent the true wind directions (at 6.00 am) data.}
\label{Fig:latent_x_for_windpair_data}
\end{figure}

\begin{figure}[htp]
\centering
\includegraphics[height=3in,width=3in]{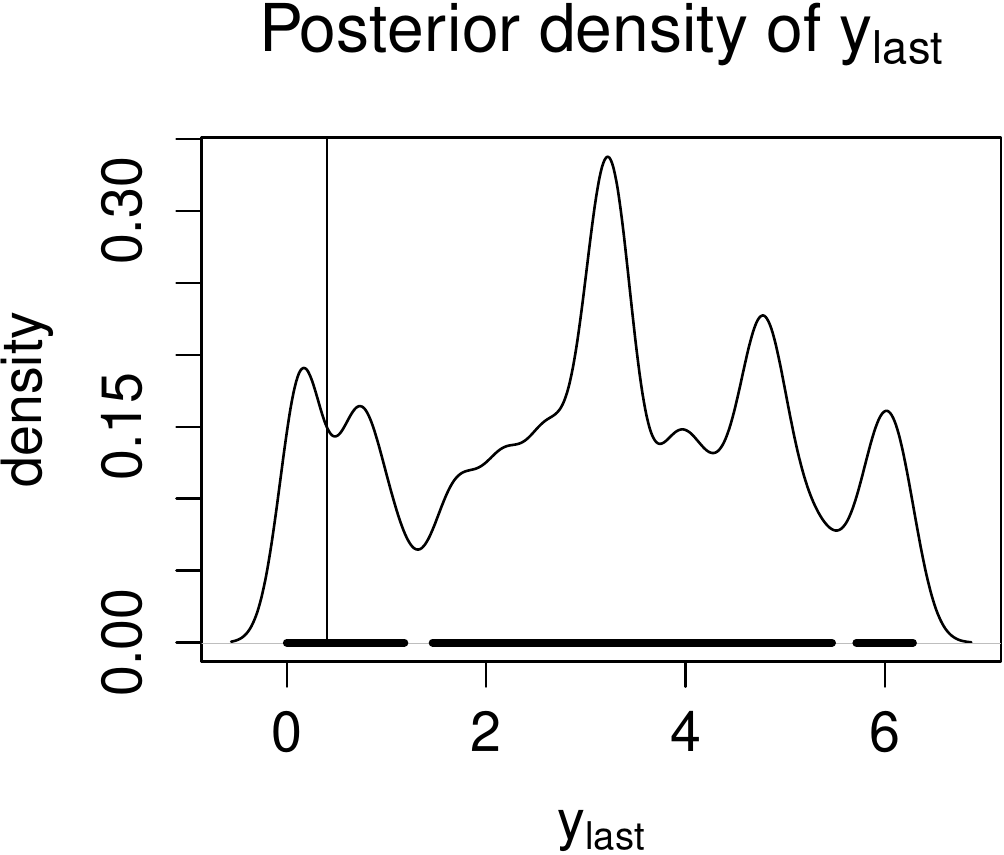}
\includegraphics[height=3in,width=3in]{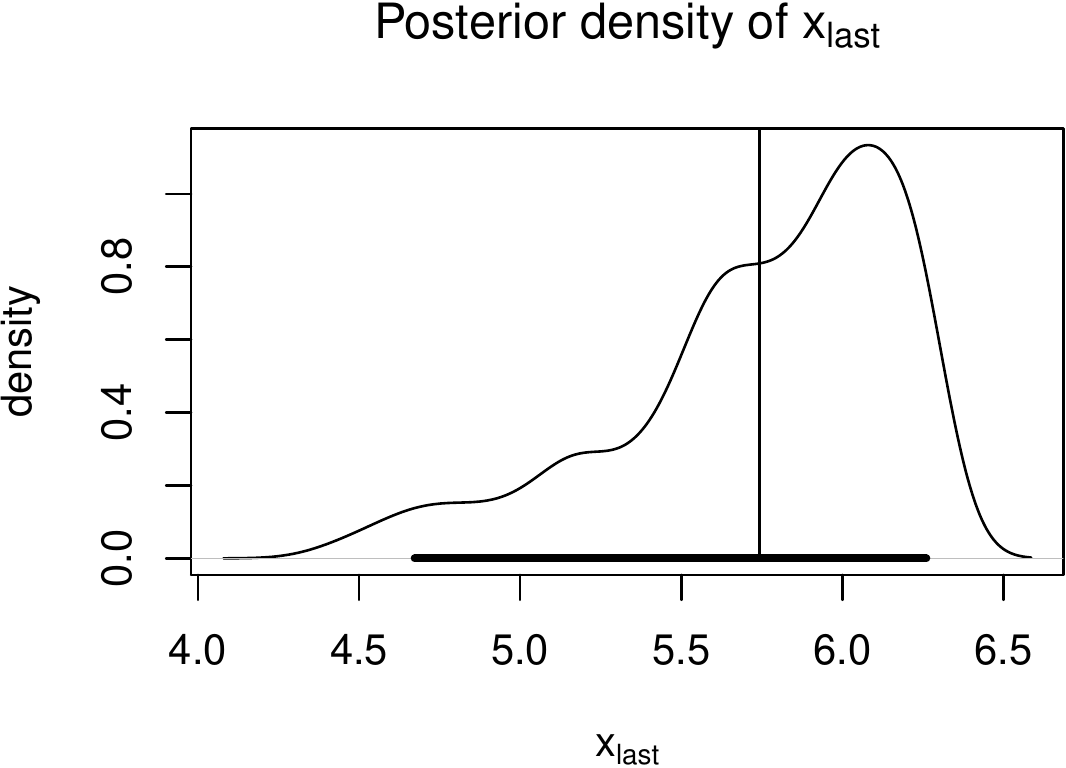}
\caption{From left\/: The first panel displays the posterior predictive density of the $21$st observation of 
wind directions at 12.00 noon for the wind pair data. The second plot represents the posterior predictive density 
of the 21st observation of wind direction at 6.00 am (latent variable). The thick 
horizontal line denotes the 95\% highest posterior density credible interval and the vertical line denotes 
the true value (in radian) for both the panels.}
\label{Fig:Post_predictive_of_y_last_of_windpair_data}
\end{figure}

The posterior densities of the four components of the vector
$\bi{\beta}_f$ and the two components of $\bi{\beta}_g$ are provided in 
Figures \ref{Fig:Post_of_beta_f_for_windpair_data} and \ref{Fig:Post_of_beta_g_for_windpair_data}, respectively,
and the posterior predictive densities of $y_{21}$ and $x_{21}$ are displayed in
Figure \ref{Fig:Post_predictive_of_y_last_of_windpair_data}.
It is observed that the true values of $y_{21}$ and $x_{21}$ 
fall well within the 95\% highest posterior density credible intervals, clearly demonstrating how well our model 
and the prior distributions of the parameters successfully describe the uncertainty present in the data. 
The marginal posterior densities of the latent variables are shown in Figure \ref{Fig:latent_x_for_windpair_data}, 
where progressively higher intensities of the color denote regions of progressively
higher posterior densities, and the blue stars 
stand for the true values of the wind directions at time $6.00$ am. It is noted that the true values of the latent variable, 
namely, the wind directions at time $6.00$ am (in radians), fall mostly in the respective high probability regions. 
Our Bayesian model and methods are 
particularly impressive in the sense that the highly nonlinear trend of the latent 
variable is adequately captured even when the data size is only 20. Since our 
nonparametric ideas permit the unknown observational and evolutionary function, based 
on Gaussian processes, to change with time, such encouraging performance is not surprising.

\subsubsection{Cross-validation results}
\label{subsubsec:cross valid wind_direction}
\begin{figure}[htp]
\centering
\includegraphics[height = 3in,width=3in]{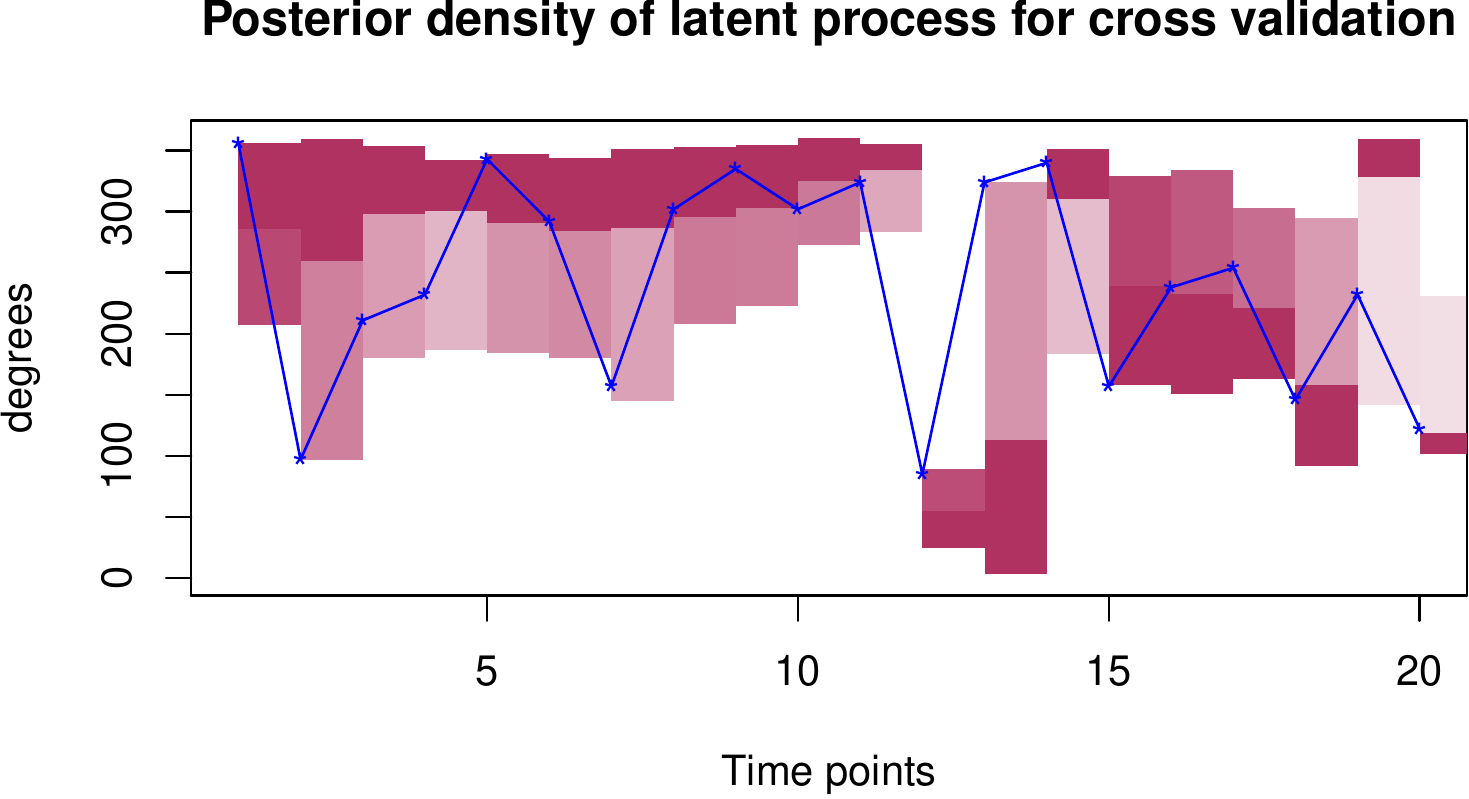}
\includegraphics[height = 3in,width=3in]{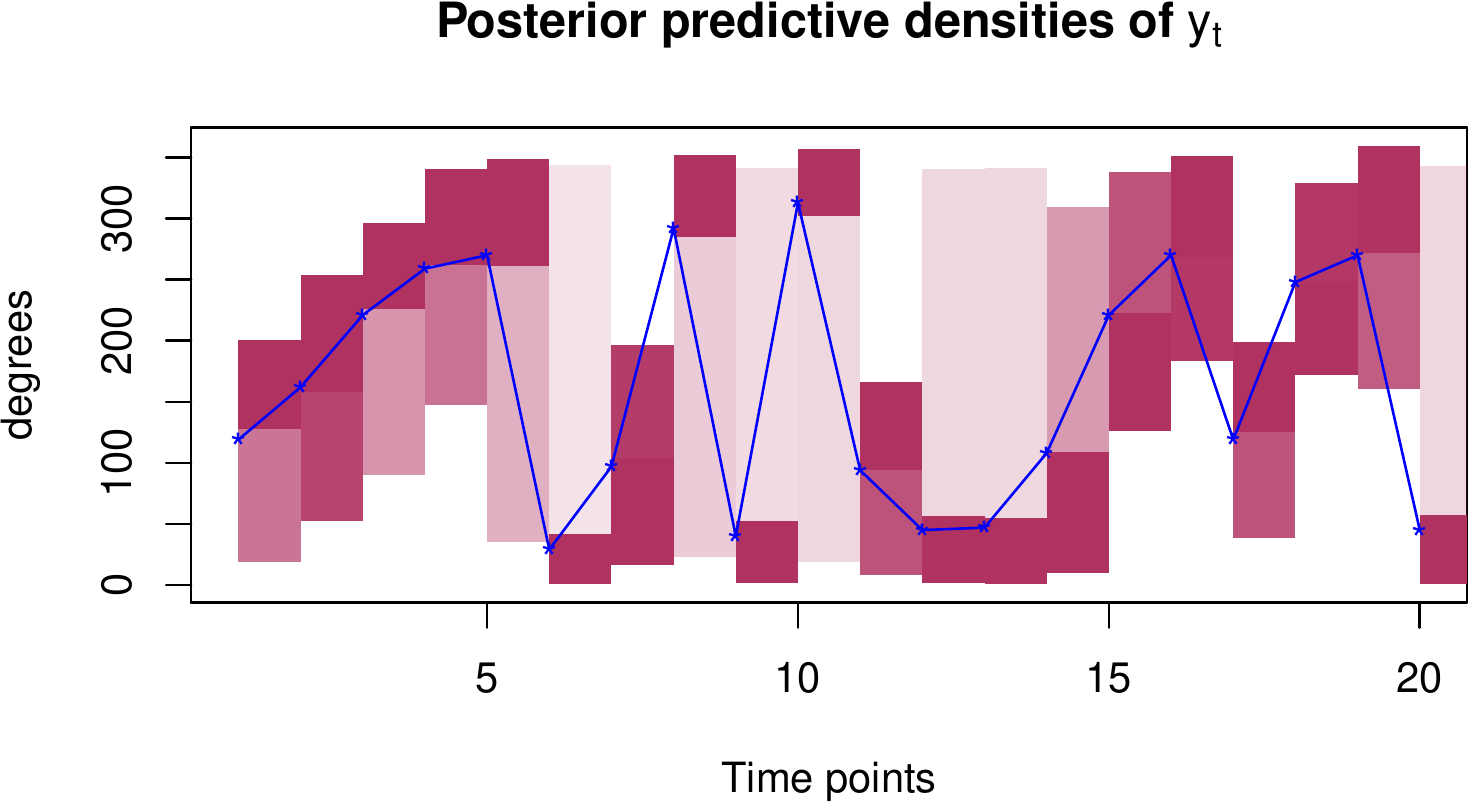}
\caption{Depiction of the posterior densities of the latent circular process wind direction at time $6.00$ am for 20 days 
and the observed circular process wind direction at time $12.00$ pm for 20 days under the leave-one-out cross-validation scheme; higher the intensity of the color, higher is the posterior density. The blue stars denote the true values.}
\label{Fig:cross valid wind pair data}
\end{figure}

We now present the results of leave-one-out cross-validation for the wind pair data. 
As in the simulation studies, here we predict wind directions at time $12.00$ pm for 20 days, assuming 
that the variable is not observed on the
$t$-th day, $t=1,\ldots, 20$. We also record the posterior distribution of the corresponding latent process,
namely, wind directions at time $6.00$ am.
The posterior densities are depicted in Figure \ref{Fig:cross valid wind pair data} for both the observed and 
latent circular processes. As expected, the true values of both the wind directions fall in the high probability 
regions in almost all the cases.    

\subsection{Spawning time and low tide time data}
\label{subsec:Spawn_time_data}
\subsubsection{A brief description of the data set}
\label{subsubsec:Spawn_data}
\begin{figure}[htp]
\centering
\includegraphics[height=3in,width=3in]{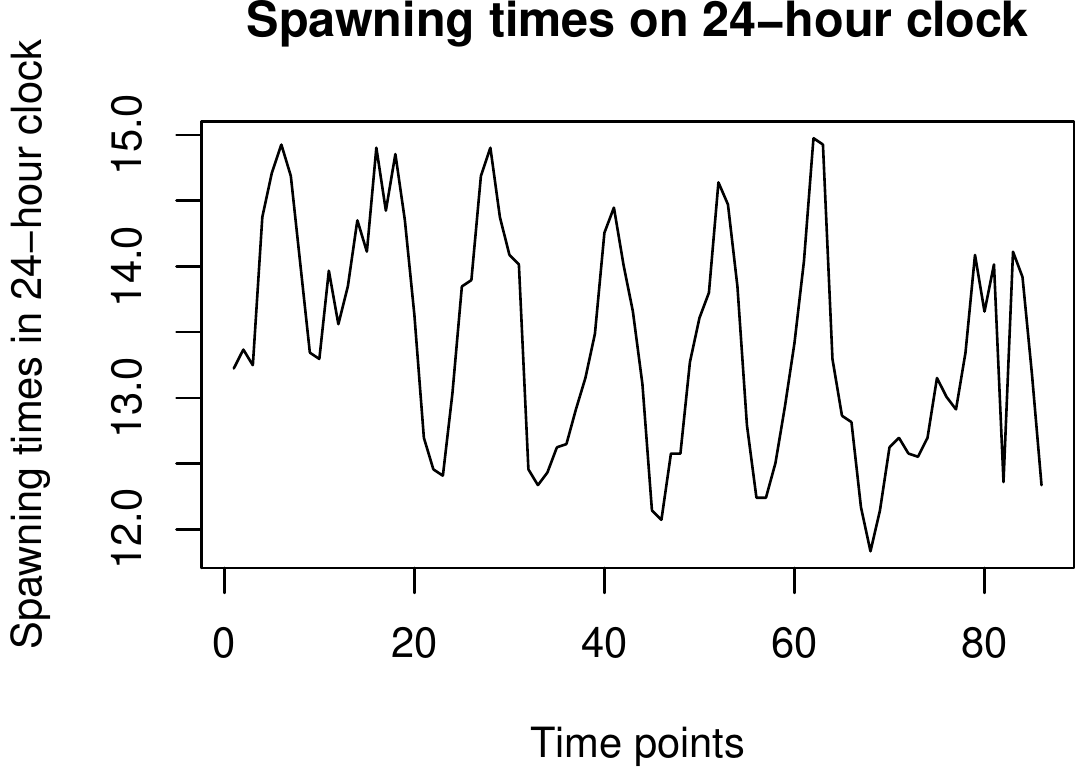}
\includegraphics[height=3in,width=3in]{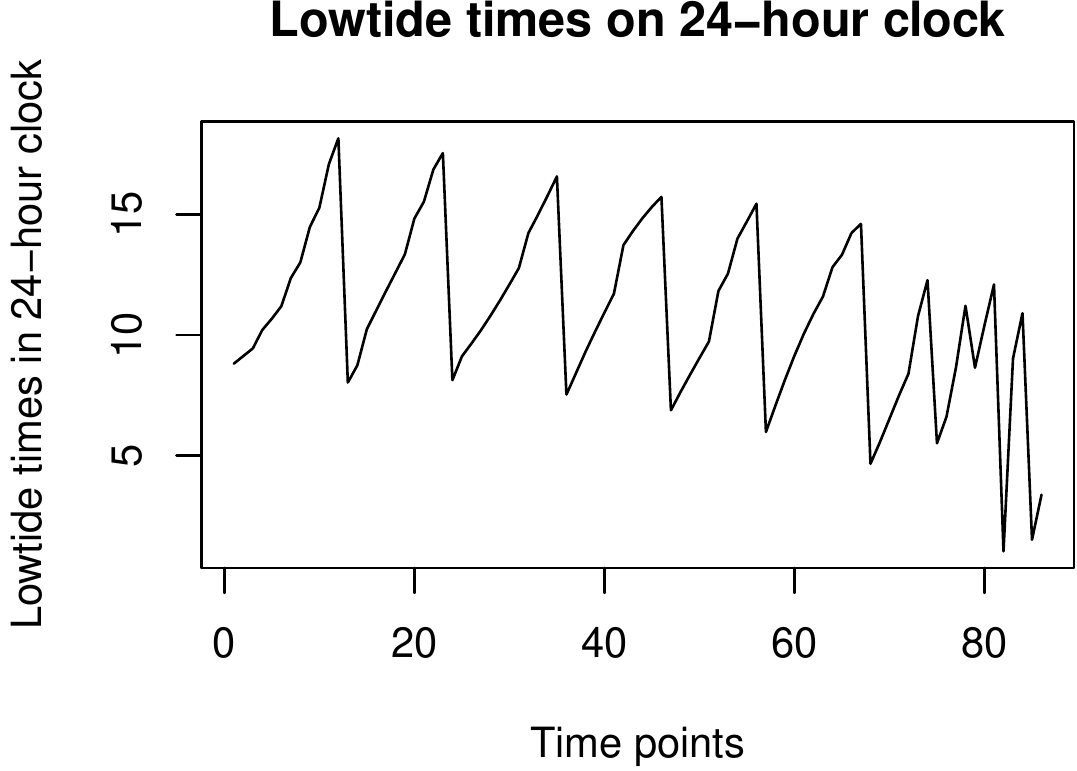}
\caption{Plots of spawning times of a fish and low tide times on the 24-hour clock.}
\label{Fig:Spawntime_and_lowtide_times_on_24_hour_clock}
\end{figure}

\begin{figure}[htp]
\centering
\includegraphics[height=3in,width=3in]{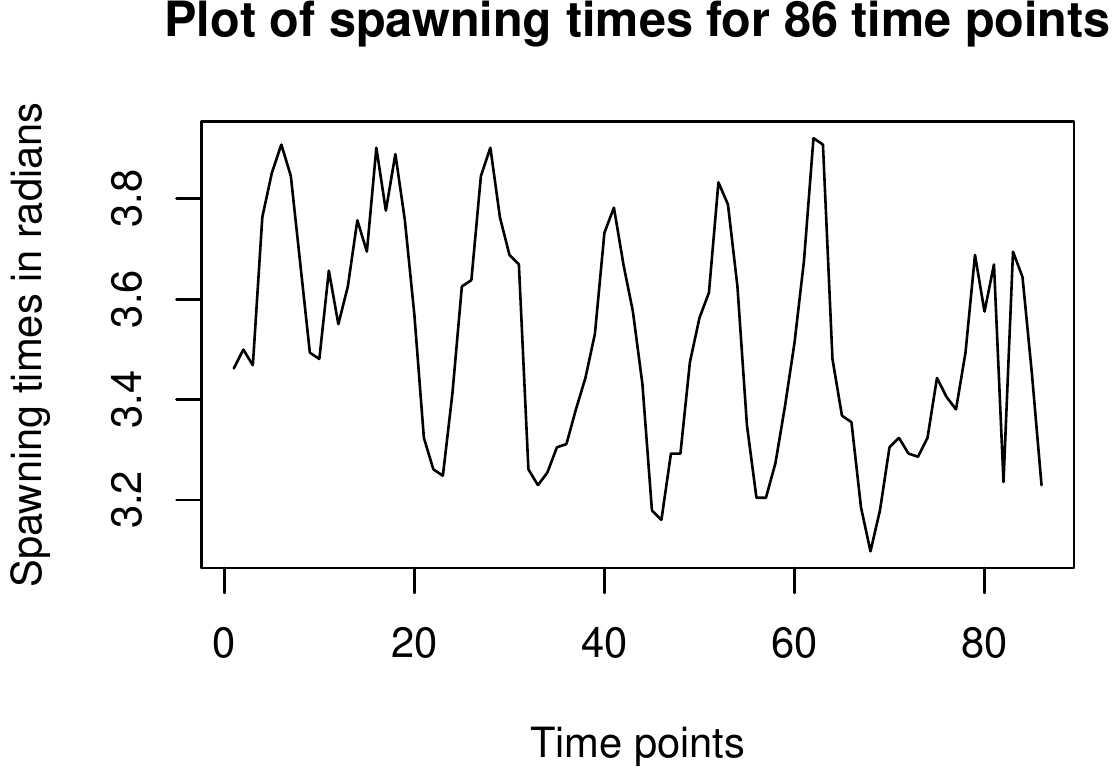}
\includegraphics[height=3in,width=3in]{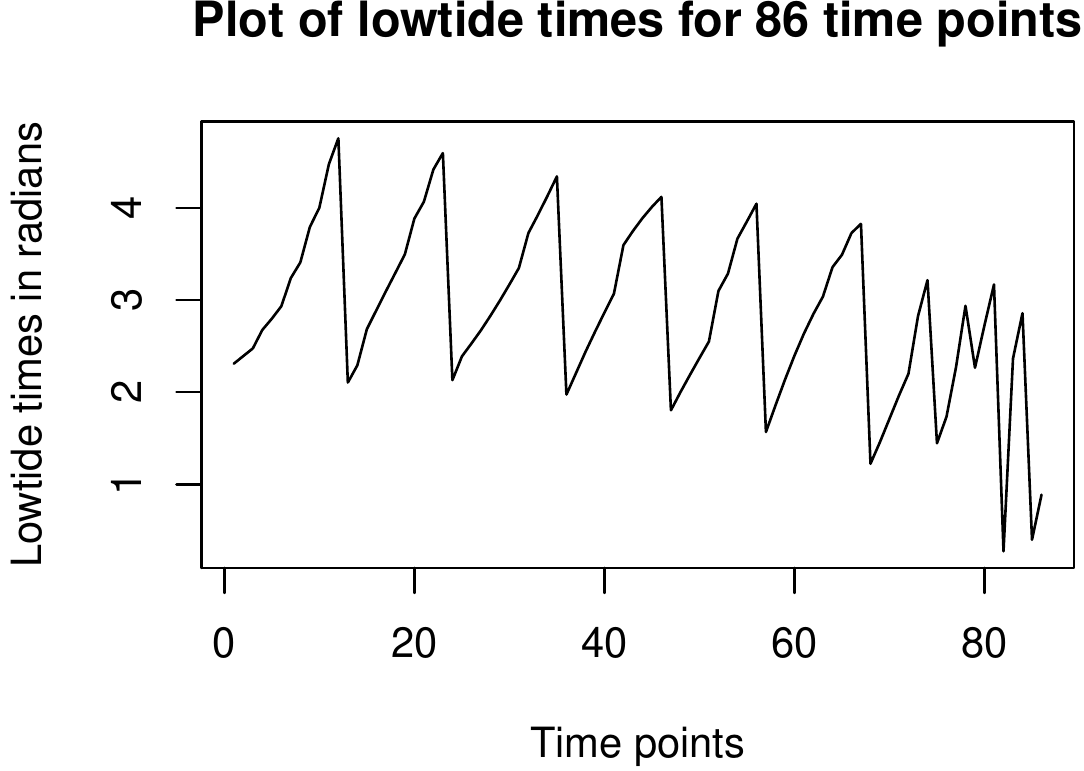}
\caption{Plots of spawning times of a fish and low tide times after wrapping on $(0,2\pi]$.}
\label{Fig:Spawntime_and_lowtide_times_in_radians}
\end{figure}

We apply our model and methodology to another real data set for the purpose of validation. 
Robert R. Warner at the University of California collected data on the spawning times of a particular 
fish in a marine biology study (see \ctn{Lund99}). It is seen that the spawning time of a fish is influenced by the tidal 
characteristic of the fish's local environment. A regression study was done by \ctn{Lund99} considering 
spawning time as dependent variable and time of low tide as the covariate. There are 86 observations for 
spawning times and low tide times. The plots of both the variables on the 24-hour clock scale are provided in 
Figure \ref{Fig:Spawntime_and_lowtide_times_on_24_hour_clock}. 
We wrap the spawning time points and low tide time points on the $(0,2\pi]$ scale 
(a corresponding plot is given in Figure \ref{Fig:Spawntime_and_lowtide_times_in_radians}) for our analysis. 
For our model we assume that the low tide times are latent and spawning times are observed. We use 85 
observations for our analysis and set aside the last observation for the prediction purpose. As in 
Section \ref{subsec:wind_pair} here also our main aim is to show that our methodology is equipped enough to capture the recorded 
low tide times, considered to be latent with respect to our model.

\subsubsection{Prior choices and MCMC implementations for spawning time data}
\label{prior_for_spawning_time_data}
In this example, we choose the prior means of $\bi{\beta}_f$ and $\bi{\beta}_g$ to be $(0,0,0,0)'$ and $(0,1.25,1,1)'$, and 
the prior covariance matrices for $\bi{\beta}_f$ and $\bi{\beta}_g$ to be the identity matrix of 
dimension 4 and a diagonal matrix with diagonal entries $(1,1,0,0)'$, respectively, the latter matrix
signifying, as usual, our attempt to resolve identifiability problems.
The MLEs of $\sigma_f$, $\sigma_{\epsilon}$, $\sigma_g$ and $\sigma_{\eta}$, obtained by simulated annealing, turned out to be 
$1.97462$, $0.37569$, $0.67960$ and $0.31686$, respectively. We implemented $2.5\times 10^5$ iterations of 
our MCMC algorithm, storing the last $50,000$ observations for inference.
The time taken by our server machine 
to implement this exercise is 18 hours 54 minutes.

\subsection{Results and discussions on the spawning time data}
\label{result_for_spawning_time_data}
\begin{figure}[htp]
\centering
\includegraphics[height=1.5in,width=1.5in]{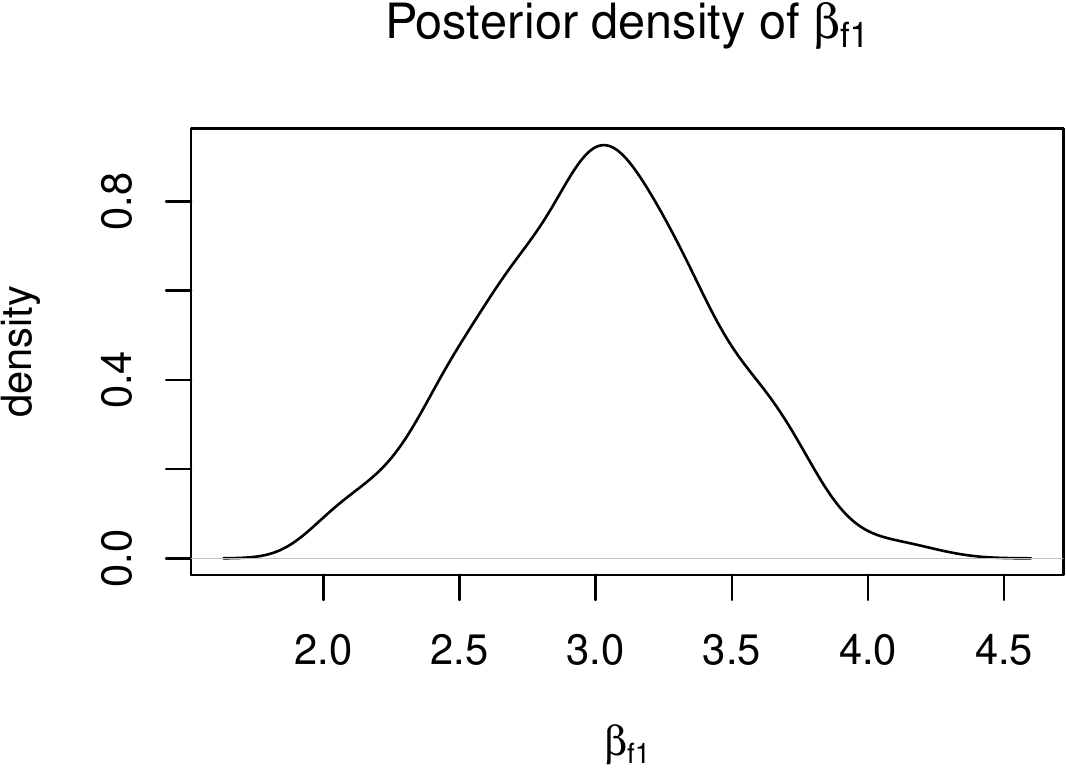}
\includegraphics[height=1.5in,width=1.5in]{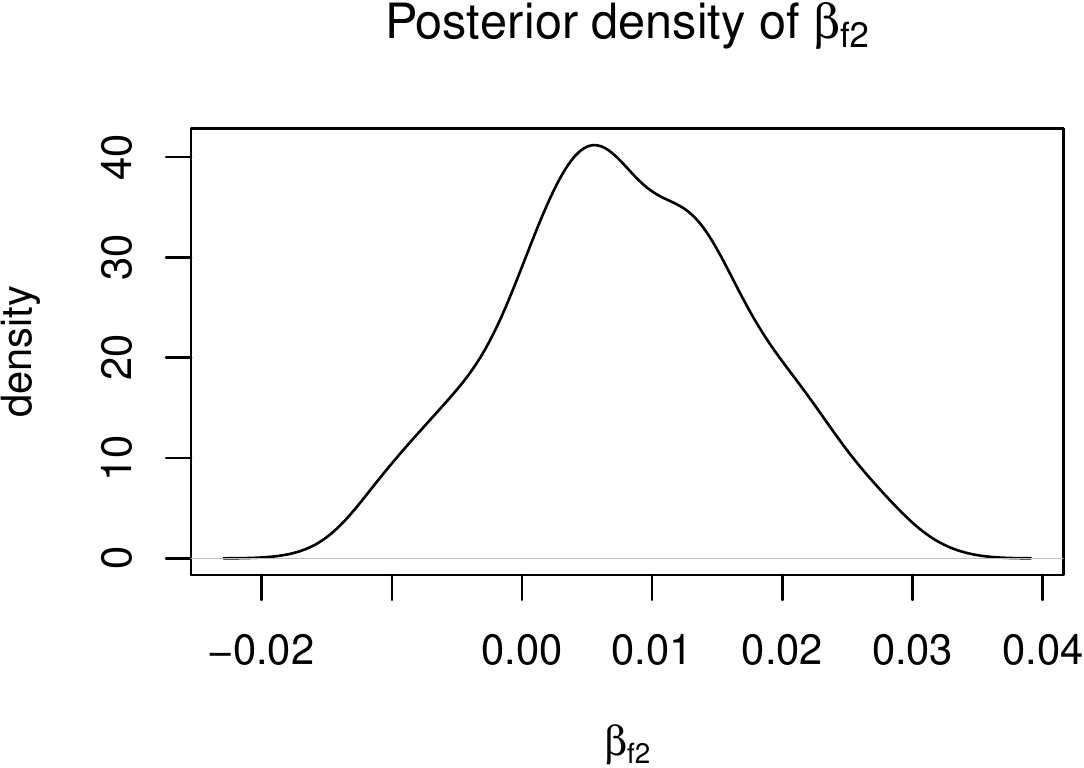}
\includegraphics[height=1.5in,width=1.5in]{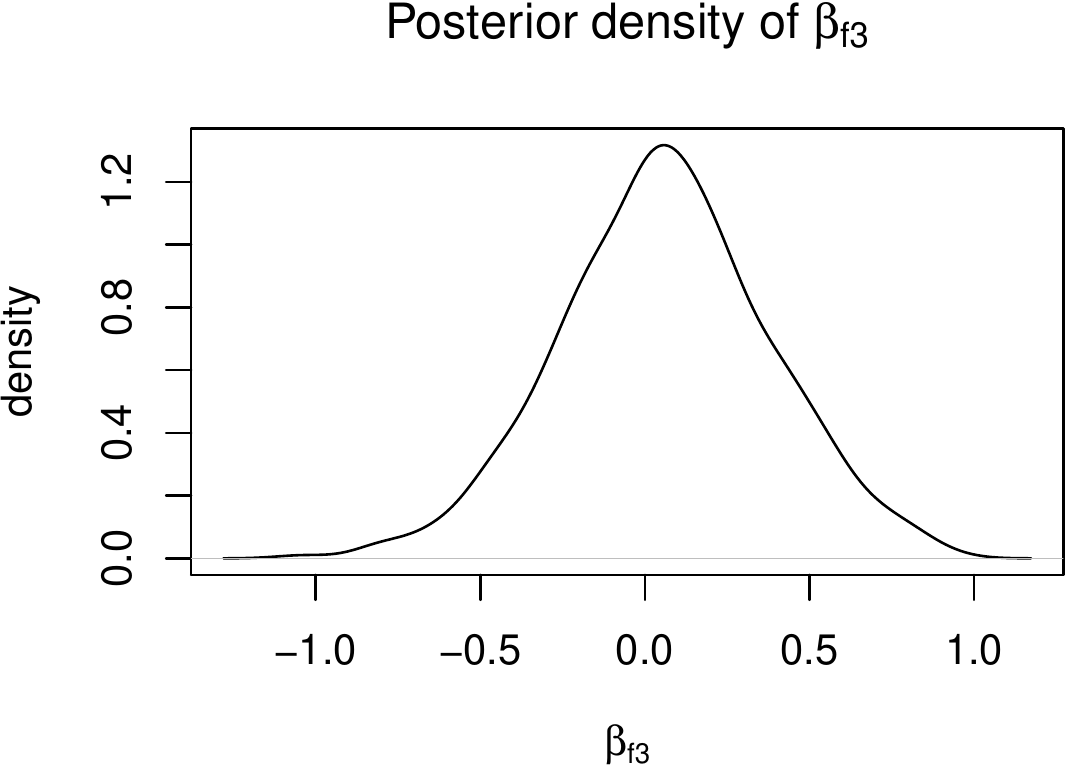}
\includegraphics[height=1.5in,width=1.5in]{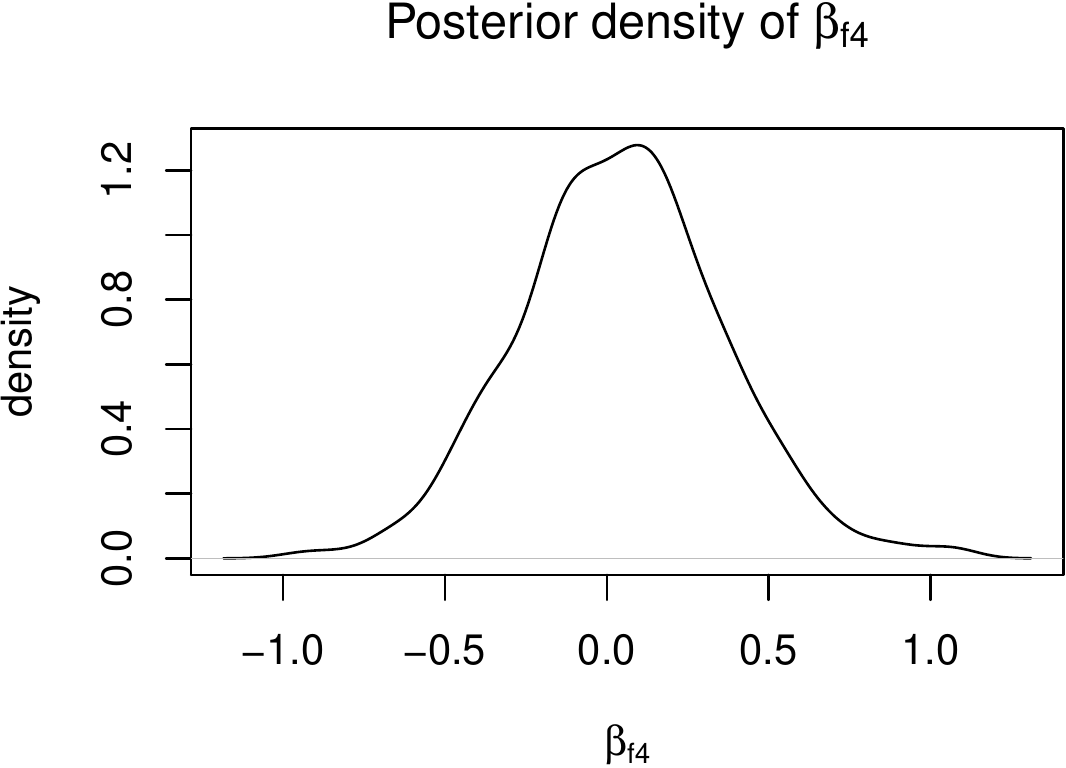}
\caption{Posterior densities of the four components of $\bi{\beta}_{f}$ for the spawning time data.}
\label{Fig:Post_of_beta_f_for_spawning_time_data}
\end{figure}

\begin{figure}[htp]
\centering
\includegraphics[height=2in,width=2in]{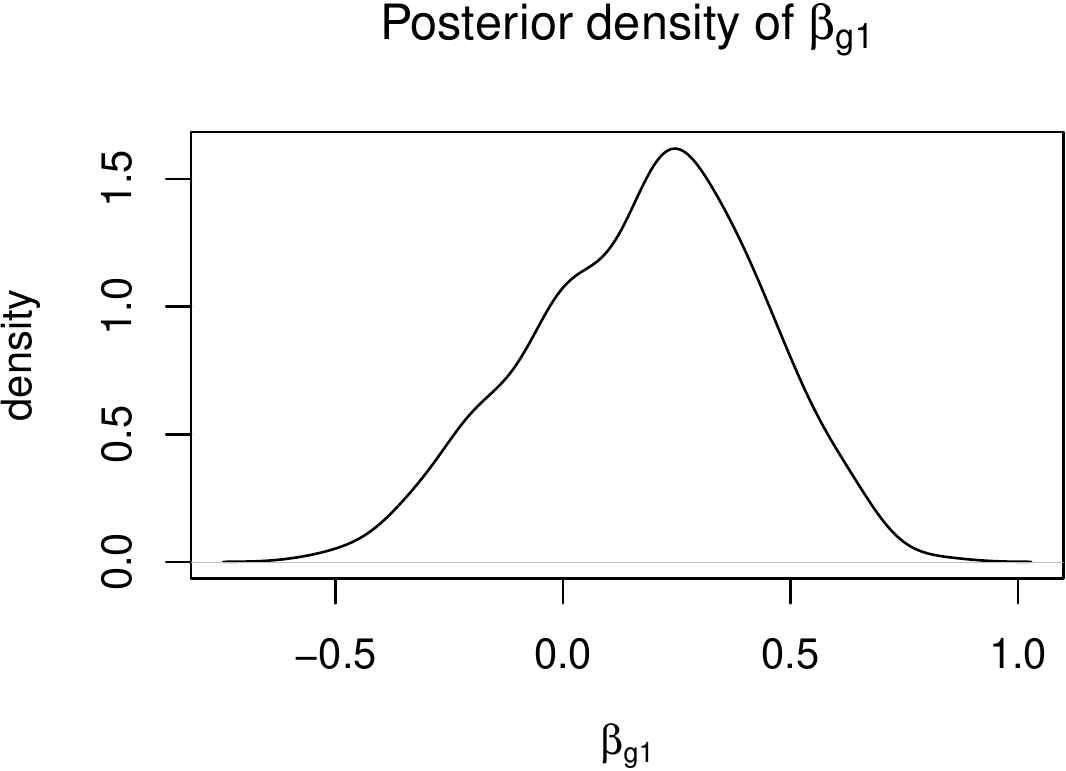}
\includegraphics[height=2in,width=2in]{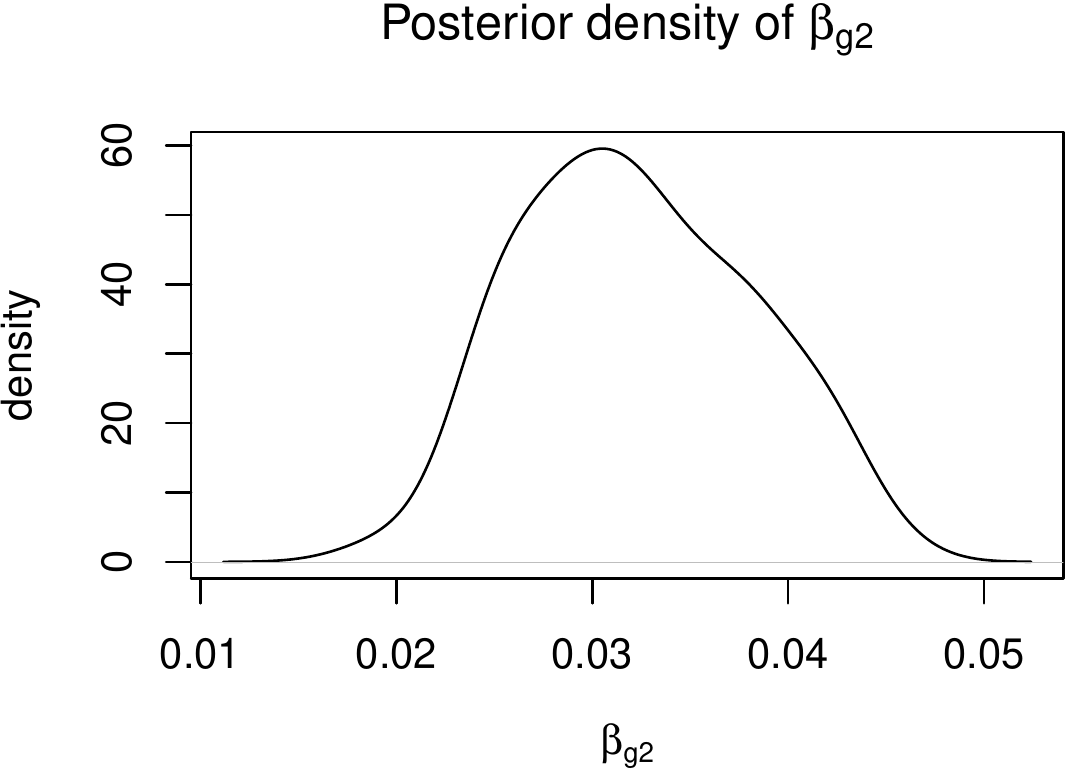}
\caption{Posterior densities of the first two components of $\bi{\beta}_{g}$ for the spawning time data.}
\label{Fig:Post_of_beta_g_for_spawning_time_data}
\end{figure}

\begin{figure}[htp]
\centering
\includegraphics[height=3in,width=5in]{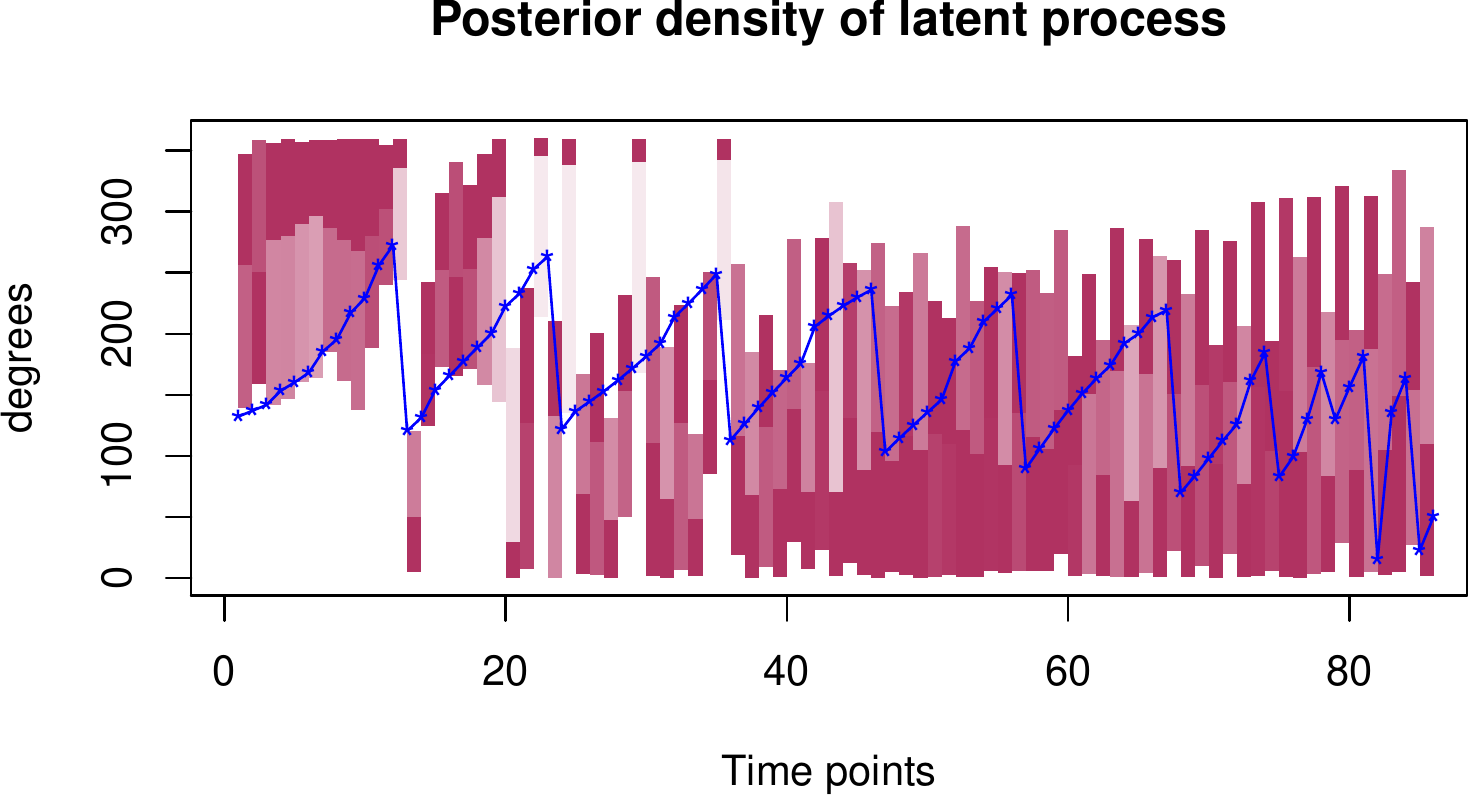}
\caption{Representation of the marginal posterior densities of the latent variables corresponding to 85 low tide times as a color plot; progressively higher densities are represented by progressively intense colors. 
The blue stars represent the true low tide timings (in degrees).}
\label{Fig:latent_x_for_spawning_time}
\end{figure}

\begin{figure}[htp]
\centering
\includegraphics[height=3in,width=3in]{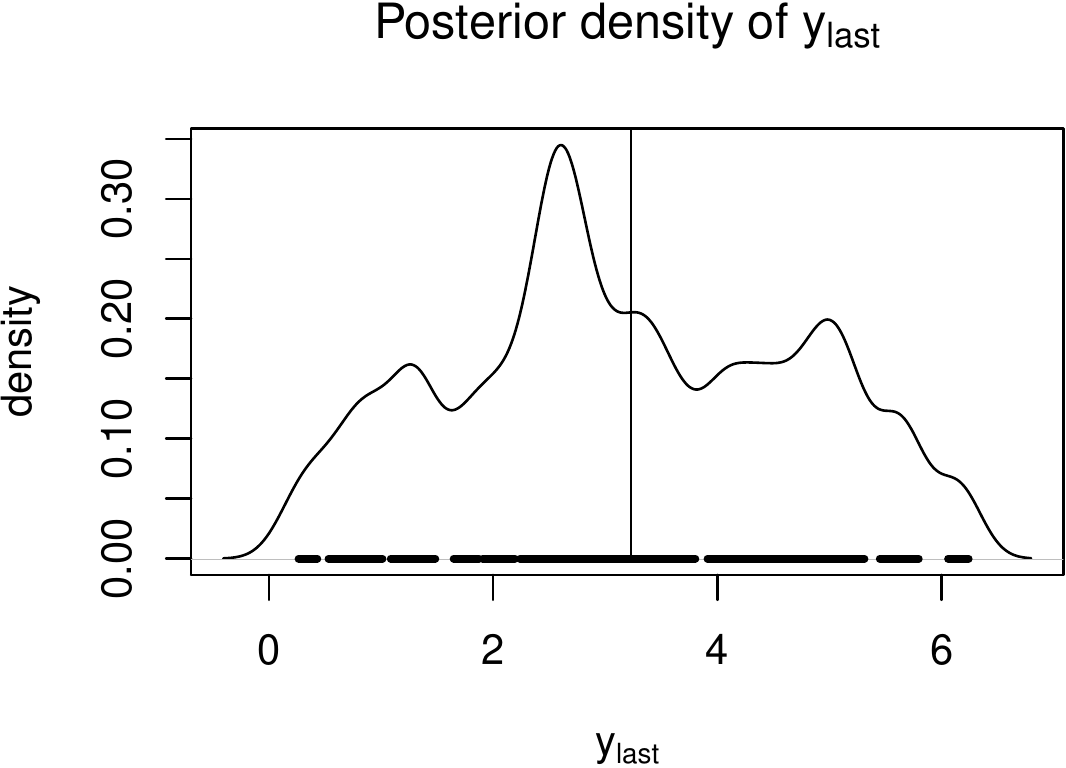}
\includegraphics[height=3in,width=3in]{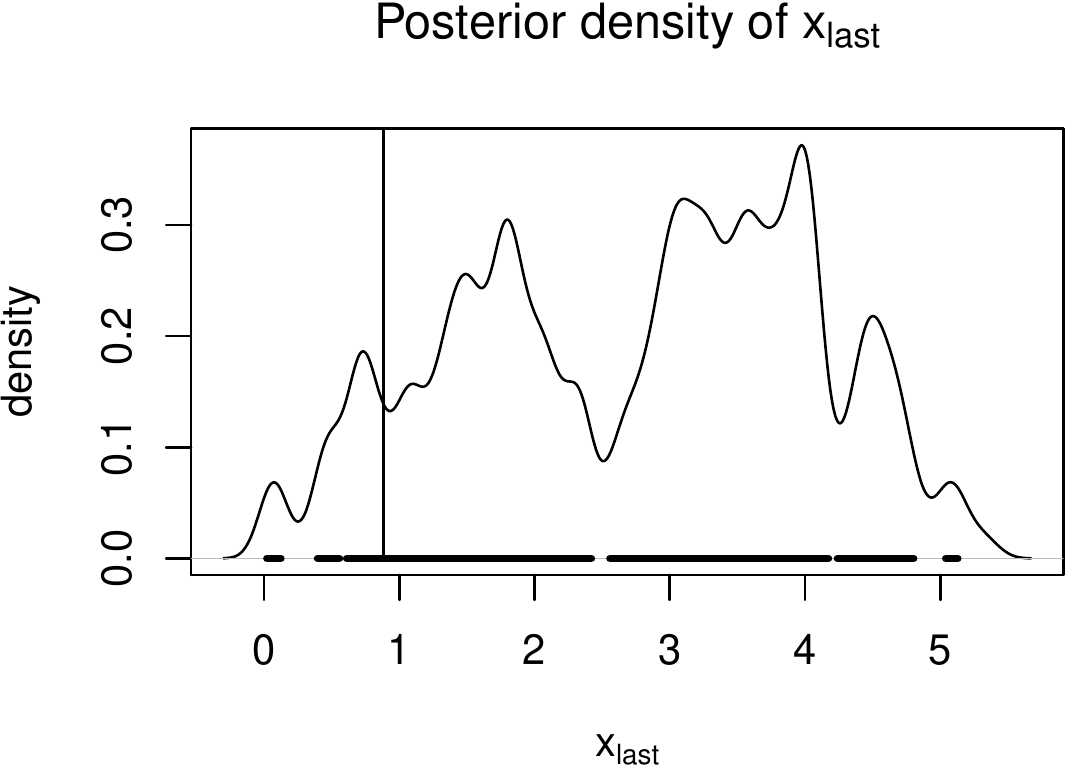}
\caption{The left panel displays the posterior predictive density of the $86$-th observation for 
the spawning time. The right panel depicts the posterior predictive density of the $86$-th observation for low tide time. 
For both the plots, the thick horizontal line denotes the 95\% highest posterior density credible interval 
and the vertical line denotes the true value (in radian).}
\label{Fig:Post_predictive_of_y_last_of_spawning_time_data}
\end{figure}

We provide the plots of the posterior densities of the four components of $\bi{\beta}_f$ for the spawning time data in 
Figure \ref{Fig:Post_of_beta_f_for_spawning_time_data} and those of the two components of $\bi{\beta}_g$ are 
displayed in Figure \ref{Fig:Post_of_beta_g_for_spawning_time_data}. The posterior densities of the latent variables, that is, 
low tide times, are depicted in Figure \ref{Fig:latent_x_for_spawning_time}. As in the previous experiment, here also 
it is seen that the posterior densities of the latent variables are highly multimodal. 
The plots of the posterior densities of the latent process for all the time points are 
presented in Figure \ref{Fig:latent_x_for_spawning_time},
where the observed values of the low tide times are denoted by the blue colored stars. Encouragingly, all the true values,
except only the first three points, fall in the high probability density regions. Finally the posterior predictive densities 
of $x_{86}$ and $y_{86}$, displayed in Figure \ref{Fig:Post_predictive_of_y_last_of_spawning_time_data}, 
show that the true values of $x_{86}$ and $y_{86}$ fall well within the 95\% highest posterior probability 
density region. We have conducted the usual tests of MCMC convergence diagnostics, all of which turned out to be satisfactory.

\subsubsection{Results of cross-validation}
\label{subsubsec:cross valid spawning data}
\begin{figure}[htp]
\centering
\includegraphics[height = 3in,width=3in]{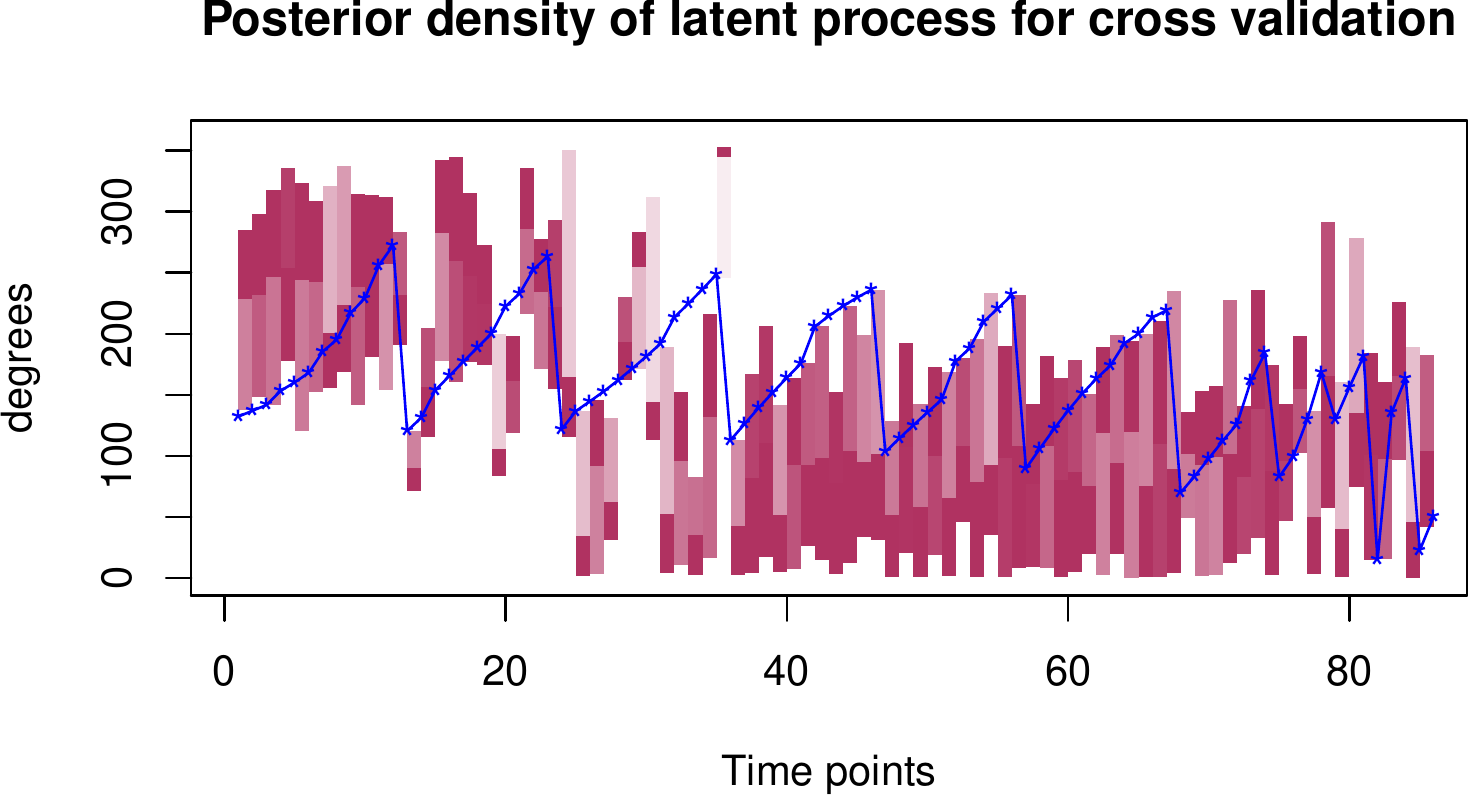}
\includegraphics[height = 3in,width=3in]{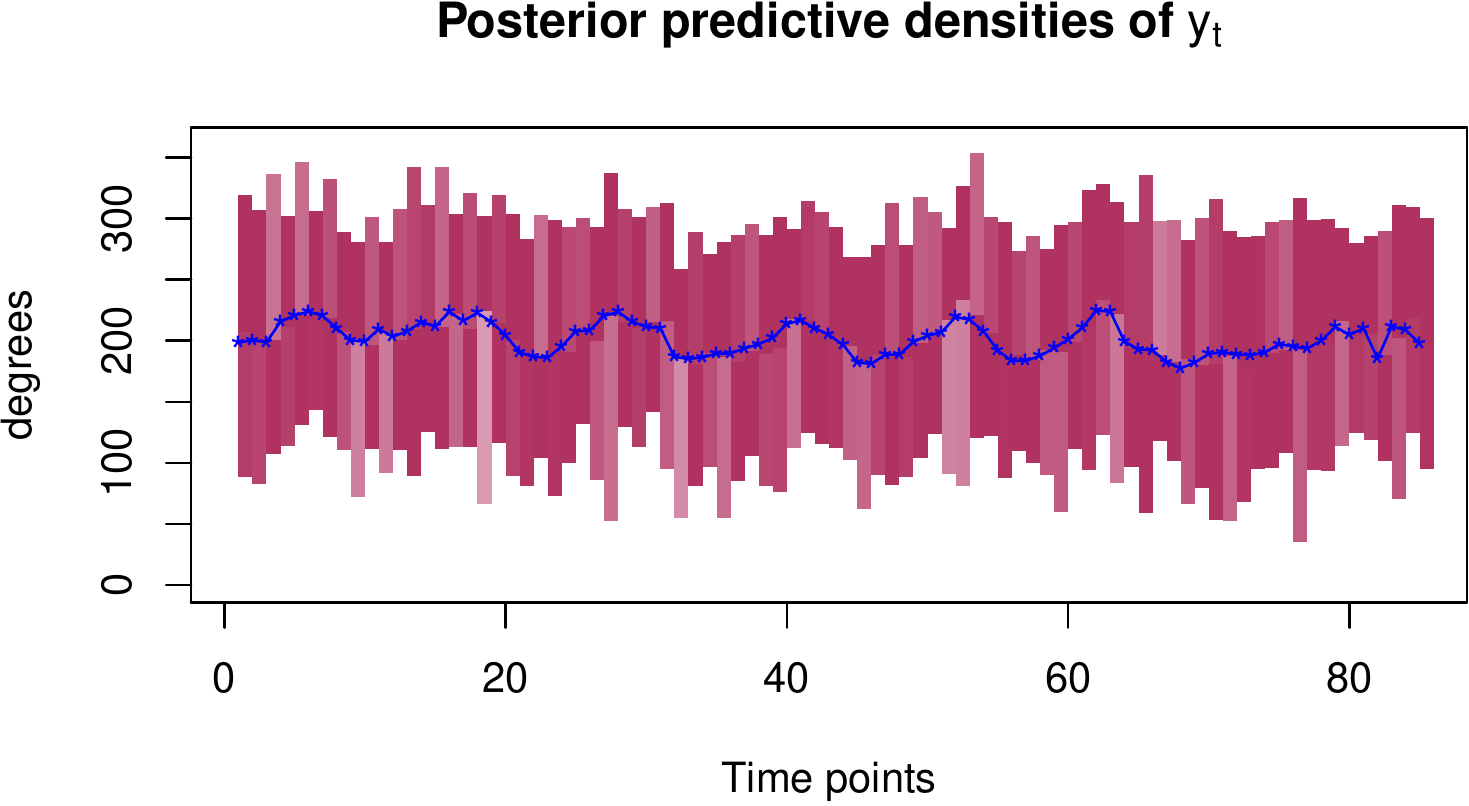}
\caption{Depiction of the posterior densities of the latent circular process, low tide times for 85 observations, and that of the observed circular process, spawning time for 85 observations, under the leave-one-out cross-validation scheme; higher the intensity of the color, higher is the posterior density. The blue stars denote the true values.}
\label{Fig:cross valid swapning_time}
\end{figure}

As earlier cases of simulation and wind pair data, here we predict the 
spawning times for 85 observations, setting aside the $t$-th spawning time, $t=1,\ldots, 85$, as unobserved; we also
record, as before, the posteriors of the $t$-th low tide time, the corresponding latent variable. 
The relevant cross-validation posterior densities along with the true time series, are 
displayed in Figure \ref{Fig:cross valid swapning_time}. Once again, the results have been highly encouraging.
%

\section{Application to whale positional data}
\label{sec:whale_data}
\subsection{A brief description of the data set}
\label{subsec:whale_data}
An interesting data set on whale movement directions is available at \url{http://nbviewer.jupyter.org/github/robertodealmeida/notebooks/blob/master/ earth_day_data_challenge/Analyzing%20whale%20tracks.ipynb}
, on which we apply our model and methodologies to predict the directions of the ocean current. 

\begin{figure}[htp]
\centering
\includegraphics[height=5in,width=5in]{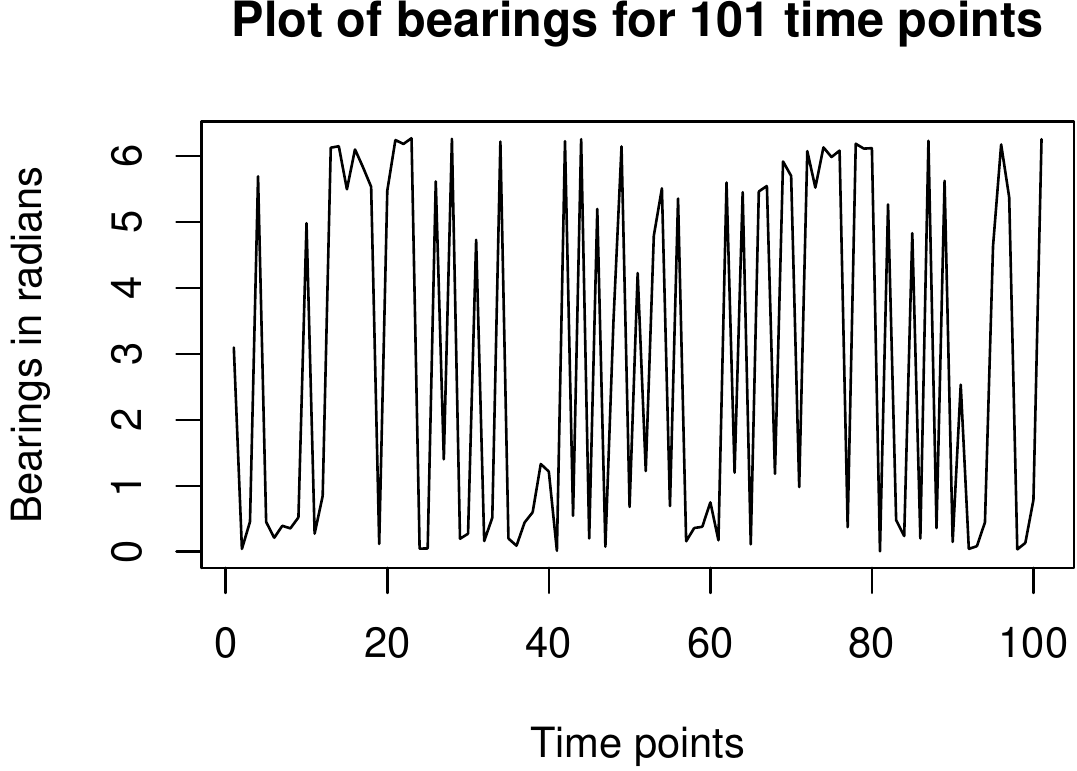}
\caption{Plots of the bearings for 101 time points of the whale positions.}
\label{Fig:Bearing_data_plot}
\end{figure}

Specifically, in this website 242 positions of a whale has been recorded in terms of 
latitudes and longitudes. The whale began its migration on 24th December of 2003 at the position 20.465 S, 40.04 W, 
and ended on 28th February of 2004 at 54.67 S, 26.261 W. This was an experimental project by 
Roberto De Almeida, who was interested whether whales could benefit from the ocean currents while migrating. 
For our purpose, first we processed the raw data by calculating the angles between these latitudes and longitudes, 
known as bearing, using the formula given in \url{http://www.movable-type.co.uk/scripts/latlong.html}. 
Finally, we obtained 241 circular univariate angular observations in radians. To reduce computational complexity 
here we have considered only the first 100 observations for our analysis setting aside the 101-th observation 
for the purpose of prediction. 
A plot of the first 101 bearings is shown in Figure \ref{Fig:Bearing_data_plot}. It is expected that the 
positions of the whale depends upon the directions of the ocean current, but the latter are not recorded. 
Hence, we anticipate that our general, nonparametric model and its corresponding methods will provide useful 
predictions of the direction of the ocean current, even though only the whale positions are observed.   

\subsection{Prior choices and MCMC implementation}
The means of the prior distributions of $\bi{\beta}_f$ and $\bi{\beta}_g$ are set as 
$(5,5,5,5)'$ and $(0,2,1,1)'$, respectively, and their covariance matrices are as before considered 
to be the $4\times 4$ identity matrix and the diagonal matrix with the diagonal entries $(1,1,0,0)'$. 
We made these choices to ensure adequate mixing of our MCMC chain. 
The MLEs of $\sigma_f$, $\sigma_{\epsilon}$, $\sigma_g$ and 
$\sigma_{\eta}$ are found to be $1.15391$, $0.24421$, $0.19264$ and $0.26019$, respectively, using simulated annealing. 
\par 
With these choices of prior parameters we implemented $2.5\times 10^5$ iterations of our 
MCMC algorithm, storing the last 50,000 observations for inference. 
The entire exercise took 25 hours 31 minutes in our server computer. 
The usual tests for convergence of MCMC indicated adequate mixing properties. 

%

\subsection{Results of the whale positional data}
\label{subsec:real_data_results}
\begin{figure}[htp]
\centering
\includegraphics[height=1.5in,width=1.5in]{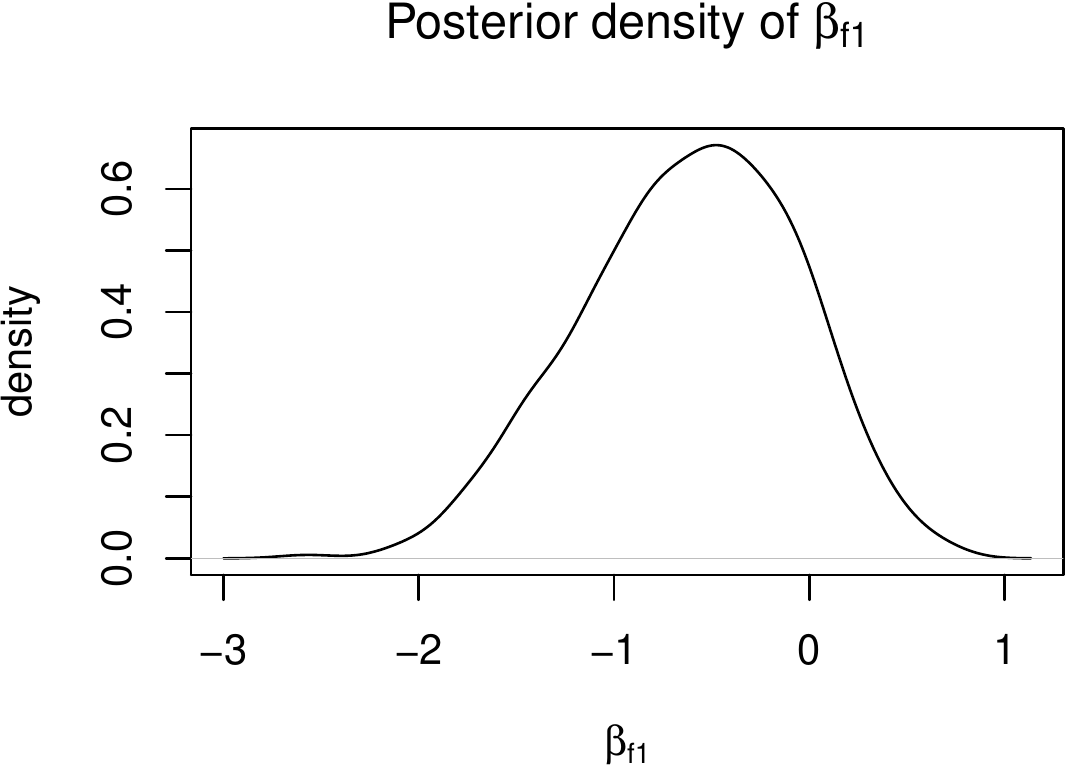}
\includegraphics[height=1.5in,width=1.5in]{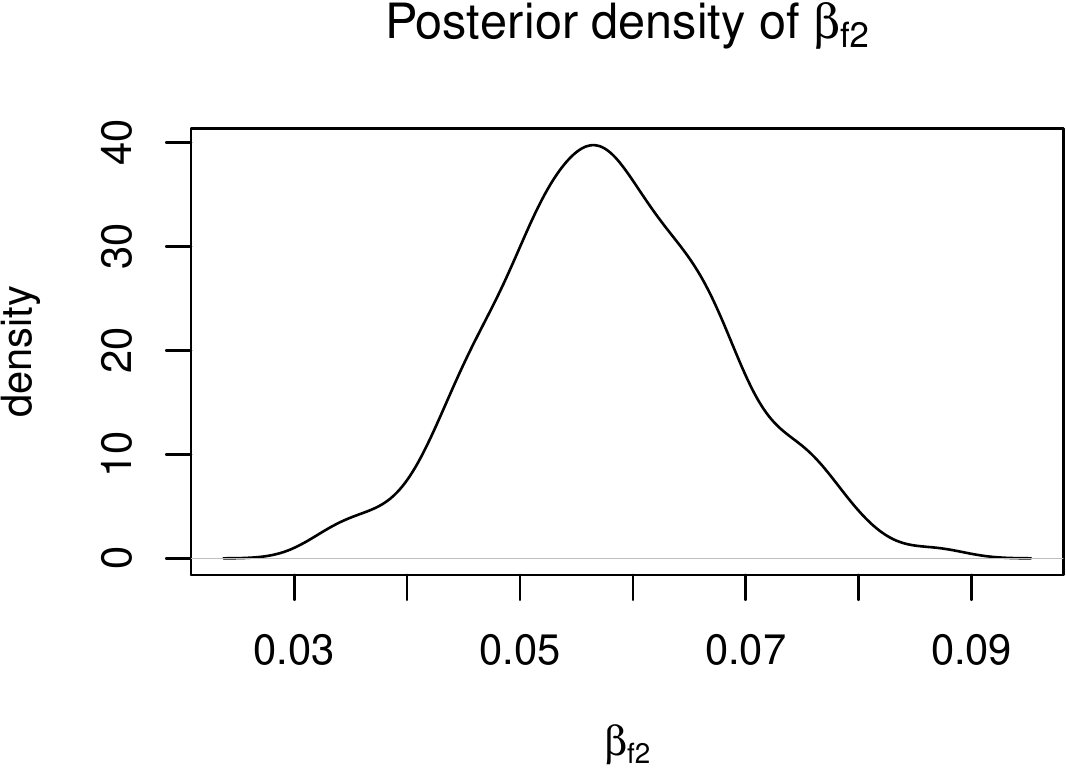}
\includegraphics[height=1.5in,width=1.5in]{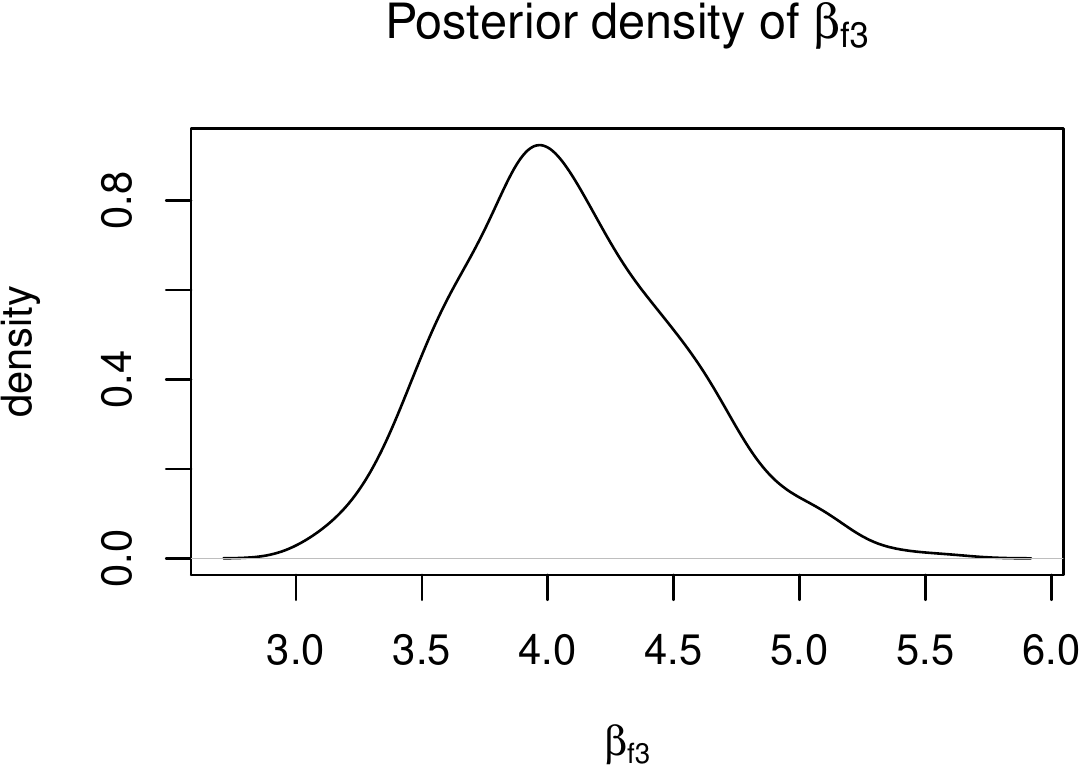}
\includegraphics[height=1.5in,width=1.5in]{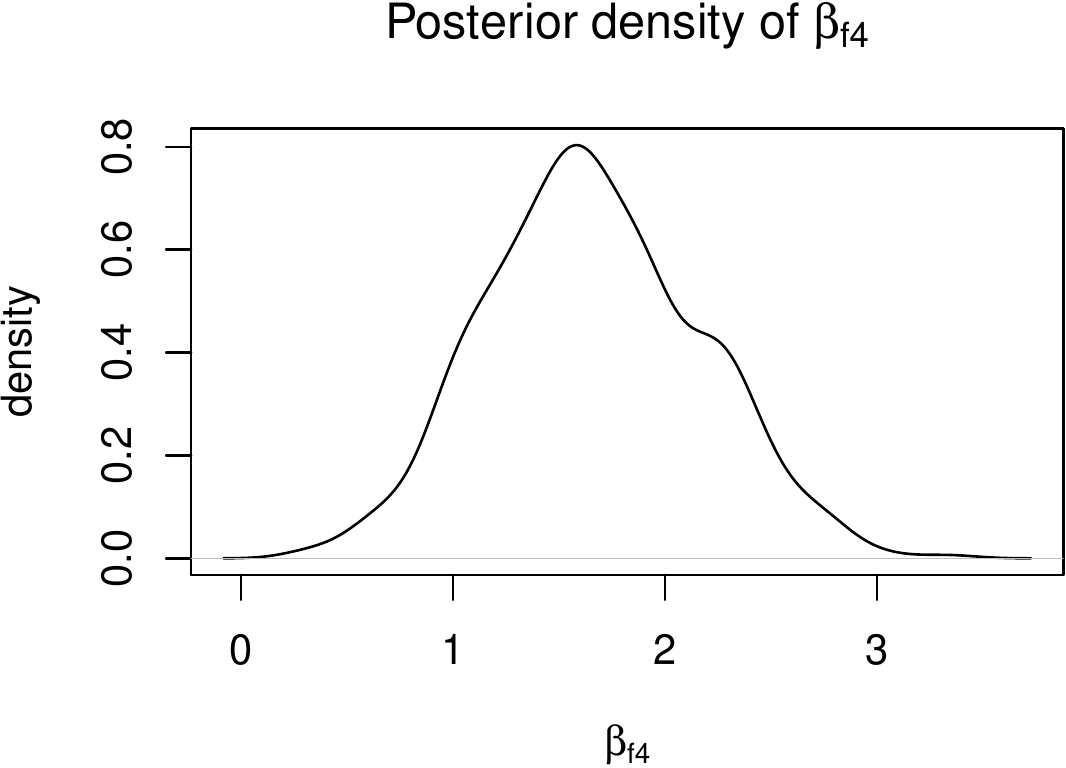}
\caption{Posterior densities of the four components of $\bi{\beta}_{f}$ for the whale positional data.}
\label{Fig:Post_of_beta_f_for_whale_data}
\end{figure}

\begin{figure}[htp]
\centering
\includegraphics[height=2in,width=2in]{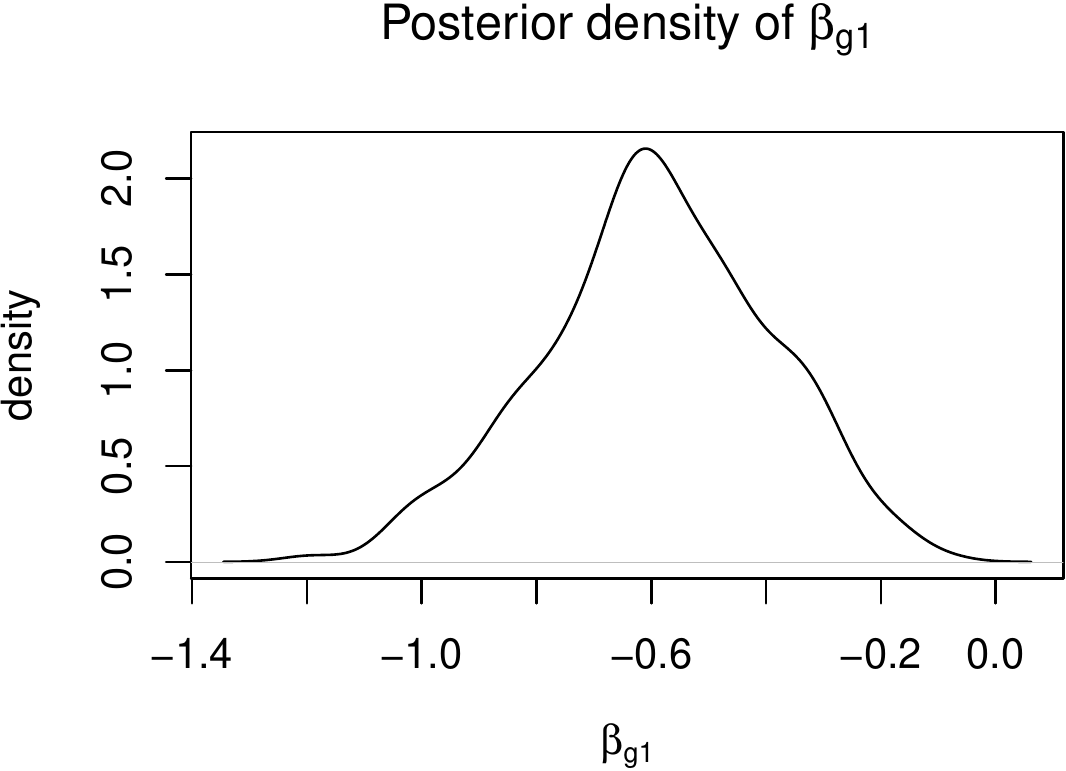}
\includegraphics[height=2in,width=2in]{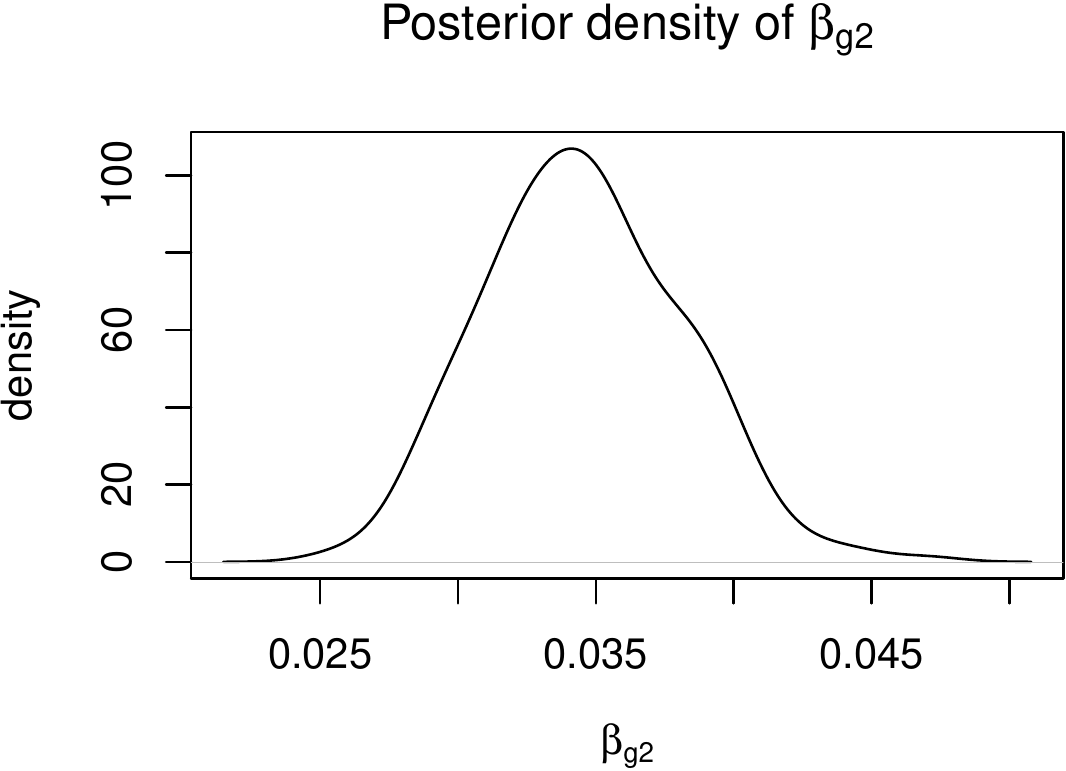}
\caption{Posterior densities of the first two components of $\bi{\beta}_{g}$ for the whale positional data.}
\label{Fig:Post_of_beta_g_for_whale_data}
\end{figure}

\begin{figure}[htp]
\centering
\includegraphics[height=3in,width=5in]{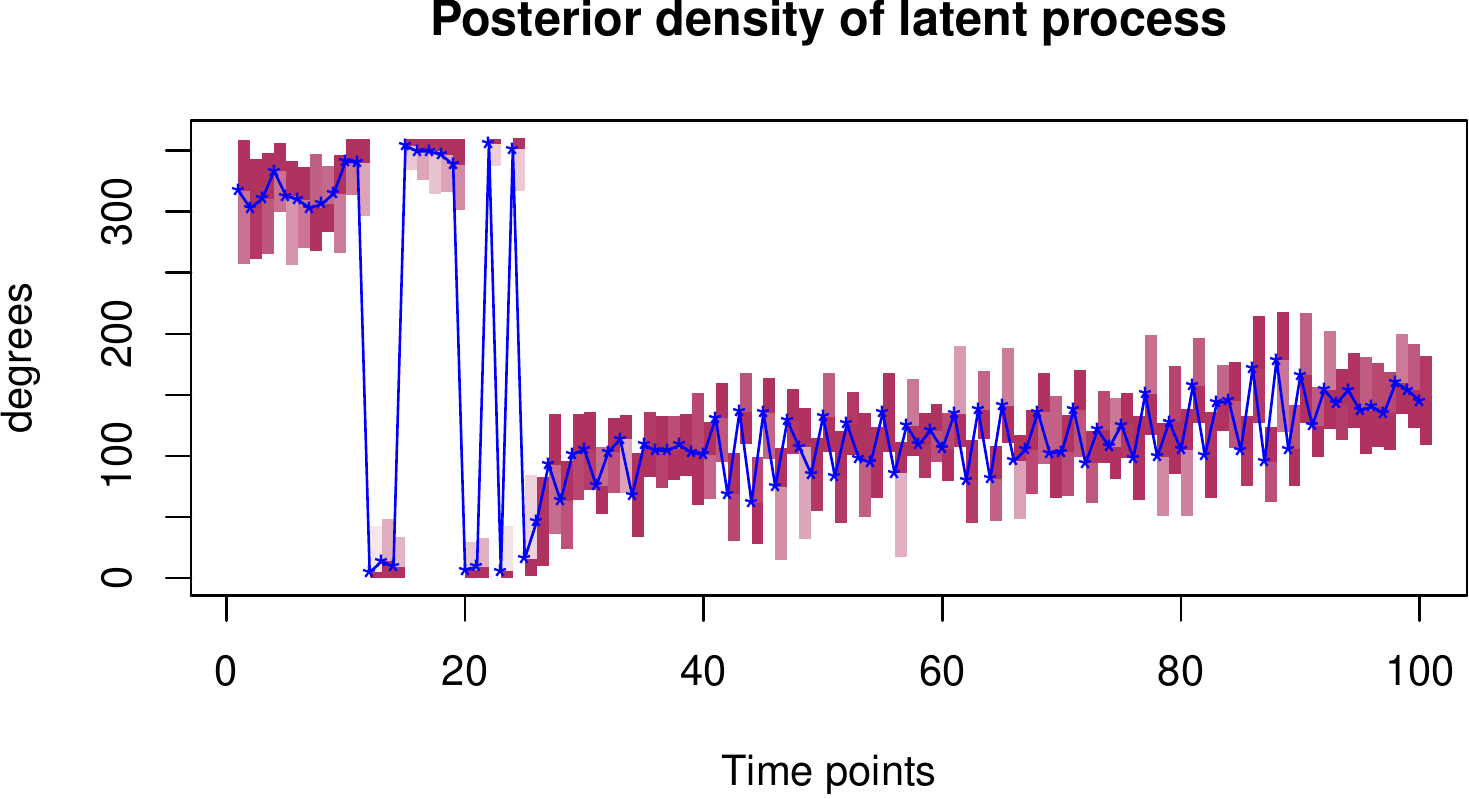}
\caption{Depiction of the marginal posterior distributions of the latent variables 
using progressively intense colors for progressively higher densities. The blue stars represent 
the posterior medians of the latent process.}
\label{Fig:latent_x_for_real_data}
\end{figure}

\begin{figure}[htp]
\centering
\includegraphics[height=3in,width=3in]{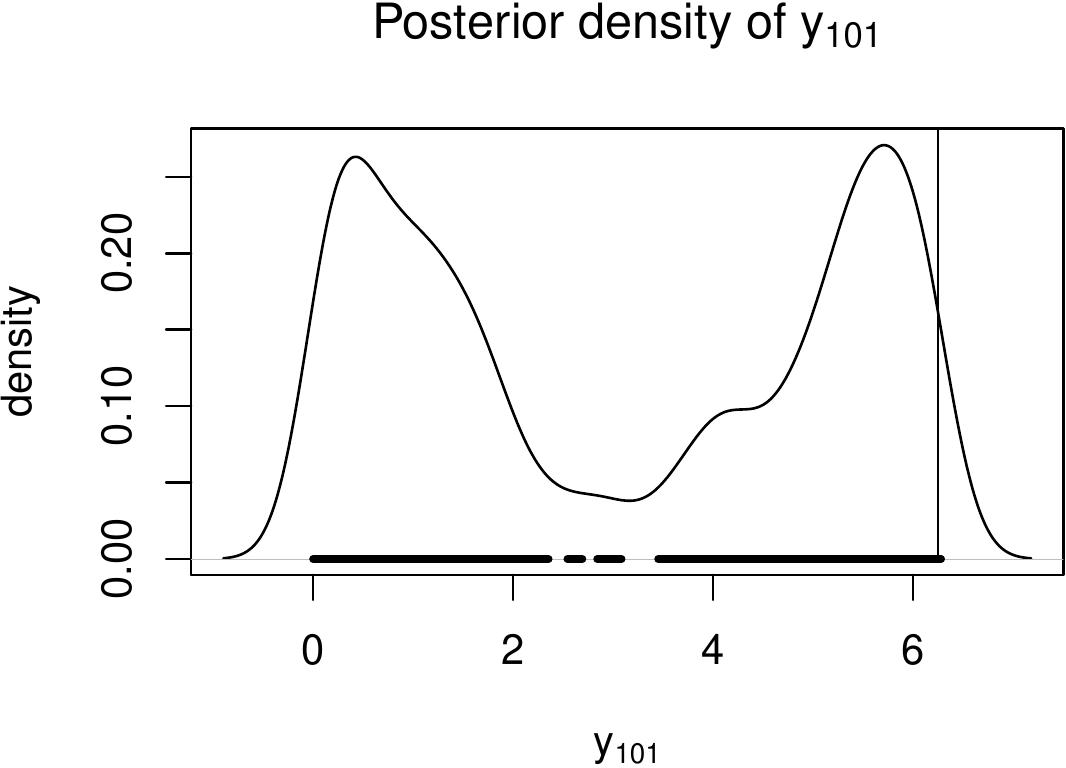}
\includegraphics[height=3in,width=3in]{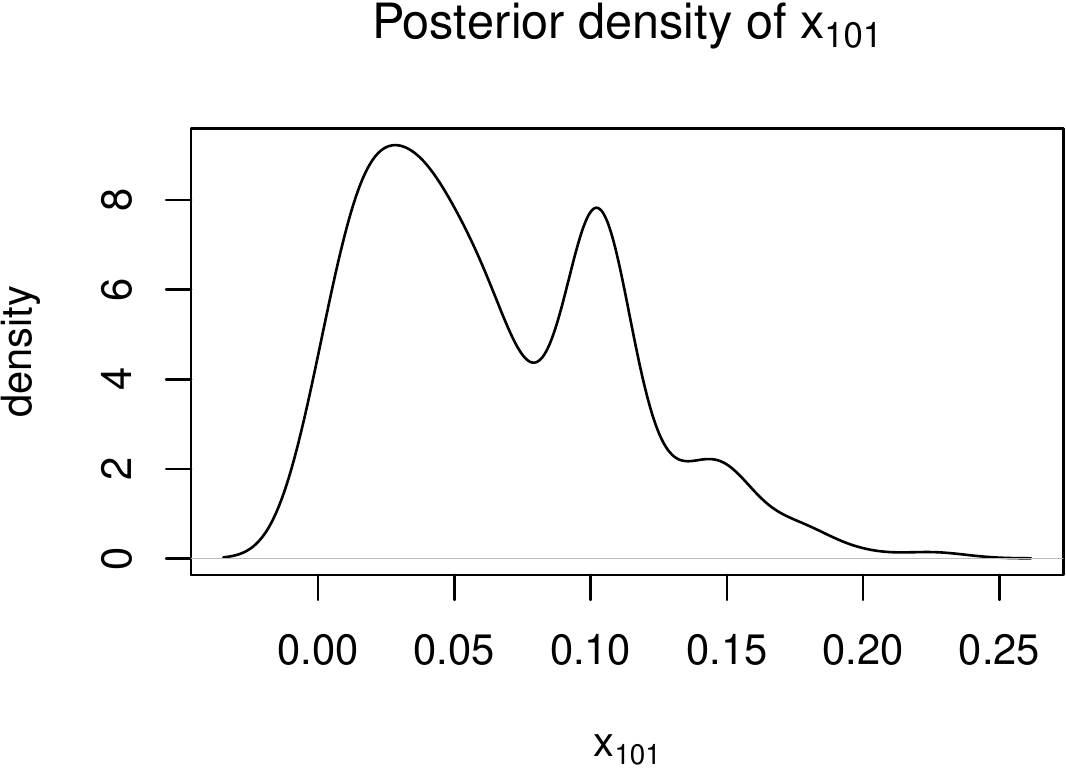}
\caption{The left panel shows the posterior predictive density of the $101$-th observation of bearing 
of the whale positional data. 
The thick horizontal line denotes the 95\% highest posterior density credible interval and 
the vertical line denotes the true value. The right panel displays the posterior predictive density 
of the 101-th observation of the direction of ocean current.}
\label{Fig:Post_predictive_of_y_last}
\end{figure}

We provide the plots of posterior densities of four components of $\bi{\beta}_f$ and two components of 
$\bi{\beta}_g$ in Figures \ref{Fig:Post_of_beta_f_for_whale_data} and 
\ref{Fig:Post_of_beta_g_for_whale_data}, respectively. 
Figure \ref{Fig:latent_x_for_real_data} shows the marginal posterior distributions  
associated with the latent circular process depicted by progressively intense colors. In 
Figure \ref{Fig:latent_x_for_real_data}, the blue stars indicate the posterior medians of the latent process.
Finally, the posterior predictive densities corresponding to $y_{101}$ and $x_{101}$ are shown in 
Figure \ref{Fig:Post_predictive_of_y_last}. 
The thin vertical line in the left panel denotes the true value of the $101$-th observation of the bearing value 
and the thick, broken line on the horizontal axis represents the 95\% highest density region 
of the posterior predictive density of $y_{101}$. 
As in our previous experiments, here also the true value 
falls well within the 95\% highest posterior density credible interval. 
Thus, overall, the performance of our model and methods seems to be quite satisfactory.

\subsection{Cross-validation results}
\begin{figure}[htp]
\centering
\includegraphics[height=3in,width=5in]{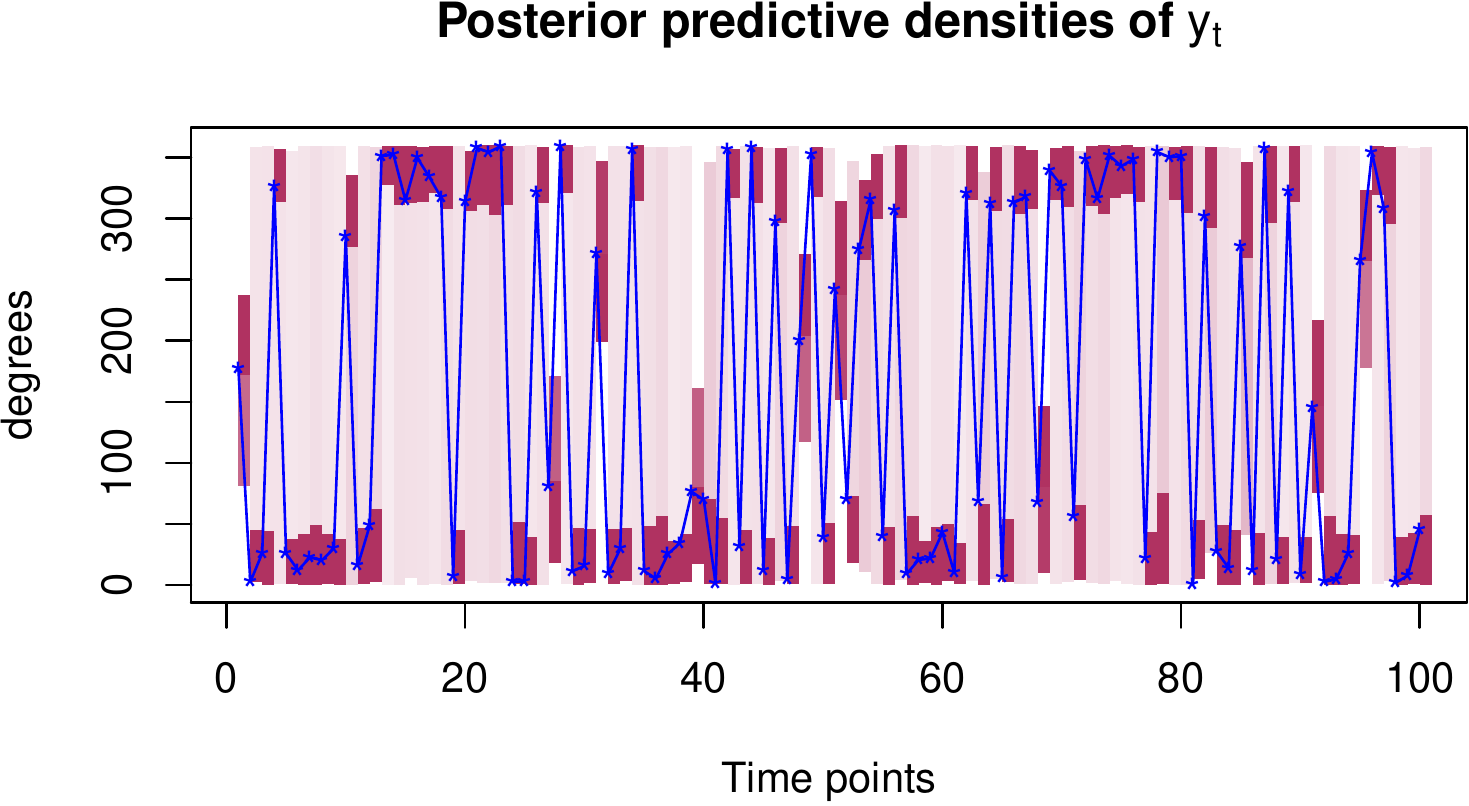}
\caption{Depiction of the posterior densities of the observed circular process, bearings of whale positions for 100 time points, under the leave-one-out cross-validation scheme; higher the intensity of the color, higher is the posterior density. The blue stars denote the true values.}
\label{Fig:cross valid whale positional data}
\end{figure}

In the leave-one-out cross-validation exercise for the whale positional data, we provide the results
only for the observed data, since, unlike the previous experiments, the true values associated with the latent
process are unknown.
The posterior densities of the 100 bearings are displayed in  
Figure \ref{Fig:cross valid whale positional data}, and as expected, the results have been highly encouraging.  

\section{Discussion and conclusion}
\label{conclusion}

In nature there exist plentiful examples of circular time series where the underlying influential circular variable
remains unrecorded. Some such examples, apart from the whale movement directions considered in this paper, 
are directions of flights of birds and missiles, where the underlying
wind directions, although affects the flight directions, generally remains unrecorded.
Modeling such circular time series using traditional parametric and nonparametric methods existing in the literature
(for example, methods proposed in \ctn{Breckling89}, \ctn{Fisher94}, \ctn{Holzmann06}, \ctn{Hughes07}, \ctn{Marzio12})
will be far less effective compared to state space approaches that treat the underlying unrecorded process as a
latent circular process. In this regard, we have introduced and developed a novel Bayesian nonparametric state space model for 
circular time series where the latent states also constitute a circular process. 

Moreover, since it is clearly
non-trivial to model general circular time series accounting for adequate non-Markov angular dependence structure, 
our endeavor seems to facilitate significant advance in this respect. Indeed,
the observed time series in our state space model has a complex and realistic dependence structure, facilitated
by the circular latent state process along with the structures induced by the wrapped Gaussian process, even though
conditionally on the latent states and the observational function we assume the circular observations to be 
independent and further assume a Markovian structure of the latent states conditional on the look-up table.  
These structures within our state space model facilitates an effective MCMC procedure, while ensuring
highly realistic dependence structure. Such realistically complex dependence structures are absent in
the traditional parametric and nonparametric circular time series models existing in the literature.

Other than our current work, the only research on state space models that considers circular latent states  
is that of \ctn{Mazumder16}; however, they assume that the observed time series is associated
with a linear process, unlike our observed circular time series. The observed circular process, as we argued, gave rise to
significantly more complexities and difficult technical challenges in our case, compared to that of
\ctn{Mazumder16}. However, the complexities of our Bayesian model and methods notwithstanding, we have amply demonstrated with
simulation studies as well as with three real data sets, the effectiveness of our proposed procedures. 

Ideally, it seems that comparison of our proposed model and methods with the existing works is desirable, but
for circular times series such comparison does not seem to be possible. We explain this as follows.  
Most linear time series admit state-space representation, 
where both the observational and evolutionary equations are defined on the real line. However, it is unclear if 
the existing circular time series admit state-space representation. Even if some of them do, it is unlikely that 
the corresponding latent states will be circular. This is because for circular latent states, it is unlikely 
that marginalization over the circular latent states is analytically possible to arrive at the known forms of the 
existing circular time series. The difficulty of marginalization stems from the need to integrate complex expressions 
over the bounded domain $(0, 2\pi]$.
We are not aware of any existing circular time series that admits a state space representation with circular 
latent states. Since the main assumption of our model is that there is an underlying circular latent structure, 
it seems that comparison of the existing methods with our method is not possible.  

Note that there perhaps exists a much larger class of circular time series in nature which depend upon
unrecorded linear processes, necessitating the development of Bayesian nonparametric
state space models with observed circular process and latent linear process. But such development is a much
simpler exercise given our current work and the developments provided in \ctn{Ghosh14}.
As more interesting alternatives, we shall focus on developing Bayesian nonparametric state space models
and methods when the observed times series and the latent states are multivariate, with several linear and several
circular components.

\section*{Acknowledgement}
The authors are thankful to Professor Ulric Lund for providing the data on spawn timings and corresponding low tide timings. The authors are also thankful to the support staff of the website \url{https://planetos.com/
marinexplore/} for supplying the details of the whale positional data that is used in this paper. In addition the authors extend their cordial thanks to two anonymous referees for their valuable comments and suggestions that helped improve our paper significantly. 

\newpage
\bibliographystyle{natbib}
\bibliography{irmcmc}

\pagebreak

\renewcommand\thefigure{S-\arabic{figure}}
\renewcommand\thetable{S-\arabic{table}} 
\renewcommand\thesection{S-\arabic{section}}

\setcounter{section}{0}
\setcounter{figure}{0}
\setcounter{table}{0}

\begin{center}
{\bf \Large Supplementary Material}
\end{center}

\section{\small Full conditional distributions of $\bi{\beta}_f$}
\allowdisplaybreaks
\begin{align}
\label{eqn4: full conditional of beta_f}
[\bi{\beta}_f|\ldots] & \propto [\bi{\beta}_f] [\bi{f}^*|x_1,\ldots,x_{T+1},\bi{\beta}_f,\sigma_f] \notag
\\[1ex]
& \propto \exp\left\{-\frac{1}{2}(\bi{\beta}_f-\bi{\beta}_{f,0})'\Sigma^{-1}_{\bi{\beta}_{f,0}} (\bi{\beta}_f-\bi{\beta}_{f,0})-\frac{1}{2}(\bi{f}^*-\bi{H}\bi{\beta}_f)'\sigma^{-2}_{f}\bi{A}_{f}^{-1}(\bi{f}^*-\bi{H}\bi{\beta}_f)\right\} \notag
\\[1ex]
& \propto \exp\left\{-\frac{1}{2}\left[\bi{\beta}_{f}'\left(\Sigma^{-1}_{\bi{\beta}_{f,0}}+\bi{H}'\sigma^{-2}_fA^{-1}_{f}\bi{H}\right)
\bi{\beta}_{f}-2\bi{\beta}_{f}'\left(\Sigma^{-1}_{\bi{\beta}_{f,0}} \bi{\beta}_{f,0}+\bi{H}'\sigma^{-2}_{f}\bi{A}_{f}^{-1}\bi{f}^*\right)\right]\right\},
\end{align}
where $\bi{H} = \left[\bi{h}(1,x_1)',\ldots, \bi{h}(T+1,x_{T+1})'\right]'$.
Therefore, the full conditional distribution of $\bi{\beta}_f$ is a four-variate normal distribution with mean
\allowdisplaybreaks
\begin{align}
\label{eqn5: mean of beta_f}
E[\bi{\beta}_f|\ldots] = \left(\Sigma^{-1}_{\bi{\beta}_{f,0}}+\bi{H}'\sigma^{-2}_fA^{-1}_{f}\bi{H}\right)^{-1} \left(\Sigma^{-1}_{\bi{\beta}_{f,0}} \bi{\beta}_{f,0}+\bi{H}'\sigma^{-2}_{f}\bi{A}_{f}^{-1}\bi{f}^*\right)
\end{align}
and variance covariance matrix
\begin{align}
\label{eqn6: cov of of beta_f}
V[\bi{\beta}_f|\ldots] = \left(\Sigma^{-1}_{\bi{\beta}_{f,0}}+\bi{H}'\sigma^{-2}_fA^{-1}_{f}\bi{H}\right)^{-1}.
\end{align}
To update $\bi{\beta}_f$ we use Gibbs sampling. 

\section{\small Full conditional distribution of $\sigma_f$}
\begin{align}
\label{eqn7: full conditional of sigma_f}
[\sigma_f^2|\ldots] &\propto [\sigma_f^2] [\bi{f}^*|x_1,\ldots,x_{T+1},\bi{\beta}_f,\sigma_f] \notag
\\[1ex]
& = (\sigma^2_f)^{-\frac{\alpha_f+2+(T+1)}{2}} \exp\left\{-\frac{1}{2\sigma^2_f}\left[ \gamma_f + (\bi{f}^*-\bi{H}\bi{\beta}_f)'\bi{A}_{f}^{-1}(\bi{f}^*-\bi{H}\bi{\beta}_f) \right]\right\}. 
\end{align}
Since $\bi{A}_f$ involves $\sigma_f^2$ (recall that $\bi{A}_f$ = ($\exp\left(-\sigma_f^4(i-j)^2\right)\cos(|x_i-x_j|)$)) so, the closed form of the above posterior density is not tractable. Therefore, we use a random walk Metropolis-Hastings step 
to update $\sigma^2_{f}$.

\section{\small Full conditional density of $\sigma_{\epsilon}$}
\begin{align}
\label{eqn8: full conditional of sigma_e}
[\sigma_{\epsilon}^2] & \propto [\sigma_{\epsilon}^2] \prod_{t=1}^{T} [N_{t}|f^*(t,x_t),\sigma_{\epsilon},x_t] \prod_{t=1}^{T} [y_t|N_t,f^*(t,x_t),\sigma_{\epsilon},x_t] \notag
\\[1ex]
 & \propto (\sigma^2_{\epsilon})^{-\frac{\alpha_{\epsilon}+2+T}{2}} \exp \left\{-\frac{1}{2\sigma^2_{\epsilon}}\sum_{t=1}^{T} (y_t+2\pi N_t - f^*(t,x_t))^2\right\}. 
\end{align}
Therefore, the full conditional distribution of $\sigma_{\epsilon}^2$ is inverse gamma with 
the parameters $\frac{\alpha_{\epsilon} + T}{2}$ and\\ $\frac{\sum_{t=1}^{T} (y_t+2\pi N_t - f^*(t,x_t))^2}{2}$. %
Thus, we use the Gibbs sampling technique to update $\sigma^2_{\epsilon}$. 

\section{\small Full conditional density of $\bi{f}_{\mbox{\scriptsize $D_{T}$}}^*$}
\begin{align}
\label{eqn9: full conditional of f}
[\bi{f}_{\mbox{\scriptsize $D_{T}$}}^*|\ldots] & \propto [\bi{f}_{\mbox{\scriptsize $D_{T}$}}^*|x_1,\ldots,x_{T+1},\bi{\beta}_f,\sigma_f] \prod_{t=1}^{T} [N_{t}|f^*(t,x_t),\sigma_{\epsilon},x_t] \prod_{t=1}^{T} [y_t|N_t,f^*(t,x_t),\sigma_{\epsilon},x_t] \notag
\\[1ex]
& \propto \exp\left\{-\frac{1}{2}(\bi{f}_{\mbox{\scriptsize $D_{T}$}}^*-\bi{H}_{\mbox{\scriptsize $D_{T}$}}\bi{\beta}_f)'\sigma^{-2}_{f}\bi{A}_{f,D_{T}}^{-1}(\bi{f}_{\mbox{\scriptsize $D_{T}$}}^*-\bi{H}_{\mbox{\scriptsize $D_{T}$}}\bi{\beta}_f)-\frac{1}{2\sigma^2_{\epsilon}}\sum_{t=1}^{T} (y_t+2\pi N_t - f^*(t,x_t))^2\right\} \notag
\\[1ex]
& \propto \exp\left\{-\frac{1}{2}\left[\bi{f}_{\mbox{\scriptsize $D_{T}$}}^*\left(\sigma^{-2}_{f}\bi{A}_{f,D_{T}}^{-1}+\sigma^{-2}_{\mbox{\scriptsize $\epsilon$}}I_{T}\right)\bi{f}_{\mbox{\scriptsize $D_{T}$}}^*
-2\bi{f}_{\mbox{\scriptsize $D_{T}$}}^*\left(\sigma^{-2}_{f}\bi{A}_{f,D_{T}}^{-1}\bi{H}_{\mbox{\scriptsize $D_{T}$}}\bi{\beta}_{f}+\sigma^{-2}_{\mbox{\scriptsize $\epsilon$}} [\bi{D}_{T}+2\pi\bi{N}]\right)\right]\right\},
\end{align}
where $\bi{D}_{T}+2\pi\bi{N}$ = $(y_1+N_1,\ldots,y_T+N_T)'$.
Hence, $[\bi{f}_{\mbox{\scriptsize $D_{T}$}}^*|\ldots]$ is a $T$-variate normal distribution with the parameters
\begin{equation}
\label{eqn10: mean of f*}
E[\bi{f}_{\mbox{\scriptsize $D_{T}$}}^*|\ldots] = \left(\sigma^{-2}_{f}\bi{A}_{f}^{-1}+\sigma^{-2}_{\mbox{\scriptsize $\epsilon$}}I_{T}\right)^{-1} \left(\sigma^{-2}_{f}\bi{A}_{f,D_{T}}^{-1}\bi{H}_{\mbox{\scriptsize $D_{T}$}}\bi{\beta}_{f}+\sigma^{-2}_{\mbox{\scriptsize $\epsilon$}} [\bi{D}_{T}+2\pi\bi{N}]\right)
\end{equation}
and
\begin{equation}
\label{eqn11: var of f^*}
\mbox{cov}[\bi{f}_{\mbox{\scriptsize $D_{T}$}}^*|\ldots] = \left(\sigma^{-2}_{f}\bi{A}_{f,D_{T}}^{-1}+\sigma^{-2}_{\mbox{\scriptsize $\epsilon$}}I_{T}\right)^{-1}.
\end{equation}
Hence, we use the Gibbs sampling technique to update $\bi{f}_{\mbox{\scriptsize $D_{T}$}}^*$.
\section{\small Full conditional distribution of $f^*(T+1,x_{T+1})$}
\begin{align}
\label{eqn0: full conditional of f*(T+1,x_T+1)}
[f^*(T+1,x_{T+1})|\ldots] \propto [f^*(T+1,x_{T+1})|\bi{f}_{\mbox{\scriptsize $D_{T}$}}^*, x_{T+1}, \bi{\beta}_f,\sigma_f].
\end{align}
Hence, $[f^*(T+1,x_{T+1})|\ldots]$ follows the univariate normal distribution with parameters 
$\mu_{\mbox{\scriptsize $f^*(T+1,x_{T+1})$}}$ and variance $\sigma_{\mbox{\scriptsize $f^*(T+1,x_{T+1})$}}^2$, 
and so, we update $f^*(T+1,x_{T+1})$ by Gibbs sampling.  

\section{\small Full conditional density of $N_t$, $t=1,\ldots,T$}

\begin{align}
[N_t|\ldots] &\propto [N_t|f^*(t,x_t),\sigma_{\mbox{\scriptsize$\epsilon$}},x_t] [y_t|N_t,f^*(t,x_t),\sigma_{\mbox{\scriptsize$\epsilon$}},x_t] \notag
\\
& \propto \exp{\left(-\frac{1}{2\sigma^2_{\mbox{\scriptsize$\epsilon$}}}(y_t+2\pi N_t - f^*(t,x_t))^2\right)} I_{\{\ldots,-1,0,1,\ldots\}}(N_t), ~ t=1,\ldots,T.
\end{align}
The full conditional distribution of $N_t$, $t=1,\ldots, T$, are not tractable, and so we update them 
by random walk Metropolis-Hastings steps. 

\section{\small Full conditional density of $\bi{\beta}_g$}

\begin{align}
\label{eqn12:full conditional of beta_g}
[\bi{\beta}_g|\cdots] &\propto [\bi{\beta}_g] [\bi{D}_z,g^*(1,x_0)|x_0,\bi{\beta}_g,\sigma^2_g] 
\prod_{t=2}^{T+1} [x_t|\bi{\beta}_g,\sigma^2_{\mbox{\scriptsize $\eta$}},\sigma^2_g, \bi{D}_z,x_{t-1},K_t] \prod_{t=2}^{T+1} \left [ K_t|\bi{\beta}_g, \sigma^2_{\mbox{\scriptsize $\eta$}} \right.
\notag
\\
&\quad 
\left. \sigma^2_g,\bi{D}_z,x_{t-1}\right ]
\end{align}

We write down the right hand side of (\ref{eqn12:full conditional of beta_g}) as follows:
\texttt{\begin{align}
\label{eqn13:rhs of full conditional for beta_g}
&[\bi{\beta}_g] [\bi{D}_z,g^*(1,x_0)|x_0,\bi{\beta}_g]\prod_{t=2}^{T+1} 
[x_t|\bi{\beta}_g,\sigma^2_{\mbox{\scriptsize $\eta$}},\bi{D}_z,x_{t-1},K_t] 
\prod_{t=2}^{T+1} [K_t|\bi{\beta}_g,\sigma^2_{\mbox{\scriptsize $\eta$}},\bi{D}_z,x_{t-1}]
\notag
\\[1ex]
&\propto \exp{\left(-\frac{1}{2}(\bi{\beta}_g-\bi{\beta}_{g,0})'\bi{\Sigma}_{\beta_{g},0}^{-1}(\bi{\beta}_g-\bi{\beta}_{g,0})\right)}
\notag
\\[1ex]
\quad
&\times\exp{\left(-\frac{1}{2}[(\bi{D}_z,g^*)'-(\bi{H}_{D_{z}}\bi{\beta}_g,\bi{h}'(1,x_0))']'\bi{A}^{-1}_{D_z,g^*(1,x_0)}
[(\bi{D}_z,g^*)'-(\bi{H}_{D_{z}}\bi{\beta}_g,\bi{h}'(1,x_0))']\right)}
\notag
\\[1ex]
&\quad \times\exp{\left\{-\sum_{i=2}^{T+1}\frac{1}{2\sigma^2_{x_{t}}}(x_t+2\pi K_t-\mu_{x_{t}})^2\right\}} \prod_{t=2}^{T+1} I_{[0,2\pi]}(x_t),
\end{align}}
where $\bi{H}_{D_{z}} = \left(\bi{h}(t_1,z_1),\ldots, \bi{h}(t_n,z_n)\right)'$, $\bi{A}_{D_z,g^*(1,x_0)}$ is $(n+1)\times (n+1)$ matrix defined as
$$\bi{A}_{D_z,g^*(1,x_0)} = \left[
\begin{array}{c c}
\bi{A}_{g,D_{z}} & \bi{s}_{g,D_{z}}(1,x_0) \\
\bi{s}_{g,D_{z}}(1,x_0)' & \sigma^2_g
\end{array}
\right], $$
$\mu_{x_{t}} = \bi{h}(t,x_{t-1})'\bi{\beta}_g + \bi{s}_{g,D_{z}} (t,x_{t-1})'\bi{A}_{g,D_{z}}^{-1}(\bi{D}_{z}-\bi{H}_{D_{z}}\bi{\beta}_g)$ and \allowdisplaybreaks$\sigma^2_{x_{t}} = \sigma^2_{\mbox{\scriptsize{$\eta$}}} + 
{\sigma_g^2}(1-(\bi{s}_{g,D_z}(t,x_{t-1}))' \bi{A}_{g,D_{z}}^{-1}\bi{s}_{g,D_z}(t,\\ 
 x_{t-1}))$, 
with $\bi{A}_{g,D_{z}}$ being a $n\times n$ matrix having $(i,j)$th element as $c_{g}((t_i,z_i),(t_j,z_j))$ and with $\bi{s}_{g,D_{z}} (\cdot,\cdot)= (c_g((\cdot,\cdot),(t_1,z_1)),\ldots,c_g((\cdot,\cdot),(t_n,z_n)))' $.
Observe that the denominator of $[x_t|\bi{\beta}_g,\sigma^2_{\mbox{\scriptsize $\eta$}},\bi{D}_z,x_{t-1},K_t]$ cancels with the density of $[K_t|\bi{\beta}_g,\sigma^2_{\mbox{\scriptsize $\eta$}},\bi{D}_z,x_{t-1}]$, 
for each $t=2,\ldots, T+1$. Also, we note that the indicator function does not involve $\bi{\beta}_g$ for any $t=2,\ldots ,T+1$. 
Therefore, after simplifying the exponent terms and ignoring the indicator function we can write
\begin{equation}
\label{eqn14:Gibbs full conditional of beta_g}
[\bi{\beta}_g|\cdots] \propto \exp{\left\{-\frac{1}{2}(\bi{\beta}_g-\mu_{\beta_g})'
\bi{\Sigma}_{\beta_g}^{-1}(\bi{\beta}_g-\mu_{\beta_g})\right\}},
\end{equation}
where
%
%
%
\begin{align}
\label{eqn15: mean of full conditional of beta_g}
& \mu_{\beta_g}= E[\bi{\beta}_g|\cdots] = \left\{ \bi{\Sigma}_{\beta_{g},0}^{-1} 
+ \frac{1}{\sigma^2_g}[\bi{H}_{D_{z}}',\bi{h}(1,x_0)]\bi{A}^{-1}_{D_z,g^*(1,x_0)}[\bi{H}_{D_{z}}',\bi{h}(1,x_0)]' \right.
\notag
\\[1ex]
& \left. 
+ \sum_{t=1}^{T} \frac{\left(\bi{H}_{D_{z}}'\bi{A}^{-1}_{g,D_z} \bi{s}_{g,D_z}(t+1,x_{t})-\bi{h}(t+1,x_{t})\right)
\left(\bi{H}_{D_{z}}'\bi{A}^{-1}_{g,D_z} \bi{s}_{g,D_z}(t+1,x_{t})-\bi{h}(t+1,x_{t})\right)'}{\sigma^2_{x_{t}}}\right\}^{-1}
\notag
\\[1ex]
&\left\{\bi{\Sigma}_{\beta_{g},0}^{-1} \bi{\beta}_{g,0} + \frac{1}{\sigma^2_g} [\bi{H}_{D_{z}}',
\bi{h}(1,x_0)]\bi{A}^{-1}_{D_z,g^*(1,x_0)} [\bi{D}_z,g^*(1,x_0)] \right.
\notag
\\[1ex]
&\left. 
+ \sum_{t=1}^{T} \frac{\left(x_{t+1}+2\pi K_{t+1}-\bi{s}_{g,D_z}(t+1,x_{t})'\bi{A}^{-1}_{g,D_z} \bi{D}_{z}\right)
\left(\bi{h}(t+1,x_{t})-\bi{H}_{D_{z}}'\bi{A}^{-1}_{g,D_z} \bi{s}_{g,D_z}(t+1,x_{t})\right)}{\sigma^2_{x_{t}}}\right\} 
\end{align}
and
\begin{align}
\label{eqn16: covariance of full conditional of beta_g} 
&\bi{\Sigma}_{\beta_g} = V[\bi{\beta}_g|\cdots] = \left\{ \bi{\Sigma}_{\beta_{g},0}^{-1} +\frac{1}{\sigma^2_g} [
\bi{H}_{D_{z}}',\bi{h}(1,x_0)]\bi{A}^{-1}_{D_z,g^*(1,x_0)}[\bi{H}_{D_{z}}',\bi{h}(1,x_0)]'\right.
\notag
\\[1ex]
& \left. 
+ \sum_{t=1}^{T} \frac{\left(\bi{H}_{D_{z}}'\bi{A}^{-1}_{g,D_z} \bi{s}_{g,D_z}(t+1,x_{t})-\bi{h}(t+1,x_{t})\right)
\left(\bi{H}_{D_{z}}'\bi{A}^{-1}_{g,D_z} \bi{s}_{g,D_z}(t+1,x_{t})-\bi{h}(t+1,x_{t})\right)'}{\sigma^2_{x_t}}\right\}^{-1}.
\end{align}
Hence $[\bi{\beta}_g|\cdots]$ follows a four-variate normal distribution with mean and variance 
$\mu_{\beta_g}$ and $\bi{\Sigma}_{\beta_g}$, respectively. Therefore, we update $\bi{\beta}_g$ by Gibbs sampling. 
%
%
\section{\small Full conditional density of $\sigma^2_{\mbox{\scriptsize $\eta$}}$}
\begin{align}
\label{eqn17:Full_conditional_density_of_sigma_eta}
[\sigma^2_{\mbox{\scriptsize $\eta$}}|\cdots] &\propto 
[\sigma^2_{\mbox{\scriptsize $\eta$}}] [x_1|g^*(1,x_0),\sigma^2_{\mbox{\scriptsize $\eta$}},K_1] 
[K_1|g^*(1,x_0),\sigma^2_{\mbox{\scriptsize $\eta$}}] \prod_{t=2}^{T+1} 
[x_t|\bi{\beta}_g,\sigma^2_{\mbox{\scriptsize $\eta$}},\sigma^2_g, \bi{D}_z,x_{t-1},K_t]\notag
\\
&\quad ~ \times\prod_{t=2}^{T+1}  [K_t|\bi{\beta}_g,\sigma^2_g, \sigma^2_{\mbox{\scriptsize $\eta$}},\bi{D}_z,x_{t-1}] \notag
\\
&\propto [\sigma^2_{\mbox{\scriptsize $\eta$}}] 
\exp{\left\{-\sum_{i=2}^{T+1}\frac{1}{2\sigma^2_{x_{t}}}(x_t+2\pi K_t-\mu_{x_{t}})^2\right\}} 
\exp{\left\{ -\frac{1}{2\sigma^2_{\mbox{\scriptsize{$\eta$}}}}(x_1+2\pi K_1 - g^*)^2\right\}}.
\end{align}
The right hand side of (\ref{eqn17:Full_conditional_density_of_sigma_eta}) does not have a closed form and hence we update 
$\sigma^2_{\mbox{\scriptsize $\eta$}}$ using a random walk Metropolis-Hastings step. 

\section{\small Full conditional density of $\sigma_g^2$}
\begin{align}
\label{eqn18:full conditional for sigma^2_g}
[\sigma^2_{g}|\cdots] &\propto [\sigma^2_{g}] [\bi{D}_z,g^*(1,x_0)|x_0,\bi{\beta}_g,\sigma^2_g] 
\prod_{t=2}^{T+1} [x_t|\bi{\beta}_g,\sigma^2_{\mbox{\scriptsize $\eta$}},\sigma^2_{g},\bi{D}_z,x_{t-1},K_t] \prod_{t=2}^{T+1} \left[ K_t|\bi{\beta}_g,\sigma^2_g, \right.
\notag
\\
&\quad ~ 
\left. \sigma^2_{\mbox{\scriptsize $\eta$}},\bi{D}_z,x_{t-1}\right].
\end{align}
It is not possible to obtain a closed form of the right hand side of above equation and therefore, 
to update $\sigma^2_{g}$, we consider a random walk Metropolis-Hastings step. 
\section{\small Full conditional distribution of $x_0$}
\begin{align}
\label{eqn19:full conditional for x_0}
[x_0|\cdots] \propto [x_0][\bi{D}_z,g^*(1,x_0)|x_0,\bi{\beta}_g,\sigma^2_g].
\end{align}
The closed form of right hand side of (\ref{eqn19:full conditional for x_0}) is not available. So, 
we update $x_{0}$ using a random walk Metropolis-Hastings step. 
\section{\small Full conditional density of $g^*(1,x_0)$}
\begin{align}
\label{eqn20:full conditional of g*}
[g^*(1,x_0)|\cdots] &\propto [g^*(1,x_0)|x_0,\bi{\beta}_g,\sigma^2_g]
[\bi{D}_z|g^*(1,x_0),x_0,\bi{\beta}_g,\sigma^2_g] [x_1|g^*(1,x_0),x_0,\sigma^2_{\mbox{\scriptsize $\eta$}},K_1] \notag
\\
&\qquad  \times[K_1|g^*(1,x_0),\sigma^2_{\mbox{\scriptsize $\eta$}}] \notag
\\
&\propto [g^*(1,x_0)|x_0,\bi{\beta}_g][\bi{D}_z|g^*(1,x_0),x_0,\bi{\beta}_g] 
\exp{\left\{-\frac{1}{2\sigma^2_{\mbox{\scriptsize{$\eta$}}}} (x_1+2\pi K_1- g^*)^2\right\}} \notag
\\
&\propto \exp{\left\{-\frac{1}{2\gamma_{g}^2}(g^*-\nu_{g})^2\right\}},
\end{align}
where 
\begin{align}
\label{eqn29: mean of full conditional of g*}
\nu_{g}= E[g^*(1,x_0)|\cdots] &= \left\{ \frac{1}{\sigma^2_{\mbox{\scriptsize $\eta$}}} 
+ \frac{1}{\sigma^2_g} (1+\bi{s}_{g,D_z}(1,x_0)'\bi{\Sigma}^{-1}_{g,D_z}\bi{s}_{g,D_z}(1,x_0))\right\}^{-1}\notag
\\[1ex]
&\qquad \left\{\frac{x_1+2\pi K_1}{\sigma_{\mbox{\scriptsize $\eta$}}^2}
+\frac{1}{\sigma^2_g}(\bi{h}(1,x_0)'\bi{\beta}_g+\bi{s}'_{g,D_z}\bi{\Sigma}^{-1}_{g,D_z}\bi{D}_z^* )\right\}
\end{align}
and
\begin{align}
\label{eqn30: variance of full conditional of g*}
\gamma_{g}^2 = V[g^*(1,x_0)|\cdots] =  \left\{ \frac{1}{\sigma^2_{\mbox{\scriptsize $\eta$}}} 
+ \frac{1}{\sigma^2_g}(1+\bi{s}_{g,D_z}(1,x_0)'\bi{\Sigma}^{-1}_{g,D_z}\bi{s}_{g,D_z}(1,x_0))\right\}^{-1},
\end{align}
with
\begin{equation}
\label{eqn31: defn of D_z^*}
\bi{D}_z^*= \bi{D}_z-\bi{H}_{D_{z}}\bi{\beta}_g + \bi{h}(1,x_0)'\bi{\beta}_g\bi{s}_{g,D_{z}}, 
\end{equation}
and
\begin{equation}
\label{eqn32: defn of Sigma_g,Dz}
\bi{\Sigma}_{g,D_z} = \bi{A}_{g,D_{z}} - \bi{s}_{g,D_z}(1,x_0) \bi{s}_{g,D_z}(1,x_0)'.
\end{equation}
Hence $[g^*|\cdots]$ follows a univariate normal distribution with mean $\nu_g$ and variance $\gamma_g$. Thus, 
we update $g^*$ using Gibbs sampling. 
\section{\small Full conditional of $\bi{D}_z$}
\begin{align}
\label{eqn21:full conditional of D_z}
[\bi{D}_z|\cdots] &\propto [\bi{D}_z|g^*(1,x_0),x_0,\bi{\beta}_g,\sigma^2_g] 
\prod_{t=2}^{T+1} [x_t|\bi{\beta}_g,\sigma^2_g, \sigma^2_{\mbox{\scriptsize $\eta$}},\bi{D}_z,x_{t-1},K_t]
\prod_{t=2}^{T+1} \left[ K_t|\bi{\beta}_g,\sigma^2_g, \sigma^2_{\mbox{\scriptsize $\eta$}}, \right.
\notag
\\
&\quad ~ 
\left. \bi{D}_z,x_{t-1}\right]
\end{align}
After simplification it turns out that the full conditional distribution of $\bi{D}_z$ is an $n$-variate normal with mean
\begin{align}
\label{eqn27:mean of full conditional of Dz}
E(\bi{D}_z|\cdots) &= \left\{\frac{\bi{\Sigma}_{g,D_z}^{-1}}{\sigma^2_g} +
\bi{A}_{g,D_z}^{-1} \left(\sum_{t=1}^{T} \frac{s_{g,D_z}(t+1,x_{t})s'_{g,D_z}(t+1,x_{t})}{\sigma_{x_{t}}^2}\right) 
\bi{A}_{g,D_z}^{-1}\right\}^{-1}\notag
\\[1ex]
\times 
&\,\left\{ \frac{\bi{\Sigma}_{g,D_z}^{-1}\bi{\mu}_{g,D_z}}{\sigma^2_g} + \bi{A}_{g,D_z}^{-1} \right.
\notag
\\[1ex]
& \left. \quad \sum_{t=1}^{T} \frac{s_{g,D_z}(t+1,x_{t}) \{x_{t+1}+2\pi K_{t+1} - \bi{\beta}'_{g} 
(\bi{h}(1,t+1,x_t)-\bi{H}'_{D_z}\bi{A}_{g,D_z}^{-1}s_{g,D_z}(t+1,x_{t}))\}}{\sigma_{x_{t}^2}}\right\}
\end{align}
and covariance matrix
\begin{align}
\label{eqn28:var of full conditional of Dz}
V(\bi{D}_z|\cdots) =  \left\{\frac{\bi{\Sigma}_{g,D_z}^{-1}}{\sigma^2_g} + 
\bi{A}_{g,D_z}^{-1} \left(\sum_{t=1}^{T} \frac{s_{g,D_z}(t+1,x_{t})s'_{g,D_z}(t+1,x_{t})}{\sigma_{x_{t}}^2}\right) 
\bi{A}_{g,D_z}^{-1}\right\}^{-1}.
\end{align}
%
We update $\bi{D}_z$ using the Gibbs sampling technique. 
\section{\small Full conditional of $x_1$}
\begin{align}
\label{eqn22:full conditional of x_1}
[x_1|\cdots] &\propto [x_1|g^*(1,x_0),\sigma^2_{\mbox{\scriptsize $\eta$}}] 
[\bi{f}_{\mbox{\scriptsize $D_{T}$}}^*|x_1,\ldots,x_{T},\bi{\beta}_f,\sigma_f] [N_{1}|f^*(1,x_1),\sigma_{\epsilon},x_1] 
\notag
\\
&\qquad [y_1|N_1,f^*(1,x_1),\sigma_{\epsilon},x_1] [x_2|\bi{\beta}_g,\sigma^2_g, \sigma^2_{\mbox{\scriptsize $\eta$}},\bi{D}_z,x_1,K_2] [K_2|\bi{\beta}_g,\sigma^2_g,\sigma^2_{\mbox{\scriptsize $\eta$}},\bi{D}_z,x_{1}] \notag
\\
& \propto \frac{1}{\sqrt{2\pi}\sigma_{\mbox{\scriptsize $\eta$}}}\exp\left(-\frac{1}{2\sigma^2_{\mbox{\scriptsize $\eta$}}}
(x_1+2\pi K_1-g^*)^2\right)I_{[0,2\pi]}(x_1) \notag
\\
& \times\exp\left\{-\frac{1}{2}(\bi{f}_{\mbox{\scriptsize $D_{T}$}}^*-\bi{H}_{\mbox{\scriptsize $D_{T}$}}\bi{\beta}_f)'\sigma^{-2}_{f}\bi{A}_{f,D_T}^{-1}(\bi{f}_{\mbox{\scriptsize $D_{T}$}}^*-\bi{H}_{\mbox{\scriptsize $D_{T}$}}\bi{\beta}_f)-\frac{1}{2\sigma^2_{\epsilon}} (y_1+2\pi N_1 - f^*(1,x_1))^2\right\} \notag
\\
& \quad \times\frac{1}{\sqrt{2\pi}\sigma_{x_{2}}}\exp\left(-\frac{1}{2\sigma^2_{x_{2}}}(x_2+2\pi K_2-\mu_{x_{2}})^2 \right),
\end{align}
where 
$\mu_{x_{2}} = \bi{h}(2,x_{1})'\bi{\beta}_g + \bi{s}_{g,D_z}((2,x_{1}))' 
\bi{A}_{g,D_{z}}^{-1} (\bi{D}_z-\bi{H}_{D_{z}}\bi{\beta}_{g})$ and \allowdisplaybreaks
$\sigma^2_{x_{2}} = \sigma^2_{\mbox{\scriptsize{$\eta$}}} + 
{\sigma_g^2}(1-(\bi{s}_{g,D_z}(2,x_{1}))' \bi{A}_{g,D_{z}}^{-1}\\ \bi{s}_{g,D_z} (2,x_{1}))$. However, 
the closed form of the above expression is not tractable. So, we update $x_1$ with a random walk Metropolis-Hastings step.

\section{\small Full conditional of $x_{t+1}$, for $t=1,\ldots,T-1$}
\begin{align}
\label{eqn24:full conditional for x_{t+1}}
[x_{t+1}|\cdots] & \propto [x_{t+1}|\bi{\beta}_g,\sigma^2_g, \sigma^2_{\mbox{\scriptsize $\eta$}},\bi{D}_z,x_{t}] [x_{t+2}|\bi{\beta}_g,\sigma^2_g, \sigma^2_{\mbox{\scriptsize $\eta$}},\bi{D}_z,x_{t+1},K_{t+2}]
\left[K_{t+2}|\bi{\beta}_g,\sigma^2_g, \sigma^2_{\mbox{\scriptsize $\eta$}},
\right. \notag
\\
&\qquad
\left. \bi{D}_z, x_{t+1}\right] [\bi{f}_{\mbox{\scriptsize $D_{T}$}}^*|x_1,\ldots,x_{T},\bi{\beta}_f,\sigma_f] [N_{t+1}|f^*(t+1,x_{t+1}),\sigma_{\epsilon},x_{t+1}] 
\notag
\\
&\qquad \times[y_{t+1}|N_{t+1},f^*(t+1,x_{t+1}),\sigma_{\epsilon},x_{t+1}] \notag
\\
& \propto \frac{1}{\sqrt{2\pi}\sigma_{x_{t+1}}}\exp\left(-\frac{1}{2\sigma^2_{x_{t+1}}}
(x_{t+1}+2\pi K_{t+1}-\mu_{x_{t+1}})^2 \right)I_{[0,2\pi]}(x_{t+1})\notag
\\
&\qquad \times\frac{1}{\sqrt{2\pi}\sigma_{x_{t+2}}}\exp\left(-\frac{1}{2\sigma^2_{x_{t+2}}}(x_{t+2}+2\pi K_{t+2}-\mu_{x_{t+2}})^2 \right)\notag
\\
&\qquad \times\exp\left\{-\frac{1}{2}(\bi{f}_{\mbox{\scriptsize $D_{T}$}}^*-\bi{H}_{\mbox{\scriptsize $D_{T}$}}\bi{\beta}_f)'\sigma^{-2}_{f}\bi{A}_{f,D_{T}}^{-1}(\bi{f}_{\mbox{\scriptsize $D_{T}$}}^*-\bi{H}_{\mbox{\scriptsize $D_{T}$}}\bi{\beta}_f)-\right. \notag
\\
&\qquad \left. \frac{1}{2\sigma^2_{\epsilon}} (y_{t+1}+2\pi N_{t+1} - f^*(t+1,x_{t+1}))^2\right\}
\end{align}
The expression of the full conditional distribution of $x_{t+1}$, $t=1,\ldots T-1$, are not tractable 
and therefore, we update $x_{t+1}$, $t=1,\ldots T-1$, using random walk Metropolis-Hastings steps.

\section{\small Full conditional of $x_{T+1}$}
\begin{align}
\label{eqn23:full conditional for x_T+1}
[x_{T+1}|\cdots] &\propto  [x_{T+1}|\bi{\beta}_g,\sigma^2_g, \sigma^2_{\mbox{\scriptsize $\eta$}},\bi{D}_z,x_T,K_{T+1}] [f^*(T+1,x_{T+1})|\bi{f}_{\mbox{\scriptsize $D_{T}$}}^*, x_{T+1}, \bi{\beta}_f,\sigma_f] \notag
\\
& \propto \frac{\frac{1}{\sqrt{2\pi}\sigma_{x_{T+1}}}
\exp\left(-\frac{1}{2\sigma^2_{x_{T+1}}}(x_{T+1}+2\pi K_{T+1}-\mu_{x_{T+1}})^2 \right)I_{[0,2\pi]}(x_{T+1})}
{\Phi\left(\frac{2\pi (K_{T+1}+1)-\mu_{x_{T+1}}}{\sigma_{x_{T+1}}}\right)
-\Phi\left(\frac{2\pi K_{T+1}-\mu_{x_{T+1}}}{\sigma_{x_{T+1}}}\right)} \notag
\\
&\qquad \times\frac{1}{\sqrt{2\pi}\sigma_{\mbox{\scriptsize $f^*(T+1,x_{T+1})$}}} \exp\left(-\frac{1}{2\sigma_{\mbox{\scriptsize $f^*(T+1,x_{T+1})$}}^2} (f^*(T+1,x_{T+1})-\mu_{\mbox{\scriptsize $f^*(T+1,x_{T+1})$}})^2\right) \notag
\\
&\propto \exp\left(-\frac{1}{2\sigma^2_{x_{T+1}}}(x_{T+1}+2\pi K_{T+1}-\mu_{x_{T+1}})^2 \right)I_{[0,2\pi]}(x_{T+1}) \notag
\\
&\qquad\times\frac{1}{\sqrt{2\pi}\sigma_{\mbox{\scriptsize $f^*(T+1,x_{T+1})$}}} \exp\left(-\frac{1}{2\sigma_{\mbox{\scriptsize $f^*(T+1,x_{T+1})$}}^2} (f^*(T+1,x_{T+1})-\mu_{\mbox{\scriptsize $f^*(T+1,x_{T+1})$}})^2\right).
\end{align}
We note that $\mu_{x_{T+1}}$ and $\sigma^2_{x_{T+1}}$ do not depend upon $x_{T+1}$ but 
$\mu_{\mbox{\scriptsize $f^*(T+1,x_{T+1})$}}$ and $\sigma_{\mbox{\scriptsize $f^*(T+1,x_{T+1})$}}$ do depend upon 
$x_{T+1}$. However, in any case the right hand side of the above equation is not tractable. Therefore, we update 
$x_{T+1}$ as well using a random walk Metropolis-Hastings step. 

\section{\small Full conditional density of $K_1$}
\begin{align}
\label{eqn25:full conditional of K_1}
 [K_1|\cdots]  &\propto [K_1|g^*(1,x_0),\sigma^2_{\mbox{\scriptsize $\eta$}}] 
 [x_1|g^*(1,x_0),\bi{\beta}_g,\sigma^2_{\mbox{\scriptsize $\eta$}},K_1] \notag
 \\
 &\propto \frac{1}{\sqrt{2\pi}\sigma_{\mbox{\scriptsize $\eta$}}}
\exp\left(-\frac{1}{2\sigma^2_{\mbox{\scriptsize $\eta$}}}(x_1+2\pi K_1-g^*)^2\right) I_{\{\ldots,-1,0,1,\ldots\}}(K_1).
\end{align}
The full conditional distribution of $K_{1}$ is not tractable and hence we update $K_1$ using a 
random walk Metropolis-Hastings step. 
\section{\small Full conditional density of $K_{t}$}
\begin{align}
\label{eqn26:full conditional of K_ts}
 [K_t|\cdots] & \propto [x_t|\bi{\beta}_g,\sigma^2_{\mbox{\scriptsize $\eta$}},\bi{D}_z,x_{t-1},K_t] 
 [K_t|\bi{\beta}_g,\sigma^2_{\mbox{\scriptsize $\eta$}},\bi{D}_z,x_{t-1}],  \notag
 \\
 & \propto
\exp\left(-\frac{1}{2\sigma^2_{x_{t}}}(x_t+2\pi K_t-\mu_{x_{t}})^2 \right)I_{\{\ldots,-1,0,1,\ldots\}}(K_t), ~ t=2,\ldots,T+1. 
\end{align}
As in the case of $K_{1}$, here also closed forms of the full conditional distributions of $K_t$, $t=2,\ldots, T+1$ 
are not available, and hence we update $K_t$, $t=2,\ldots, T+1$, using random walk Metropolis-Hastings steps.

\end{document}